\documentclass{article}
\usepackage{arxiv}

\usepackage[utf8]{inputenc}
\usepackage{soul}
\usepackage{cancel}
\usepackage{standalone}
\usepackage{booktabs}
\usepackage{multirow}
\usepackage[dvipsnames]{xcolor}

\usepackage{tikz}
\usetikzlibrary{tikzmark}
\usepackage{siunitx} 
\usepackage{physics}
\usepackage{float}
\usepackage{tabularray}
\usepackage{graphicx}

\usepackage{amsmath, amssymb}
\usepackage{bbold} 
\usepackage{scalerel} 

\DeclareMathOperator{\interior}{int}
\DeclareMathOperator{\diag}{diag}

\DeclareMathOperator*{\argmax}{arg\,max}
\DeclareMathOperator*{\assembly}
{\scalerel*{\mathbb{A}}{\sum}} 

\usepackage{bm}
\newcommand{\bs}[1] {\bm{\mathit{#1}}} 

\usepackage{subcaption}
\usepackage[version=4]{mhchem}

\usepackage{multirow}

\definecolor{clear}{HTML}{ebebeb}
\definecolor{selected}{HTML}{bcbec0}
\definecolor{darker}{HTML}{d6d6d6}
\definecolor{redline}{HTML}{be1e2d}
\definecolor{myred}{HTML}{c1272d}
\definecolor{myblue}{HTML}{0071bc}

\graphicspath{{./figures/}{./final_designs/}}

\makeatletter
\def\input@path{{./figures/}}
\makeatother

\usetikzlibrary{%
calc,hobby,positioning,patterns,decorations.pathreplacing,arrows.meta
}

\usepackage{pgfplots}
\usepackage{pgfplotstable}
\usepgfplotslibrary{groupplots,fillbetween,colorbrewer}
\pgfplotsset{compat=1.18}

\newsavebox{\imagebox}
\usepackage{arydshln}

\makeatletter
  \renewcommand*\env@matrix[1][*\c@MaxMatrixCols c]{%
    \hskip -\arraycolsep
    \let\@ifnextchar\new@ifnextchar
  \array{#1}}
\makeatother

\title{Dissipation Dilution-Driven Topology Optimization for Maximizing the $Q$ Factor of Nanomechanical Resonators}

\usepackage{hyperref}

\author{\href{https://orcid.org/0009-0001-5955-8856}{\includegraphics[scale=0.06]{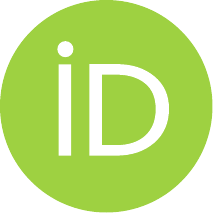}}\hspace{1mm}Hendrik J.~Algra
	 \And
	 \href{https://orcid.org/0000-0002-3625-7081}{\includegraphics[scale=0.06]{orcid.pdf}}\hspace{1mm}Zichao Li
	 \And
	 \href{https://orcid.org/0000-0003-2106-2246}{\includegraphics[scale=0.06]{orcid.pdf}}\hspace{1mm}Matthijs Langelaar
	 \And
	 \href{https://orcid.org/0000-0002-7439-8483}{\includegraphics[scale=0.06]{orcid.pdf}}\hspace{1mm}Farbod Alijani
	 \And
	 \href{https://orcid.org/0000-0003-2275-6207}{\includegraphics[scale=0.06]{orcid.pdf}\hspace{1mm}}Alejandro M.~Aragón\thanks{Corresponding author. E-mail address: \texttt{\href{mailto:a.m.aragon@tudelft.nl}{a.m.aragon@tudelft.nl}} (A.M.~Aragón)}
	 \\
	 \AND
	 \\[-1em] 
	 Faculty of Mechanical Engineering, Delft University of Technology\\
	 Mekelweg 2, 2628 CD, Delft, Zuid-Holland,
	 The Netherlands\\
}

\renewcommand{\shorttitle}{Dissipation Dilution-Driven TO for Maximizing the $Q$ Factor of Resonators}

\begin{document}
\maketitle

\begin{abstract}
The quality factor ($Q$ factor) of nanomechanical resonators is influenced by geometry and stress, a phenomenon called dissipation dilution. Studies have explored maximizing this effect, leading to softly-clamped resonator designs. This paper proposes a topology optimization methodology to design two-dimensional nanomechanical resonators with high $Q$ factors by maximizing dissipation dilution. A formulation based on the ratio of geometrically nonlinear to linear modal stiffnesses of a prestressed finite element model is used, with its corresponding adjoint sensitivity analysis formulation. Systematic design in square domains yields geometries with comparable $Q$ factors to literature. We analyze the trade-offs between resonance frequency and quality factor, and how these are reflected in the geometry of resonators. We further apply the methodology to optimize a resonator on a full hexagonal domain. By using the entire mesh---i.e., without assuming any symmetries---we find that the optimizer converges to a two-axis symmetric design comprised of four tethers.
\end{abstract}


\section{Introduction}
Nanomechanical resonators are ubiquitous in modern electronics and sensing devices, as they function as extremely precise time-keeping oscillators~\cite{nguyuen07} and excellent force~\cite{mamin01} and mass~\cite{chaste12, burg07} sensors. They are also used for a variety of other applications, which include conducting optomechanical quantum experiments~\cite{norte16, oconnell10}, investigating fundamental nonlinear dynamic effects ~\cite{zichao24} and studying electric and magnetic phase transitions~\cite{steeneken20, ata23}. In many of these applications, a limiting factor to their usefulness is the influence of thermal noise, which results in a random force on the oscillating body~\cite{fedorov21}. This thermal noise force limits the precision in detecting quantities such as forces, masses, or pressures~\cite{shin21}. According to the fluctuation-dissipation theorem, the influence of thermal noise increases with resonator mass, temperature, and dissipation rate~\cite{fedorov21}. Apart from reducing the mass and operation temperature of these oscillators, minimizing the amount of energy they dissipate is a critical aspect when it comes to mitigating the effect of thermal noise~\cite{fedorov21,zichao23}.

An important figure of merit commonly used to quantify the performance of nanomechanical resonators is the quality factor, or simply the $Q$ factor. 
A high $Q$ factor implies a low dissipation rate, which enables the oscillator to function at low power~\cite{zichao23}. Additionally, it results in a sharper resonance peak, which allows the resonance frequency of the oscillator to be determined more precisely~\cite{farbod23}.
Resonators with higher $Q$ factors can be achieved through a process known as \textit{dissipation dilution}~\cite{schmid11}---i.e., by reducing the effective energy dissipation (or internal friction) in a system. This is achieved by redistributing the stored energy, such as by tensioning components, to minimize the relative contribution of material losses, effectively ``diluting'' the dissipation~\cite{pratt22}.
Over the years, researchers have developed resonators with increasingly higher dissipation dilution, employing geometric designs specifically tailored to minimize the effects of specific dissipation mechanisms.
One notable example includes hierarchical fractal-shaped resonators~\cite{fedorov20, beccari21}, which aim to reduce the gradient of the fundamental mode near the clamping points in order to decrease bending losses. Furthermore, integrating phononic crystals (PnCs) into resonator geometry has enabled researchers to isolate a high-order vibration mode (beyond the fundamental mode) in a central defect---a deliberate disruption in the periodic structure of the PnC within an otherwise regular lattice---thereby strongly reducing its acoustic radiation losses. This technique has been successfully implemented for both string-like resonators~\cite{fedorov19, beccari22, ghadimi17, gisler22} and 2-D membrane structures~\cite{tsaturyan17}.

An important factor considered in the design of all resonators is the presence of pretension in the system, which is used to increase the energy stored in the structure, and enhance overall $Q$ factor. The dependency of the $Q$ factor on geometry and stress has encouraged the use of (structural) optimization techniques to design high-$Q$ resonators. An example is the so-called ``spiderweb'' resonator~\cite{shin21}, which was realized by optimizing a parameterized layout using Bayesian optimization. However, Bayesian optimization demands vast computational resources as it needs to invert a dense covariant matrix, thereby only being useful to optimize in low-dimensional search spaces (six variables in the spiderweb design, including the length, width, and number of tethers). Topology optimization (TO)~\cite{sigmund2004} overcomes this limitation as it has been used in studies with over a billion design parameters~\cite{aage17}. A recent study successfully applied TO for the design of nanomechanical resonators, showing promising numerical~\cite{gao20} and experimental~\cite{hoj21article} results. 
The studies in question used an empirical model based on both bending and boundary losses to estimate the $Q$ factor from experimental data.

In this work we propose a topology optimization procedure to design high-$Q$ resonators without relying on prior knowledge such as empirical models or predefined parameterized structures. This is achieved by computing the dissipation dilution simply as the ratio between geometrically nonlinear and linear modal stiffnesses. An adjoint sensitivity formulation is derived accordingly. The methodology is then used to design a range of square resonators by taking advantage of symmetry and targeting different fundamental mode frequencies. Based on the results, we discuss the effects the nonlinear and linear modal stiffnesses exert on the $Q$ factor. We draw attention to challenges inherent in the optimization problem itself. Specifically, we discuss the emergence of void regions under compressive forces and the non-convex objective function landscape, whereby the optimizer is likely to discover (suboptimal) local minima, further hindering the design process in pursuit of optimal resonator topologies.

We also apply our methodology to the optimization of a resonator on a hexagonal computational domain. Such geometry is fascinating for a number of reasons. Hexagons not only optimize the tessellation of material in a plane (they have the smallest perimeter for a given area), but also distribute forces evenly across their sides, which makes them highly resistant to deformation. Therefore, the tessellated hexagon is the structure chosen by bees for their honeycombs. Hexagons also arise in nature, for instance in the atomic structure of graphene, which makes it incredibly strong and stable. Inspired by these aspects, and also by the high-Q factor resonators attained in a defect of a phononic crystal with a hexagonal lattice structure~\cite{tsaturyan17}, we optimize on a hexagonal computational domain using an entire finite element mesh---i.e.,  without assuming any symmetry. Noteworthy, the optimizer still converges to a two-axis symmetric design comprised of four tethers that connect to the central pad.

\section{Finite element evaluation of dissipation dilution}
\label{sec:dissipation_dilution}
%
For a resonator under mechanical stress, the $Q$ factor is defined as
\begin{equation}\label{eq:DQ defintion}
    Q = D_Q \, Q_0 \ , 
\end{equation}

where $Q_0$ represents the \textit{intrinsic $Q$ factor} of the unstressed system (limited by intrinsic material damping) and $D_Q$ is the \textit{dilution factor}, which quantifies the extent of dissipation dilution due to additional stress. In general, $Q_0$ can be treated as a constant, depending on the material properties and thickness of the membrane. On the other hand, the dilution factor strongly depends on geometry, as Schmid et al.~\cite{schmid11} showed that it could be approximated as $D_Q \approx W_{t}/W_{b}$, where $W_t$ and $W_b$ denote the tension and bending energies, respectively. This ratio can be computed as \cite{shin21, fedorov19}
\begin{equation}\label{eq:comsolQ}
    D_Q \approx\frac{W_{t}}{W_{b}} = \frac{12(1-\nu^2)}{Eh^2}  \frac{\int \alpha \dd{\varOmega}}{\int \beta \dd{\varOmega}}, 
\end{equation}
where $E$ and $\nu$ are the material's Young's modulus and Poisson's ratio, respectively. In Cartesian coordinates, the quantities $\alpha$ and $\beta$ are quantities integrated over the surface of the structure defined as
\begin{equation} \label{eq:alpha_beta}
    \begin{split}
        \alpha &=  \sigma_{xx} \left(\frac{\partial w}{\partial x}\right)^2+\sigma_{yy} \left(\frac{\partial w}{\partial y}\right)^2+2\sigma_{xy} \frac{\partial w}{\partial x}\frac{\partial w}{\partial y}\ , \\
        \beta &=  \left(\frac{\partial^2 w}{\partial x^2}\right)^2 + \left(\frac{\partial^2 w}{\partial y ^2}\right)^2 + 2 \nu \frac{\partial^2 w}{\partial x ^2} \frac{\partial^2 w}{\partial y ^2} + 2 \left(1-\nu\right)\left(\frac{\partial^2 w}{\partial x \partial y}\right)^2,
    \end{split}
\end{equation}
where $w$ denotes the transverse displacement of the resonator, and $\sigma_{xx}, \sigma_{yy}$, and $\sigma_{xy}$ are the components of the Cauchy's stress tensor.
In~\eqref{eq:alpha_beta} $\alpha$ is proportional to the elastic energy stored in tension, while $\beta$ is proportional to the bending energy.
Although experiments have shown that this equation approximates the effects of dissipation dilution~\cite{zichao23}, and the equation has successfully been used as an objective function in conjunction with Bayesian optimization~\cite{shin21}, its use as an objective for TO is not straightforward. This is due to the intrinsic gradient-based nature of TO, which requires the calculation of the sensitivities of the objective function---in this case Eq.~\eqref{eq:comsolQ}---with respect to changes in the element-level density design variables.

We propose an alternative way to write~\eqref{eq:comsolQ}~\cite{thesishendrik}, which is more appealing for finite element analysis (FEA). For a geometrically nonlinear finite element model of a resonator, the energy stored in tension (low-loss) for a particular vibration mode $\bs{\Phi}$---not necessarily the fundamental resonance mode---is proportional to $\bs{\Phi} ^{\intercal} \widetilde{\bs{K}} \bs{\Phi}$, where $\widetilde{\bs{K}}$ is the geometrically nonlinear stiffness matrix. Similarly, the energy stored in bending (lossy) for the same mode is $\bs{\Phi} ^{\intercal} \overline{\bs{K}} \bs{\Phi}$, where $\overline{\bs{K}}$ the linear stiffness matrix. We therefore approximate the dilution factor the ratio between geometrically nonlinear and linear modal stiffnesses:
\begin{equation}\label{eq:DQ objective}
    D_Q \approx\frac{W_{t}}{W_{b}} = \frac{\bs{\Phi} ^{\intercal} \widetilde{\bs{K}} \bs{\Phi} }{\bs{\Phi}^{\intercal} \overline{\bs{K}} \bs{\Phi}}.
\end{equation}
This equation, which is purely based on the stiffness matrices and mode shapes of the prestressed eigenvalue problem, enables a straightforward computation of sensitivity information at element level (i.e., on arrays computed per element).
While~\eqref{eq:DQ objective} is general and thus it could be applied to any resonance mode, herein we focus on optimizing the fundamental mode of vibration. One reason for this is that in density-based topology optimization, higher-order modes can become occluded by the appearance of intermediate spurious modes in low-density regions---a phenomenon that is further discussed in Section~\ref{sec:material interpolation}.

\begin{figure}[b]
	\centering
	\includegraphics{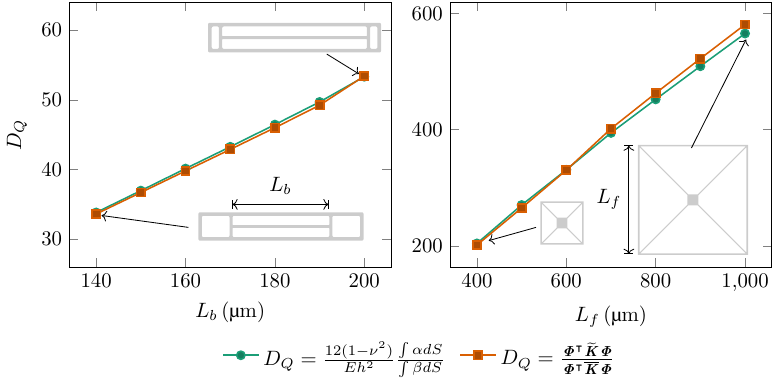}
  \caption{Dissipation dilution computed by Eqs.~\eqref{eq:comsolQ} and~\eqref{eq:DQ objective} as a function of geometric parameters that define the geometry of the resonators. (left) H-beam resonators by Li et al.~\cite{zichao23}, for which the length of the central beam $L_b$ is varied; (right) Trampoline resonators by Norte et al.~\cite{norte16}, for which the length of the outer frame $L_f$ is varied.}
  \label{fig:DQ comparison}
\end{figure}

We first test the behavior of \eqref{eq:DQ objective} on two well-known resonator geometries~\cite{thesishendrik}, namely H-beam resonators by Li  et al.~\cite{zichao23} and trampoline-like resonators by Norte et al.~\cite{norte16}.
The results are summarized in Figure~\ref{fig:DQ comparison}, which compares Eqs.~\eqref{eq:comsolQ} and~\eqref{eq:DQ objective} across various geometric parameters---$L_b$ for H-beam resonators and $L_f$ for trampoline resonators.
A thickness of $\SI{340}{\nano\metre}$ and a pretension of $\SI{1.1}{\giga\pascal}$ were used for all silicon nitride (\ce{Si3N4}) resonators, following the work of Li et al.~\cite{zichao23}.

Figure~\ref{fig:DQ comparison} demonstrates that the numerical values of these two equations are remarkably similar, thereby verifying the application of~\eqref{eq:DQ objective} to model dissipation dilution. Consequently, we use this equation as the objective function in a topology optimization study aimed at designing resonators with high $Q$ factors.

\section{Problem description and analysis formulation}\label{sec:prestressed modelling}
\begin{figure}[b]
	\centering
	\includegraphics{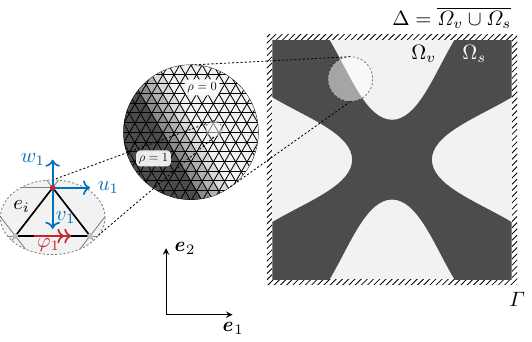}
	\caption{Schematic representation of a finite element model based on shell elements, used in combination with a generic density-based topology optimization model. The latter prescribes a (density) field of values between 0 and 1, to indicate material and void areas respectively. A shape represented by this density field is shown on the right, with a cutout showing the individual elements and the corresponding density field that defines the shape. In this work, the element formulation by Van Keulen \cite{keulen93} was used, for which the degrees of freedom corresponding to a single node can be found in the cutout on the left.
	}
	\label{fig:resonator_schematic}
\end{figure}

High-$Q$ nanomechanical resonators are usually realized from \ce{Si3N4} membranes grown on silicon substrate, for a very high intrinsic in-plane prestress (around $\SI{1}{\giga\pascal} $ or even more)~\cite{zichao23}.
Consider in Figure~\ref{fig:resonator_schematic} the schematic representation of a resonator. Without loss of generality, consider the square prism $\Delta \times h  \in \mathbb{R}^3$, where $\Delta \equiv l \times l$ is the square base and $h$ is the thickness that is considered constant henceforth. We are thus interested in solving a 2-D topology optimization problem, whereby the objective is to find the optimized placement of the resonator phase $\varOmega_s$ (or alternatively the void phase $\varOmega_v$) such that $ \Delta = \overline{\varOmega_s \cup \varOmega_v} $ and $\varOmega_s \cap \varOmega_v = \varnothing $ (the phases are mutually exclusive). With a chosen Cartesian basis $\left\{ \bs{e}_1, \bs{e}_2 \right\}$, for each spatial coordinate $\bs{x} \in \Delta$, we use an indicator function $\iota \left( \bs{x} \right)$ to determine whether there is material or void, i.e., $\iota \left( \bs{x} \right) = 1$ for $\bs{x} \in \varOmega_s$ and $\iota \left( \bs{x} \right) = 0$ for $\bs{x} \in \varOmega_v$.  Therefore, we can write any integral of a function $f \left( \bs{x} \right) $ over the resonator domain as $ \int_{\varOmega_s} f \left( \bs{x} \right) \dd{\varOmega} \equiv \int_{\Delta} \iota \left( \bs{x} \right)  f \left( \bs{x} \right) \dd{\varOmega} $. We omit the dependence of the indicator function on position henceforth for brevity.

The boundary of the domain is $\partial \Delta \equiv \varGamma = \overline{\Delta} \setminus \Delta$. Similarly, resonator and void boundaries are $\varGamma_s = \overline{\varOmega_s} \setminus \varOmega_s$ and $\varGamma_v = \overline{\varOmega_v} \setminus \varOmega_v$, respectively. Homogeneous Dirichlet boundary conditions on the displacement field are prescribed throughout this boundary, so we are only interested in pure Dirichlet problems (i.e., $\varGamma \equiv \varGamma^D$). We note, however, that the domain boundary may contain both resonator and void.

As the in-plane dimensions of the resonators are orders of magnitude larger than their thickness, we use shell theory to model their behavior.
The dynamic equilibrium of the plate is determined by D'Alembert's principle:
\begin{equation}\label{eq:d'Alembert}
	\int_{\Delta} \left( \iota \, N_{\alpha\beta} \ \delta \gamma_{\alpha\beta} + \iota \, M_{\alpha\beta} \ \delta\kappa_{\alpha\beta} \right) \dd{\varOmega} +  \int_{\Delta} \iota \, \sigma_0 \, \delta u_{\alpha} \dd{\varOmega} - \int_{\Delta} \iota \, \rho \, \ddot{u}_i \, \delta u_i \dd{\varOmega} = 0,  
\end{equation}
for sufficiently smooth membrane strain $\delta \gamma_{\alpha\beta}$, curvature $\delta\kappa_{\alpha\beta}$, and displacement $\delta u_i$ variations. In~\eqref{eq:d'Alembert}  $i \in \left\{ 1, 2, 3 \right\}$, $\alpha, \beta \in \left\{ 1, 2 \right\}$, $\sigma_0$ is a constant pretension, $\rho$ is the mass density, and $\ddot{u}_i$ the acceleration; the membrane forces $N_{\alpha\beta}$ and bending moments $M_{\alpha\beta}$ are defined as
\begin{equation}
	N_{\alpha\beta} = \int_{-h/2}^{h/2} \sigma_{\alpha\beta} \dd{z} \text{~~and~~} M_{\alpha\beta} = \int_{-h/2}^{h/2} z \, \sigma_{\alpha\beta} \dd{z},
\end{equation}
respectively.
The middle surface strain $\gamma_{\alpha \beta}$, curvature $\kappa_{\alpha\beta}$, and membrane rotation $\varphi_{\alpha}$ are
\begin{equation} \label{eq:strain_curvature_rotation}
	\gamma_{\alpha \beta} = \frac{1}{2}\left( \pdv{u_{\alpha}}{x_{\beta}} + \pdv{u_{\beta}}{x_{\alpha}} + \pdv{u_i}{u_{\alpha}}\pdv{u_i}{u_{\beta}} \right), \quad  \kappa_{\alpha \beta} = \frac{1}{2}\left( \pdv{\varphi_{\alpha}}{x_\beta} + \pdv{\varphi_{\beta}}{x_{\alpha}} \right),  \text{~~and~~} \varphi_{\alpha} = \pdv{u_3}{x_{\alpha}}.
\end{equation}
In our formulation, Eqs.~\eqref{eq:strain_curvature_rotation} are incorporated into~\eqref{eq:d'Alembert} using Lagrange multipliers, which yields our final variational form~\cite{keulen93}
\begin{equation}\label{eq:variaitonal}
\begin{split}  
	\int_{\Delta} \iota \, \delta & \left\{ N_{\alpha \beta} \left[ \frac{1}{2}\left( \pdv{u_{\alpha}}{x_{\beta}} + \pdv{u_{\beta}}{x_{\alpha}} + \pdv{u_i}{u_{\alpha}}\pdv{u_i}{u_{\beta}} \right) \right]\right\} - \iota \, \delta N_{\alpha \beta} \gamma_{\alpha \beta} + \iota \, \delta \left( M_{\alpha \beta u_3} \right) - \iota \, \delta M_{\alpha \beta} \kappa_{\alpha \beta}  \dd{\varOmega} \\ 
	  & \qquad + \int_{\Delta} \iota \, \sigma_0 \, \delta u_{\alpha} \dd{\varOmega} - \int_{\Delta} \iota \, 
	  \rho \,  \ddot{u}_i \, \delta u_i \dd{\varOmega} = 0. 
\end{split}
\end{equation}

\subsection{Finite element discretization}

We now discretize~\eqref{eq:variaitonal} using finite element spaces, whereby the domain $\Delta$ is subdivided by a set of finite elements $\left\{ e_1, \ldots, e_{n_E} \right\}$ such that $\Delta^h = \interior{\left(  \cup_i^{n_E} \overline e_i \right)}$ and $e_i \cap e_j = \varnothing \; \forall i \neq j$. One such finite element $e_i$ is shown in the inset of Figure~\ref{fig:resonator_schematic}.

We use the triangular shell element formulation proposed by Van Keulen~\cite{keulen93}, which is valid for small strains and moderate rotations only $\left(\varphi^2 \ll 1\right)$.
A triangular element with area $A$ is defined by its coordinates $\left\{ \bs{x}_1, \bs{x}_2, \bs{x}_3 \right\}$. The three side vectors are $\bs{s}_i = a_i \bs{e}_1 + b_i \bs{e}_2$, where $a_i, b_i \in \mathbb{R}$ are geometric scalar parameters.
The lengths of the sides are denoted as $\lambda_i \equiv \norm{\bs{s}_i}$.

While we assume constant rotations along the element sides, the displacement field is linearly interpolated between the corner nodes as 
\begin{equation} \label{eq:interpolation}
  u \left( \bs{x} \right) = \sum_{i=1}^{3} \xi_i \left( \bs{x} \right) \tilde{u}_i, \quad v \left( \bs{x} \right) = \sum_{i=1}^{3} \xi_i \left( \bs{x} \right) \tilde{v}_i, \text{~~and~~} w \left( \bs{x} \right) = \sum_{i=1}^{3} \xi_i \left( \bs{x} \right) \tilde{w}_i,
\end{equation}
where $\xi_i$ is the $i$th Lagrange shape function associated with node $\bs{x}_i$, and $\tilde{u}_i, \tilde{v}_i$ and $\tilde{w}_i$ the corresponding degrees of freedom (DOFs).
The element therefore contains 12 DOFs: Three translational DOFs are associated with the three vertices of the element, and three rotational DOFs associated with the sides (in the left inset of Figure \ref{fig:resonator_schematic}, $ \tilde{u}_1, \tilde{v}_1 $ and $\tilde{w}_1$ for the top vertex $\bs{x}_1$ and the rotation $\varphi_1$ associated with its opposite edge). The DOF vector for the element can thus be written as $\bs{u}_e^\intercal = \begin{bmatrix} \tilde{\bs{u}} & \tilde{\bs{v}} & \tilde{\bs{w}} & \tilde{\bs{\varphi}} \end{bmatrix}$, where $\tilde{\bs{u}}^\intercal = \begin{bmatrix} \tilde{u}_1 & \tilde{u}_2 & \tilde{u}_3 \end{bmatrix}$, and similarly for the other directions, and $\bs{\varphi}$ contains the three rotations multiplied by their corresponding side lengths.

\paragraph{\textbf{Stiffness matrix}} We write the generalized strain tensor as 
\begin{equation}
  \bs{\varepsilon} = \begin{bmatrix}
   \bs{\varepsilon}_m \\
   \bs{\varepsilon}_b  
 \end{bmatrix},
\end{equation}
where $\bs{\varepsilon}_m$ is the \textit{membrane strain} defined (in Voigt notation) as
\begingroup
\renewcommand*{\arraystretch}{1.5}
\begin{equation}\label{eq:membrane_strain_components}
	\bs{\varepsilon}_m = \begin{Bmatrix}
	  \gamma_{11} \\ \gamma_{22} \\ 2 \gamma_{12}
	\end{Bmatrix} =  \left\{ \vphantom{\begin{matrix}
	  \bs{d}_1^\intercal \\ \bs{d}_2^\intercal \\ \bs{d}_1^\intercal \end{matrix}} \right.  \underbrace{ \begin{matrix}
	  \bs{d}_1^\intercal \tilde{\bs{u}} \\
	  \bs{d}_2^\intercal \tilde{\bs{v}} \\
	  \bs{d}_2^\intercal \tilde{\bs{u}} + \bs{d}_1^\intercal \tilde{\bs{v}}
	\end{matrix}}_{\text{linear}}
	\underbrace{
	\begin{matrix}
	  +\frac{1}{2}\left( \tilde{\bs{u}}^\intercal \bs{C}_1 \tilde{\bs{u}} + \tilde{\bs{v}}^\intercal  \bs{C}_1  \tilde{\bs{v}} + \tilde{\bs{w}}^\intercal  \bs{C}_1 \tilde{\bs{w}}  \right) \\
	  +\frac{1}{2}\left( \tilde{\bs{u}}^\intercal  \bs{C}_2  \tilde{\bs{u}} + \tilde{\bs{v}}^\intercal  \bs{C}_2 \tilde{\bs{v}} + \tilde{\bs{w}}^\intercal  \bs{C}_2  \tilde{\bs{w}} \right) \\
	   +\frac{1}{2}\left( \tilde{\bs{u}}^\intercal  \bs{C}_3  \tilde{\bs{u}} + \tilde{\bs{v}}^\intercal  \bs{C}_3 \tilde{\bs{v}} + \tilde{\bs{w}}^\intercal  \bs{C}_3  \tilde{\bs{w}} \right)
	\end{matrix}}_{\text{nonlinear}}
	\left. \vphantom{\begin{matrix}
	  \bs{d}_1^\intercal \\ \bs{d}_2^\intercal \\ \bs{d}_1^\intercal \end{matrix}} \right\},
\end{equation}
\endgroup
with $\bs{d}_1 = \frac{1}{2A} \begin{bmatrix} b_1 & b_2 & b_3 \end{bmatrix}^\intercal$, $\bs{d}_2 = \frac{1}{2A} \begin{bmatrix} a_1 & a_2 & a_3 \end{bmatrix}^\intercal$, $\bs{C}_1 = \bs{d}_1 \bs{d}_1^\intercal$, $\bs{C}_2 = \bs{d}_2 \bs{d}_2^\intercal$, and $\bs{C}_3 = \bs{d}_1 \bs{d}_2^\intercal + \bs{d}_2 \bs{d}_1^\intercal$.

The \textit{bending strain} $\bs{\varepsilon}_b$ is 
\begin{equation}\label{eq:bending_strain_components}
	\bs{\varepsilon}_b = \bs{B}_{bw} \cdot \tilde{\bs{w}} + \bs{B}_{b\varphi} \cdot \tilde{\bs{\varphi}},
\end{equation}
where
\begingroup
\renewcommand*{\arraystretch}{1.5}
\begin{align}
  \bs{B}_{bw} & = \begin{bmatrix}
    \frac{a_3 b_3}{\lambda_3^2} - \frac{a_2 b_2}{\lambda_2^2} & \frac{a_1 b_1}{\lambda_1^2} - \frac{a_3 b_3}{\lambda_3^2} & \frac{a_2 b_2}{\lambda_2^2} - \frac{a_1 b_1}{\lambda_1^2} \\
    \frac{a_2 b_2}{\lambda_2^2} - \frac{a_3 b_3}{\lambda_3^2} & \frac{a_3 b_3}{\lambda_3^2} - \frac{a_1 b_1}{\lambda_1^2} & \frac{a_1 b_1}{\lambda_1^2} - \frac{a_2 b_2}{\lambda_2^2} \\
    \left( \frac{b_2^2}{\lambda_2^2} - \frac{a_2^2}{\lambda_2^2} \right) - \left( \frac{b_3^2}{\lambda_3^2} - \frac{a_3^2}{\lambda_3^2} \right) & %
    \left( \frac{b_3^2}{\lambda_3^2} - \frac{a_3^2}{\lambda_3^2} \right) - \left( \frac{b_1^2}{\lambda_1^2} - \frac{a_1^2}{\lambda_1^2} \right) & %
    \left( \frac{b_1^2}{\lambda_1^2} - \frac{a_1^2}{\lambda_1^2} \right) - \left( \frac{b_2^2}{\lambda_2^2} - \frac{a_2^2}{\lambda_2^2} \right)
  \end{bmatrix},
\intertext{and}
  \bs{B}_{b\varphi} & = \begin{bmatrix}
    -\frac{b_1^2}{\lambda_1^2} & -\frac{b_2^2}{\lambda_2^2} & -\frac{b_3^2}{\lambda_3^2} \\ 
    -\frac{a_1^2}{\lambda_1^2} & -\frac{a_2^2}{\lambda_2^2} & -\frac{a_3^2}{\lambda_3^2} \\ 
    -\frac{2 a_1 b_1}{\lambda_1^2} & -\frac{2 a_2 b_2}{\lambda_2^2} & -\frac{2 a_3 b_3}{\lambda_3^2}
  \end{bmatrix}.
\end{align}
\endgroup

\noindent
The final $6 \times 12$ strain-displacement matrix is written as \\[5pt]
\begin{equation}\label{eq:B nonlinear}
  \bs{B} = \begin{bmatrix}[ccc:c]
     \tikzmark{a} \bs{d}_1^\intercal + \tilde{\bs{u}}^\intercal \bs{C}_1 & \tilde{\bs{v}}^\intercal \bs{C}_1 & \tilde{\bs{w}}^\intercal \bs{C}_1 \tikzmark{b} & \bs{0}_{1 \times 3} \\
    \tilde{\bs{u}}^\intercal \bs{C}_2 & \bs{d}_2^\intercal + \tilde{\bs{v}}^\intercal \bs{C}_2 & \tilde{\bs{w}}^\intercal \bs{C}_2 & \bs{0}_{1 \times 3} \\
    \bs{d}_2^\intercal + \tilde{\bs{u}}^\intercal \bs{C}_3 & \bs{d}_1^\intercal + \tilde{\bs{v}}^\intercal \bs{C}_3 & \tilde{\bs{w}}^\intercal \bs{C}_3 & \bs{0}_{1 \times 3} \\ \hdashline[4pt/4pt]
    \bs{0}_{3 \times 3} & \bs{0}_{3 \times 3} & \bs{B}_{bw} & \bs{B}_{b \varphi}
  \end{bmatrix}.
\end{equation}

\begin{tikzpicture}[remember picture, overlay]
\draw[decorate,decoration={brace}] ([shift={(0ex,2ex)}]{pic cs:a}) -- ($([shift={(0ex,2ex)}]{pic cs:a})!0.5!([shift={(0ex,2ex)}]{pic cs:b})$) node[above] {$\bs{B}_m$} --   ([shift={(0ex,2ex)}]{pic cs:b});
\end{tikzpicture}


\noindent
The generalized stress (also expressed in Voigt notation) is also composed of membrane and bending components, i.e., 
\begin{equation}
  \bs{\sigma} = \begin{bmatrix}
   \bs{\sigma}_m \\
   \bs{\sigma}_b  
 \end{bmatrix} = \bs{S} \bs{\varepsilon} = \begin{bmatrix}
	\bs{S}_{m} & \bs{0}_{3 \times 3} \\
	\bs{0}_{3 \times 3} & \bs{S}_{b}
\end{bmatrix} \begin{bmatrix}
   \bs{\varepsilon}_m \\
   \bs{\varepsilon}_b  
 \end{bmatrix},
\end{equation}
where we assume linearly isotropic elastic behavior. The constitutive matrix $\bs{S}$  contains membrane and bending components that are defined as 
\begin{equation}
\bs{S}_{m} = \frac{E h}{1 - \nu^2} 
\begin{bmatrix}
	1 & \nu & 0 \\
	\nu & 1 & 0 \\
	0 & 0 & \frac{1 - \nu}{2}
\end{bmatrix}  \text{~~and~~} 
\bs{S}_{b} = \frac{E h^3}{12 A^2 (1 - \nu^2)} 
\begin{bmatrix}
	1 & \nu & 0 \\
	\nu & 1 & 0 \\
	0 & 0 & \frac{1 - \nu}{2}
\end{bmatrix},
\end{equation}
where $E$ is Young's modulus and $\nu$ is Poisson's ratio.

The element stiffness matrix for the $e$th element is computed by adding both material\footnote{We note that here $\overline{\bs{k}}'_e$ is used to indicate that the material stiffness matrix in fact depends on the displacements through~\eqref{eq:B nonlinear}, which would require an iterative solver. In our implementation, outlined in the next subsection, we compute the material stiffness using a linearized version of \eqref{eq:B nonlinear}.} $\overline{\bs{k}}'_e$ and geometric $\widetilde{\bs{k}}_e$ stiffness matrices, given respectively by
\begin{equation}
   \overline{\bs{k}}'_e = \bs{B}^\intercal \bs{S} \bs{B} \text{~~and~~} \widetilde{\bs{k}}_e = \diag \left(  \bs{G}_d, \bs{G}_d, \bs{G}_d, \bs{0} \right),
\end{equation}
with $ \left( \bs{G}_d \right)_{ij} = \sum_{k} \sigma_{k} \ \frac{\partial ^2 \varepsilon_{k}}{\partial u_i \partial u_j} =  \sigma_{11} \bs{C}_1 + \sigma_{22} \bs{C}_2 + \sigma_{12} \bs{C}_3$ (here we express the in-plane stress components $\sigma_{k}$ in Voigt notation as the components of a vector).

\paragraph{\textbf{Force vector}} The only contributions to the force vector are due to the prestress. The local force vector is computed as
\begin{equation}\label{eq:force_vector}
  \bs{f}_e = - \bs{B}^{\intercal} \bs{\sigma}_0.
\end{equation}

\paragraph{\textbf{Mass matrix}} The mass matrix is $\bs{m}_e = \diag \left( \bs{m}_{uu}, \bs{m}_{vv}, \bs{m}_{ww}, \bs{0}_{3 \times 3} \right)$ with
\begin{equation}
  \bs{m}_{uu} = \bs{m}_{vv} = \bs{m}_{ww}  = \frac{\rho h A}{12}
  \begin{bmatrix}
    2 & 1 & 1 \\ 1 & 2 & 1 \\ 1 & 1 & 2
  \end{bmatrix},
\end{equation}
so inertia is added only to translational DOFs.

\paragraph{\textbf{Final discretized equation}}
By accounting for the contributions of the local arrays in all finite elements, the corresponding global arrays are obtained through the standard assembly procedure. To wit,
\begin{equation}\label{eq:assembly}
	\overline{\bs{K}}' = \assembly_{e=1}^{n_E} \overline{\bs{k}}'_e, \quad \widetilde{\bs{K}} = \assembly_{e=1}^{n_E} \widetilde{\bs{k}}_e,  \quad \bs{F}_{\sigma} = \assembly_{e=1}^{n_E} \bs{f}_e, \text{~~and~~} \bs{M} = \assembly_{e=1}^{n_E} \bs{m}_e,
\end{equation}
where $\assembly{}$ denotes the standard finite element assembly operator.

The final discretized form of~\eqref{eq:variaitonal}, which describes the dynamic equilibrium of the resonator, is given by
\begin{equation} \label{eq:discrete_form}
  \bs{M} \bs{\ddot{U}} + \left( \overline{\bs{K}}' + \widetilde{\bs{K}} \right) \bs{U} = \bs{F}_{\sigma}.
\end{equation}

\subsection{Prestress analysis}
 The finite element formulation described above is general as it includes a nonlinear strain-displacement relation for the membrane strain components in \eqref{eq:membrane_strain_components}. However, in our case the isotropic pretension is in-plane only, so a linear analysis suffices for determining the equilibrium state of the system, in similar fashion to two-step approach presented by Pedersen \cite{pedersen01}. Therefore, the higher-order terms from the displacement-strain relation are not taken into account in the computation of the linear stiffness matrix.
 Based on the linear strain-displacement matrix $\overline{\bs{B}}$ (i.e., where we neglect the displacement-dependent terms in \eqref{eq:B nonlinear}), we compute the linear stiffness matrix and force vector as
\begin{equation}
	\overline{\bs{K}} = \assembly_{e=1}^{n_E} \overline{\bs{B}}^\intercal \bs{S} \overline{\bs{B}} \text{~~and~~} \bs{F}_\sigma = \assembly_{e=1}^{n_E} - \overline{\bs{B}}^{\intercal} \bs{\sigma}_0.
\end{equation}

After prescribing clamped boundary conditions---i.e., zero displacement and zero rotation along the boundary---we solve the linear system of equations
\begin{equation} \label{eq:static analysis}
  \overline{\bs{K}} \bs{U}_{\sigma_0} = \bs{F}_\sigma.
\end{equation}
The global displacement vector $\bs{U}_{\sigma_0}$ caused by pretension is then used to determine the stress at the element level as 
\begin{equation}\label{eq:element stress level}
	\widehat{\bs{\sigma}} = \bs{S}\overline{\bs{B}} \bs{L} \bs{U}_{\sigma_0} - \bs{\sigma}_0 = \bs{S}\left( \overline{\bs{B}} \bs{L} \bs{U}_{\sigma_0} - \bs{\varepsilon}_0\right),
\end{equation}
where $\bs{L}$ selects, from the global displacement vector, the DOFs corresponding to the element; we note again that $\overline{\bs{B}}$ again denotes the linear strain-displacement matrix, and $\bs{\varepsilon}_0$ denotes the strain produced by pretension, i.e., $\bs{\sigma}_0 = \bs{S} \bs{\varepsilon}_0$.

\subsection{Eigenvalue analysis}

To determine the out-of-plane vibration modes we now perform an eigenvalue analysis, whereby we consider the high-order nonlinear effects of the geometric nonlinearity caused by the prestress.
To compute the geometric stiffness contribution we compute the derivatives of the nonlinear strain-displacement relation. At element level, this contribution is computed as
\begin{equation}\label{eq: def G}
	\widetilde{K}_{ij} = \sum_{k} \widehat{\sigma}_{k} \ \frac{\partial ^2 \varepsilon_{k}}{\partial u_{i} \partial u_{j}}, 
\end{equation}
where the components of $\widehat{\bs{\sigma}}$ (in Voigt notation) are multiplied with coefficients that depend on the geometry of the element; the latter can be derived by differentiating the displacement-dependent terms in \eqref{eq:membrane_strain_components}. It is worth noting that, although the higher-order displacement-dependent terms are omitted for the static analysis (as the prestress is in-plane only with very small deformation), we include them in the computation of the geometric stiffness matrix because vibration modes are out-of-plane.

The free-vibration modes are determined by assuming a harmonic solution to~\eqref{eq:discrete_form}, i.e., of the The following eigenvalue problem is then solved to determine the prestressed eigenfrequencies ($\omega_i^2$) and corresponding vibration modes ($\bs{\Phi}_i$) of the system,
\begin{equation}\label{eq:modal problem}
	\left( \overline{\bs{K}} + \widetilde{\bs{K}} - \omega_i^2 \bs{M} \right) \bs{\Phi}_i = \bs{0}.
\end{equation}

\section{Topology optimization}\label{sec:topology_optimization}

This work employs density-based topology optimization (TO)~\cite{sigmund2004}, which assigns each finite element a design variable that influences its (local) material properties. For the $e$th element this parameter is denoted by $\rho_e$, and it is bounded by $0 \leq \rho_e \leq 1$, where the upper bound represents ``solid'' material and the lower bound ``void'' material. The optimization formulation presented herein is based on the approach by Gao et al.~\cite{gao20}, so the same notation has been used for the design $(\bs{\rho})$, filtered $(\bs{\tilde{\rho}})$ and projected $(\bs{\bar{\rho}})$ density fields, which are discussed later on in this section.

We are interested in finding an optimized topology that maximizes the $Q$ factor. Formally, we can write this topology optimization problem as
\begin{equation}\label{eq:optimization_problem_original}
\begin{alignedat}{3}
& \bs{\rho}^\star = && \argmax_{\bs{\rho} \in \mathcal{D}} && f \left( \bs{\rho} \right) =  D_Q \\
&                   && \text{subject to} \quad             && \overline{\bs{K}}\bs{U}_{\sigma_0}  = \bs{F},\\
&                   &&              && \left( \overline{\bs{K}} + \widetilde{\bs{K}} - \omega_i^2 \bs{M} \right) \bs{\Phi}_i = \bs{0},\\
 &                   &&                                     && V_m \leq V_c, \\
		&                   &&                                     && f_0 \leq f_{\min}.
\end{alignedat}
\end{equation}
where the design space is $\mathcal{D} = \left\{ \left. \bs{\rho} \right| \, \bs{\rho} \in \mathbb{R}^{n_e}, 0 \leq \rho_e \leq 1 \, \forall e \in \left\{ 1, \ldots , N_e \right\} \right\} $ and thus $\rho$ is a vector containing the densities of all $N_e$ finite elements in the discretization. Furthermore, $V_m$ is the fraction of material used in the domain, for which an upper bound $V_c$ is prescribed. A minimum frequency constraint is also imposed to the problem.


In the following sections we explain how the design parameters influence the local material properties in Section~\ref{sec:material interpolation}, we derive the analytical sensitivity formulation of the objective in Section~\ref{sec:sensitivity}, and further develop the overall optimization problem with a robust formulation in Section~\ref{sec:robust optimization}.

\subsection{Material Interpolation}
\label{sec:material interpolation}

The material properties of each finite element, such as Young's modulus and mass density, are indirectly controlled by the design parameters. Before altering the local material properties, the field of design variables is first filtered and then projected in order to promote designs with smooth and clear boundaries between void and solid regions. In this work we use a simple distance filter \cite{BRUNS20013443}:
\begin{equation}\label{eq:filter}
    \Tilde{\rho}_e = \frac{\sum_{i\in \iota_e}  \left(r_f - \norm{\bs{x}_e - \bs{x}_i} \right) \rho_i }{\sum_{i\in \iota_e} \left(r_f - \norm{\bs{x}_e - \bs{x}_i} \right) },
\end{equation}
where the set $\iota_e$ includes all elements whose barycenter lie inside a circle with radius $r_f$ centered at the element considered $\bs{x}_e$ (including the element itself). Note that the term $\left(r_f - \norm{\bs{x}_e - \bs{x}_i} \right)$ is always positive.
The filtering operation prevents the emergence of checkerboard patterns---which can be beneficial to the optimizer but are not physical---and results in a gradual transition between solid and void regions. 
A projection $\mathcal{P}: \mathbb{R} \to \mathbb{R}$ is then performed on the filtered density field in order to push the density of intermediate elements to either $0$ or $1$, i.e. \cite{guest04}, 
\begin{equation}\label{eq:projection}
    \Bar{\rho}_e = \mathcal{P} \left( \tilde{\rho}_e \right) \equiv  \frac{\tanh{\beta \eta } + \tanh{\beta\left(\tilde{\rho}_e - \eta \right)}}{\tanh{\beta \eta } + \tanh{\beta\left(1 - \eta \right)}},
\end{equation}
where $\beta$ and $\eta$ denote, respectively, the projection slope and mid-point (i.e., the value of $\tilde{\rho}_e$ for which $\bar{\rho}_e$ is equal to 0.5).
The projected densities penalize the Young's modulus and mass density at the element level. The Young's modulus is penalized by means of the RAMP function \cite{stolpe01} as

\begin{equation}\label{eq:ramp}
    E_e\left(\Bar{\rho}_e\right) = E_0 \left[  \frac{E_{\min}}{E_0} + \frac{\left(1-\frac{E_{\min}}{E_0}\right)  \Bar{\rho}_e }{1+q\left(1 -\Bar{\rho}_e \right)} \right],
\end{equation}
where $E_0$ is the Young's modulus of the solid material, and $E_{\min}$ is a small lower bound that prevents the global stiffness matrix from becoming singular (although its condition number can increase substantially with low-density regions). In this work, the ratio of $E_{\min}/E_0$ was set to $10^{-6}$. The mass density is penalized linearly as 
\begin{equation}\label{eq:mass penalization}
    m_e(\bar{\rho}_e) =m_0\left[\frac{m_{\min}}{m_0} + \left(1-\frac{m_{\min}}{m_0}\right)\Bar{\rho}_e  \right].
\end{equation}	
In this work, the ratio between the mass of a void element ($m_{\min}$) and a solid element ($m_0$) was set to $10^{-7}$.

Both the mass and Young's modulus ratios were set, not only to prevent singular matrices, but also to prevent the appearance of \textit{spurious modes}, which are non-physical vibration modes that emerge due to extremely low stiffness-to-mass ratios in the void regions~\cite{pedersen00}.
Prescribing minimum ratios for mass and Young's modulus of elements prevents these modes from dominating the fundamental mode of the design, but they can still appear in higher-order modes.

Additionally, the static prestress analysis may result in void regions under compression, thereby inducing instabilities in the eigenvalue analysis. In order to prevent this, a projection similar to Eq.~\eqref{eq:projection} is used on the static displacement field, based on the work of Wang et al. \cite{wang14}. A penalization parameter $\alpha_e$ for the $e$th element is computed based on the local Young's modulus of the element as $\alpha_e(E_e) = \mathcal{P} \left( E_e/E_0 \right)$,
where projection slope $\beta = 30$ and mid-point $\eta = 0.05$ are typically used parameters (see Gao et al.~\cite{gao20}).
The projected displacement and resulting element stress are computed as 
\begin{align}\label{eq:static_projection}
  \begin{split}
    \bs{\tilde{U}}_\sigma &= \alpha_e  \bs{U}_{\sigma_0},\\  
    \widehat{\bs{\sigma}} &=  \bs{S}\left( \overline{\bs{B}} \bs{L} \bs{\tilde{U}}_{\sigma_0} - \bs{\varepsilon}_0\right).
  \end{split}
\end{align}
Applying this filter prior to the computation of equilibrium stresses ensures that low-density elements remain in tension (i.e., with the prestressed tension $\widehat{\bs{\sigma}}^{(e)} =  - \bs{S}_e \bs{\varepsilon}_0$)~\cite{wang14}. Due to the high slope, only elements with a projected density at or below the mid-point are affected. As this projection influences the computation of element-level stresses, it also has an effect on the relation between the geometric stiffness matrix \eqref{eq: def G} and the design variables $\bs{\rho}$. This is included in the sensitivity analysis of the matrix $\widetilde{\bs{K}}$, which is detailed in Appendix~\ref{ap:Gmatrix}.

\subsection{Sensitivity analysis}\label{sec:sensitivity}
We use an adjoint formulation to determine the sensitivity of Eq.~\eqref{eq:DQ objective} with respect to a design density $\rho_e$. Consequently, we write a Lagrangian, whereby the objective function $D_Q$ is augmented with constraint terms involving static equilibrium and prestressed modal analyses. The Lagrange multipliers are then determined such that computationally-expensive terms in the derivative of the Lagrangian are conveniently omitted from the analysis. To wit,
\begin{align}\label{eq:lagrangian}
    \begin{split}
    L = &\frac{\bs{\Phi} ^{\intercal} \widetilde{\bs{K}} \bs{\Phi} }{\bs{\Phi}^{\intercal} \overline{\bs{K}} \bs{\Phi}} + \bs{\lambda}_1^{\intercal} \left(\bs{F} - \overline{\bs{K}} \bs{U}_{\sigma_0} \right) + \bs{\lambda}_2^{\intercal} \left( \overline{\bs{K}}+\widetilde{\bs{K}}-\omega^2 \bs{M}  \right) \bs{\Phi} .
    \end{split}
\end{align}

The sensitivity of the Lagrangian with respect to the $e$th density design parameter is
\begin{align}\label{eq:DQ sens main expression}
    \begin{split}
            \frac{d L}{d \rho_e} =&\frac{\bs{\Phi}^{\intercal} \frac{\partial \widetilde{\bs{K}} }{\partial \rho_e}\bs{\Phi}}{\bs{\Phi}^{\intercal} \overline{\bs{K}} \bs{\Phi}} - \frac{ \bs{\Phi}^{\intercal} \widetilde{\bs{K}} \bs{\Phi} \cdot  \bs{\Phi}^{\intercal} \frac{d \overline{\bs{K}} }{d \rho_e}\bs{\Phi} }{\left( \bs{\Phi}^{\intercal} \overline{\bs{K}} \bs{\Phi} \right)^2} + \bs{\lambda}_1^{\intercal} \left( \frac{d \bs{F}}{d \rho_e} -  \frac{d \overline{\bs{K}}}{d \rho_e}\bs{U}_{\sigma_0}\right) \\ 
            & \qquad +\bs{\lambda}_2^{\intercal} \left( \frac{d\overline{\bs{K}}}{d \rho_e} + \frac{\partial \widetilde{\bs{K}} }{\partial \rho_e}  - \omega^2 \frac{d\bs{M}}{d \rho_e} \right) \bs{\Phi},
    \end{split}
\end{align}
in which $\bs{\lambda}_2$ is expressed as $ \bs{\lambda}_2 \equiv  c \bs{\Phi} + \bs{\psi}$ (following the approach by Tsai and Cheng~\cite{tsai13}), with parameters $c$ and $\bs{\psi}$ defined as
\begin{align}
    \begin{split}
         \left( \overline{\bs{K}}+\widetilde{\bs{K}}-\omega^2 \bs{M}  \right)\bs{\psi} & = \left(- \frac{2   }{ \bs{\Phi}^{\intercal} \overline{\bs{K}} \bs{\Phi}} \widetilde{\bs{K}} + \frac{2\bs{\Phi}^{\intercal} \widetilde{\bs{K}} \bs{\Phi}  }{\left( \bs{\Phi}^{\intercal} \overline{\bs{K}} \bs{\Phi}\right)^2} \overline{\bs{K}}\right) \bs{\Phi}, \\
        c & = - \bs{\psi}^\intercal \bs{M} \bs{\Phi},
    \end{split}
\end{align}
The Lagrange multiplier $\bs{\lambda}_1$ is obtained by solving
\begin{equation}\label{eq: adjoint u}
       \overline{\bs{K}}\bs{\lambda}_1  = \left( \frac{\bs{\Phi}^{\intercal} }{ \bs{\Phi}^{\intercal} \overline{\bs{K}} \bs{\Phi}}  + \bs{\lambda}_2^{\intercal}  \right)  \frac{\partial \widetilde{\bs{K}} }{\partial \bs{U}_{\sigma_0}}\bs{\Phi}.
\end{equation}
The partial derivatives of $\overline{\bs{K}}$, $\widetilde{\bs{K}}$, $\bs{M}$ and $\bs{F}$ are found by differentiating Eqs.~\eqref{eq:ramp} and \eqref{eq:mass penalization}, as they penalize the matrices at the element level. Here, the chain rule must be carefully applied to include the filtering and projection operations to relate the projected $\bs{\bar{\rho}}$ density to the design density $\bs{\rho}$. The geometric stiffness matrix $\widetilde{\bs{K}}$ has an additional dependence on the design densities through a dependency on the static displacement $\bs{U}_{\sigma_0}$. The computation of this derivative is discussed in detail in Appendix~\ref{ap:Gmatrix}.
The analytical sensitivity given by~\eqref{eq:DQ sens main expression} was verified using a finite difference (FD) check. Details about this verification are given in Appendix~\ref{ap:FDcheck}.

\subsection{Robust optimization formulation}\label{sec:robust optimization}
Our preliminary results indicated that designs can rely on elements with intermediate densities to obtain unrealistically high $Q$ factors, a fact that was also discussed by Gao et al.~\cite{gao20}.
Although regions with intermediate density values are acceptable during the optimization procedure, we prefer black-and-white final designs since the physical properties of intermediate densities are not well defined. Herein, we use the robust formulation described by Wang et al.~\cite{wang11} to promote the convergence to fully black-and-white designs. This approach, which extends the TO formulation described in Section~\ref{sec:material interpolation}, considers different mid-points for the projection function \eqref{eq:projection}. At every iteration of the optimization, the vector of filtered design variables $\bs{\tilde{\rho}}$ is projected using three slightly different mid-points (denoted by $\eta_D = 0.45, \eta_I = 0.5$, and $\eta_E = 0.55$ ), which results in three slightly different projected density fields ($\bs{\bar{\rho}}_D,\bs{\bar{\rho}}_I$, and $\bs{\bar{\rho}}_E$, respectively). These fields correspond to the \textit{dilated}, \textit{intermediate} and \textit{eroded} designs, which are all analyzed for their performance. Then the sensitivities of the design with the worst objective function value are used to update the vector of design densities $\bs{\rho}$. Shifting the projection mid-point changes the location of elements with intermediate densities, and using the worst-performing projection to update the design variables ensures that if there is a design direction that strongly depends on intermediate densities to obtain a high objective, it is not preferred by the optimizer. Furthermore, a lower bound is prescribed for the fundamental resonance frequency of the design, which avoids poorly-connected designs by ensuring a minimum level of connection to the boundaries of the system \cite{gao20}. Including the robust approach requires us to reformulate problem~\eqref{eq:optimization_problem_original} as
\begin{equation}\label{eq:optimization_problem_robust}
	\begin{alignedat}{3}
		& \bar{\bs{\rho}}^\star = && \argmax_{\bar{\bs{\rho}} \in \mathcal{D}} &&   \min_{\bar{\bs{\rho}}_i \in \left\{ \bar{\bs{\rho}}_D,\bar{\bs{\rho}}_I,\bar{\bs{\rho}}_E \right \} } D_Q \left(  \bs{\bar{\rho}}_i  \left(\bs{\rho}\right)  \right) \ \\
		&                   && \text{subject to} \quad             && \overline{\bs{K}}\bs{U}_{\sigma_0}  = \bs{F},\\
		&                   &&              && \left (\overline{\bs{K}} + \widetilde{\bs{K}} - \omega_i^2 \bs{M} \right) \bs{\Phi}_i = \bs{0},\\
		&                   &&                                     && V_m \leq V_c, \\
		&                   &&                                     && f_0 \geq f_{\min}.
	\end{alignedat}
\end{equation}

\def\imagetop#1{\raisebox{-\height+2\baselineskip}{#1}}

\newcommand{\dummyfigure}{\tikz \fill [NavyBlue] (0,0) rectangle node [black] {Q} (1.5,1.5);}
\newcommand{\addresonator}[1]{\imagetop{\includegraphics[width=1.5cm]{#1}}}

\section{Numerical examples}\label{sec:numerical}

In this section we showcase the capability of the methodology in maximizing the $Q$ factor. We discuss thoroughly the optimization of a resonator on a square domain, and we later showcase the generality of our procedure by optimizing on an hexagonal domain.
For all examples we use the Method of Moving Asymptotes (MMA) by Svanberg \cite{svanberg87} as the optimizer. A prestress $\sigma_0 = \SI{1}{\giga \pascal}$ and a thickness $h = \SI{340}{\nano\metre}$ were used in all optimizations. The latter follows the work by Li et al.~\cite{zichao23}, who successfully modeled, fabricated, and tested high-$Q$ \ce{Si3N4} resonators. In all examples, a upper bound $V_m=0.5$ was prescribed for the volume fraction. The latter was applied to the dilated design only (i.e., to $\bs{\bar{\rho}}_D$), since by definition it has the highest volume. Finally, the frequency constraint $f_{\min}$ (which varied between simulations) was prescribed for all three designs of the robust formulation, (i.e. $\bs{\bar{\rho}}_D$, $\bs{\bar{\rho}}_I$ and $\bs{\bar{\rho}}_E$). We now discuss results of our optimization methodology applied to both square and hexagonal computational domains, and highlight the non-convex property of the optimization problem at hand.

\subsection{Square computational domain}\label{sec:square simulations}
\begin{figure}[b]
	\centering
    \includegraphics{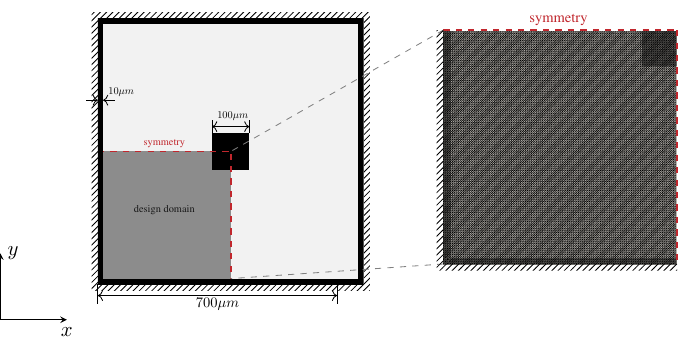}
	\caption{Schematic of the problem considered in Section~\ref{sec:square simulations}. The inset shows the finite element mesh used to model a quarter of the domain to reduce computational cost, where symmetry conditions are indicated by red dotted lines. Darker regions indicate the fixed non-design domain regions where the design density was fixed as $\rho_e = 1$ (these correspond to the black inner square and outermost boundary). Dashed lines indicate the clamped boundary where both the displacement and rotation fields are homogeneous. }
	\label{fig:domains}
\end{figure}

Consider in Figure~\ref{fig:domains} a $700 \times \SI{700}{\micro \metre ^2}$ square domain, with a non-design  $100\times\SI{100}{\micro \metre ^2}$ area at the center and an outer fixed region of $\SI{10}{\micro\metre}$, inspired by the trampoline-like resonators of Norte et al.~\cite{norte16} and previous optimization studies by Gao et al.~\cite{gao20} and H{\o}j et al.~\cite{hoj21article}. We take advantage of symmetry to model a quarter of the domain, thereby reducing  significantly the overall computational cost. Clamped boundary conditions (i.e., zero displacement and zero rotation) are prescribed on the outer boundary of the domain.
The filter radius was set to $r_f = \SI{29.8}{\micro\metre} $ (approximately 6 elements).

\subsection*{Dependence on target frequency}
Because there exists a trade-off between the frequency of the fundamental mode and the $Q$ factor of a resonator~\cite{norte16}, we first investigate how the target frequency influences the optimized geometry. We therefore maximize the $Q$ factor while setting a lower bound on the fundamental resonance frequency with values $f_{\min} (\si{\kilo\hertz}) = \left\{ 300, 350, 400\right\}$. 
As commonly done in density-based topology optimization, we choose an initial design with a homogeneous material distribution equal to the volume fraction, i.e., $\rho_e = 0.5$ for all elements in the design domain. During the optimization, a continuation scheme increases the projection slope $\beta$ in~\eqref{eq:projection} by 1\% every 3 iterations, starting at a slope  $\beta = 1$ and ending with $\beta = 110$. For the first and last projection slope, 50 iterations were allowed instead of 3. This resulted in a total of \num{1519} iterations.

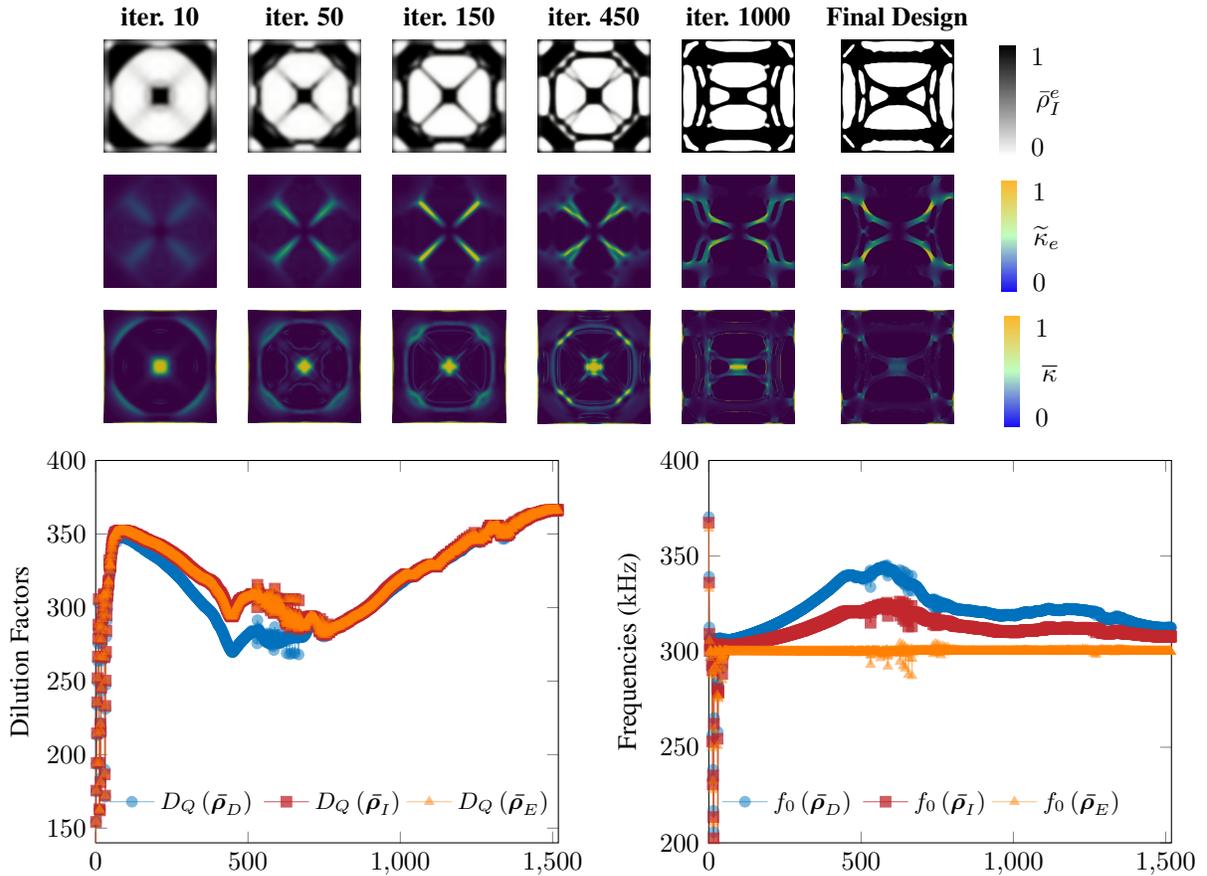
\begin{figure}[!b]
	\centering
	\begin{tblr}{
			cells={valign=m,halign=c},
			row{1}={font=\bfseries,rowsep=0pt},
			colspec={QQQQQQQ},
			column{1}={valign=m},
		}
		iter. 10  & iter. 50 & iter. 150 & iter. 450& iter. 1000 &Final Design &  & \\
		\addresonator{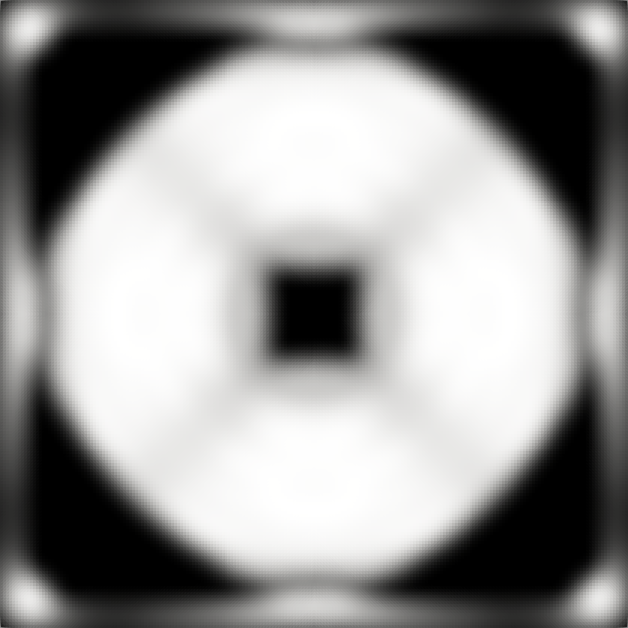} & \addresonator{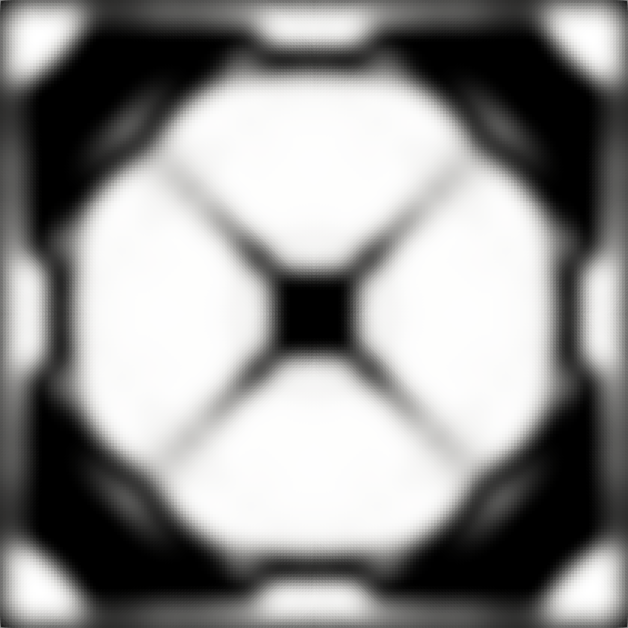} & \addresonator{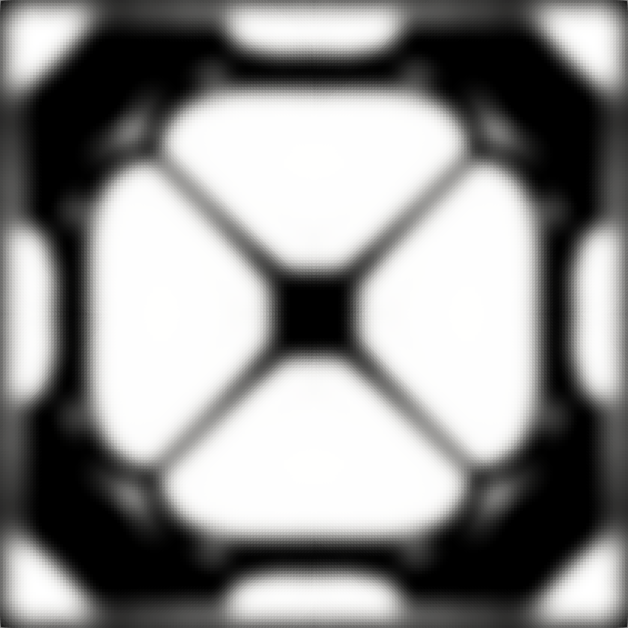} & \addresonator{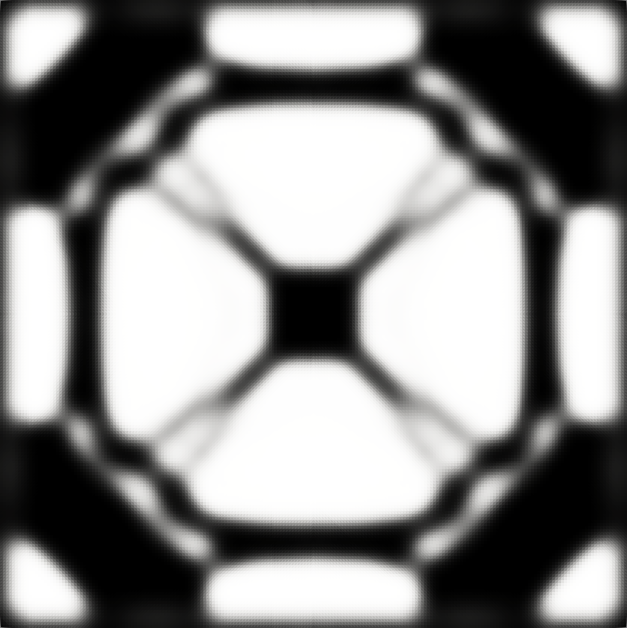} & \addresonator{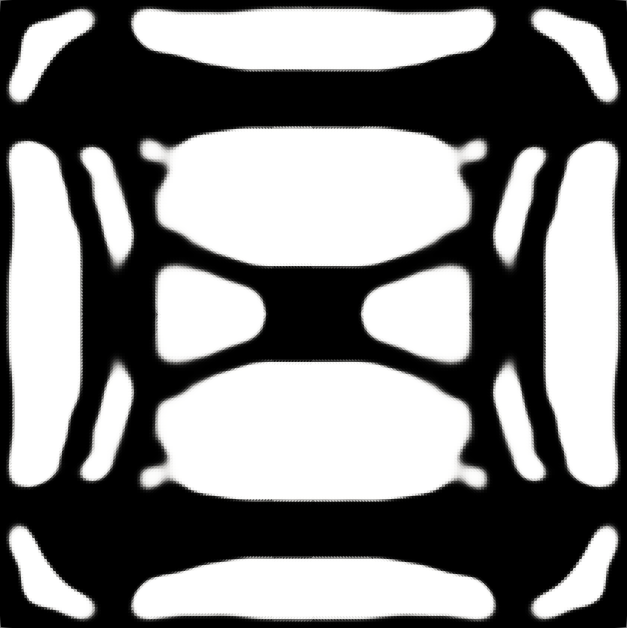} & \addresonator{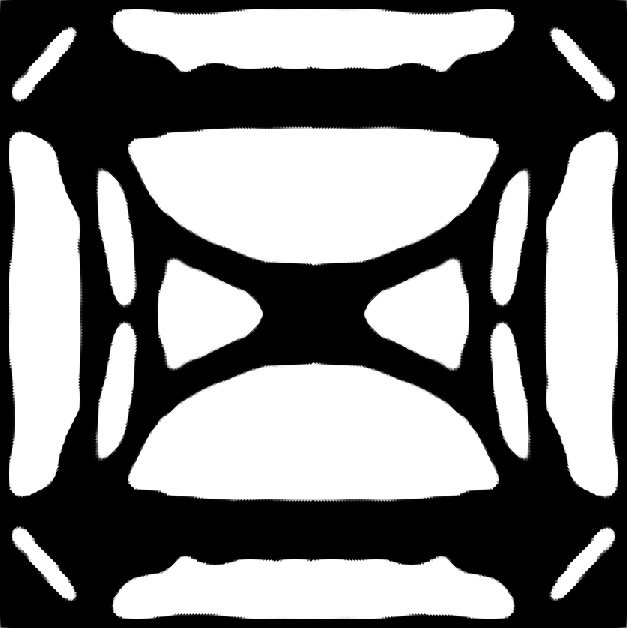} & \imagetop{\begin{tikzpicture} 		
				\fill[top color=black,bottom color=white]
				(-0.1,-0.72) rectangle (0.1,0.75); 
				\node [xshift=0.55cm] (0.0,0.0)  {$\bar{\rho}^{e}_I$};
				\node [xshift=0.4cm, yshift=0.6cm] (0.0,0.0)  {$1$};
				\node [xshift=0.4cm, yshift=-0.6cm] (0.0,0.0) {$0$};
		\end{tikzpicture}} \\
		\addresonator{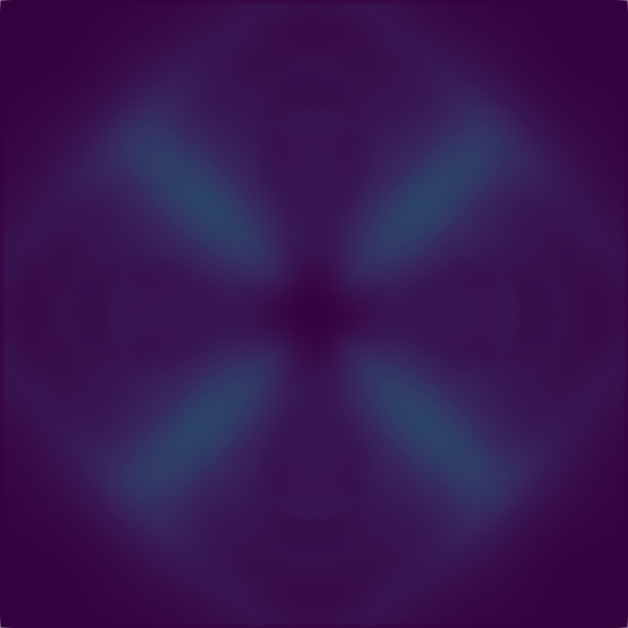} & \addresonator{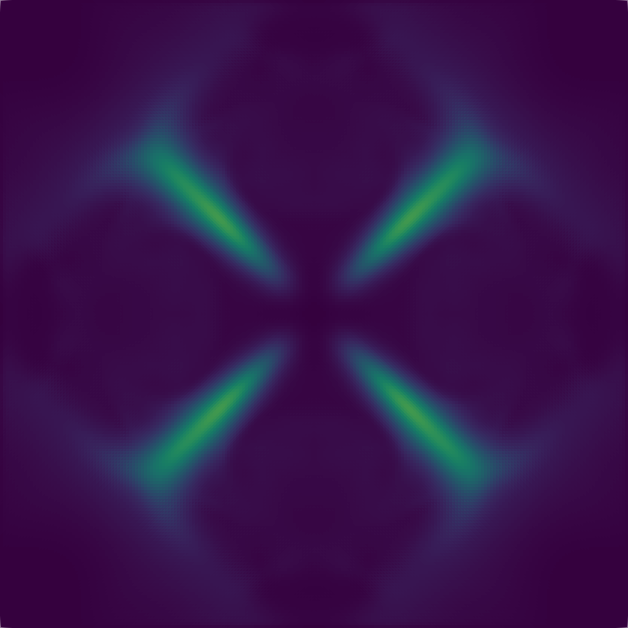} & \addresonator{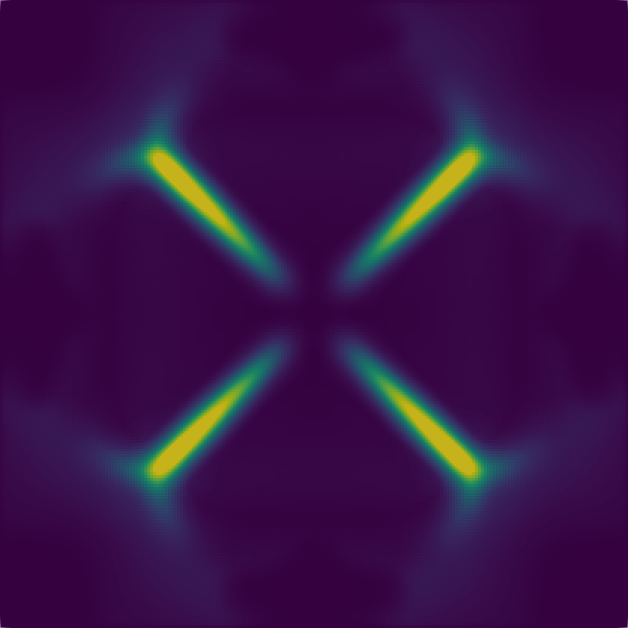} & \addresonator{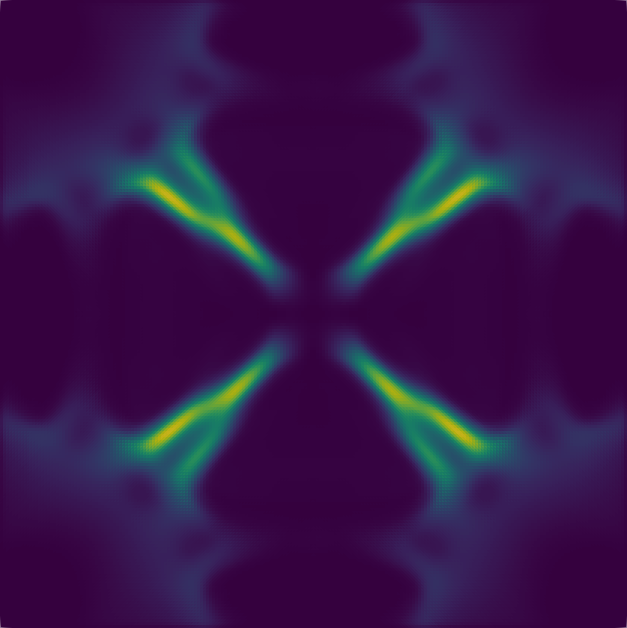} & \addresonator{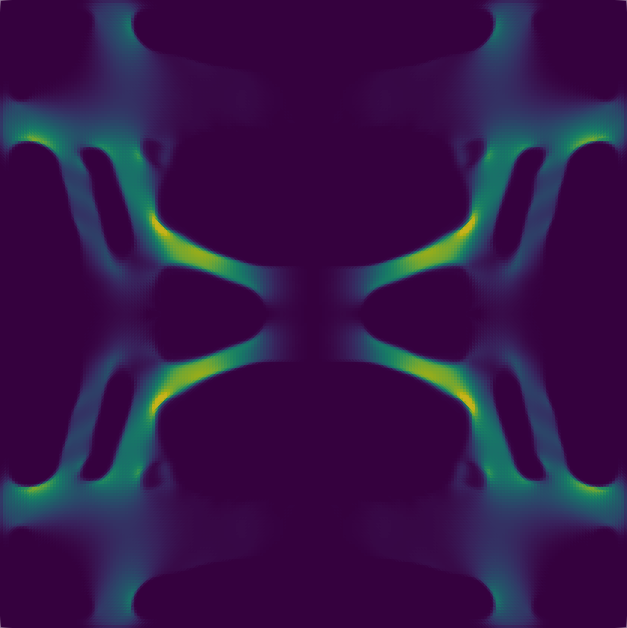}& \addresonator{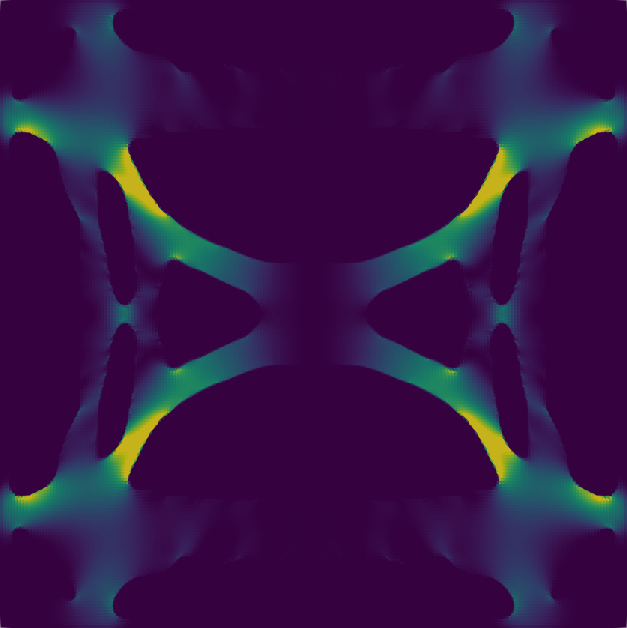} & \imagetop{\begin{tikzpicture} 		
				\node [xshift=0.5cm] (0.0,0.0) {$\widetilde{\kappa}_e$};
				\node [xshift=0.4cm, yshift=0.6cm] (0.0,0.0)  {$1$};
				\node [xshift=0.4cm, yshift=-0.6cm] (0.0,0.0) {$0$};
				\fill[top color=Dandelion,bottom color=BlueViolet, middle color = Green!25]
				(-0.1,-0.72) rectangle (0.1,0.75); 
		\end{tikzpicture}} \\
		\addresonator{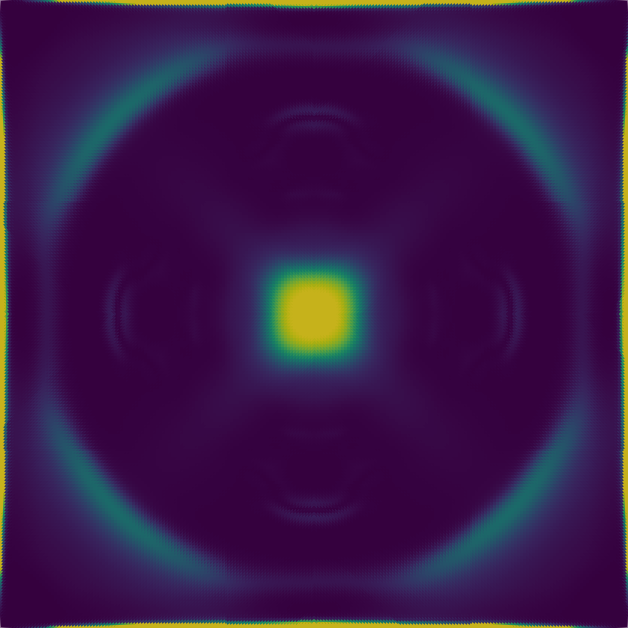} & \addresonator{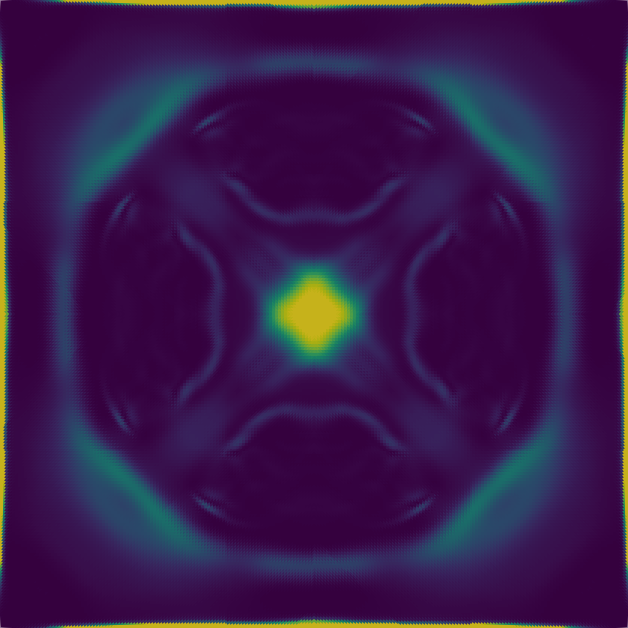} & \addresonator{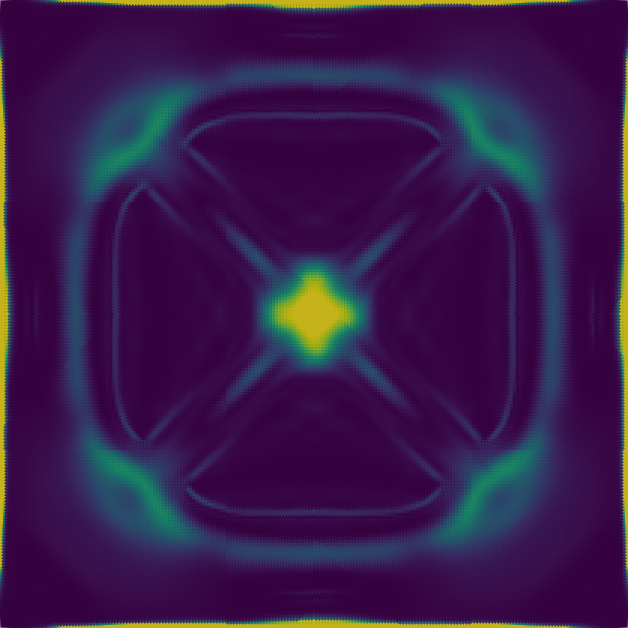} & \addresonator{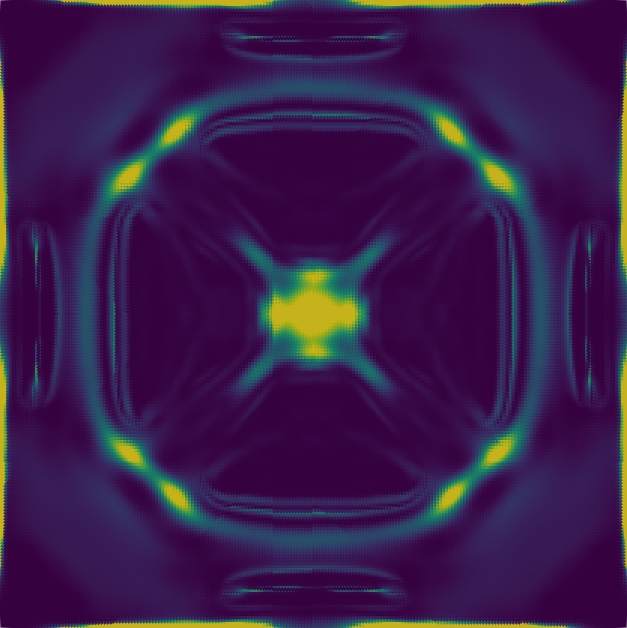} & \addresonator{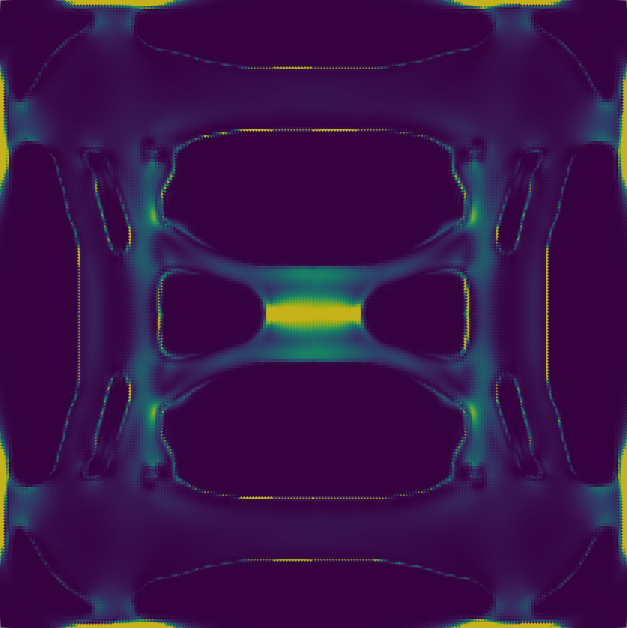}& \addresonator{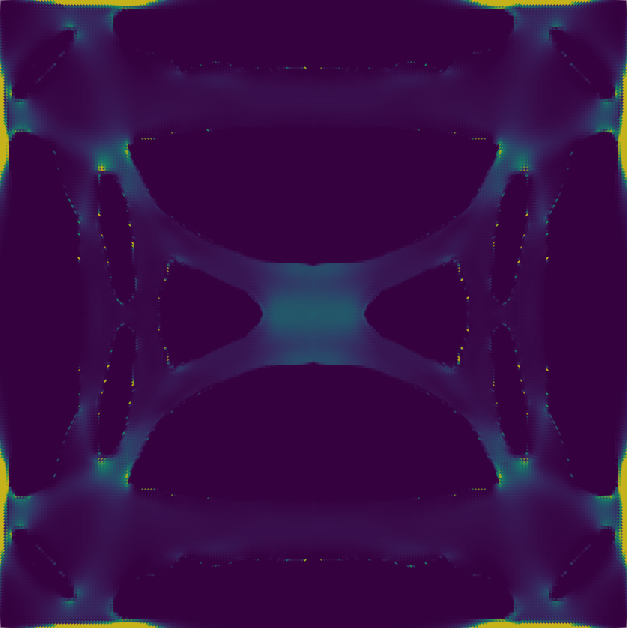} &\imagetop{\begin{tikzpicture} 		
				\fill[top color=Dandelion,bottom color=BlueViolet, middle color = Green!25]
				(-0.1,-0.72) rectangle (0.1,0.75); 
				\node [xshift=0.5cm] (0,0) {$\overline{\kappa}$};
				\node [xshift=0.4cm, yshift=0.6cm] (0,0.0)  {$1$};
				\node [xshift=0.4cm, yshift=-0.6cm] (0,0.0) {$0$};
		\end{tikzpicture}} 
	\end{tblr}
	
	\includegraphics{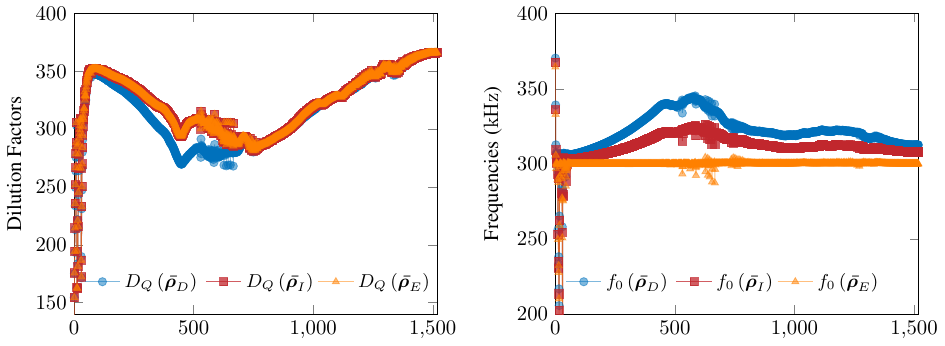}

	\hfill
	\caption{Optimization of a square resonator with a prescribed lower frequency $f_{\min}= \SI{300}{\kilo\hertz}$. The top row shows the evolution of the intermediate projected density, while the second and third rows depict, respectively, the element-level contributions to the geometrically nonlinear and linear modal stiffnesses; these are computed, respectively, as $\widetilde{\kappa}_e = \bs{\Phi}_e^\intercal \tilde{\bs{k}}_e \bs{\Phi}_e/2000$, and $\overline{\kappa}_e = \bs{\Phi}_e^\intercal \bar{\bs{k}}_e \bs{\Phi}_e / 2$ (they are also normalized). The color bar has been capped for both quantities to give a clearer depiction of these fields (as this limit was exceeded locally). The bottom two graphs show the evolution of the dilution factor (left) and fundamental resonance frequency (right) throughout the optimization for all three fields of the robust projection.}
	\label{tbl:f300}
\end{figure}

Figures \ref{tbl:f300}, \ref{tbl:f350} and \ref{tbl:f400} summarize the results for each target frequency. The designs at different iterations are shown on the top row, while the second and third rows show, respectively and at the element level, their corresponding geometrically nonlinear and linear modal stiffnesses. The last row shows the dilution factor and frequency as a function of iteration for all three fields used in the robust formulation. 

The mode shapes and fundamental frequencies of the final designs were evaluated with COMSOL Multiphysics, and have been summarized in Figure \ref{tbl:comsol}. The $Q$ factors for all designs were computed based on the $\SI{340}{\nano\metre}$ thickness used in the simulations. We notice the typical behavior that designs with higher fundamental mode frequencies tend to have lower $Q$ factors. We also note that our design for the target frequency of $\SI{350}{\kilo\hertz}$ shows strong similarity to that obtained by H{\o}g et al.~\cite{hoj21article}, who also fabricated it with thicknesses between $\SI{12}{\nano\metre}$ to $\SI{50}{\nano\metre}$, and showed experimentally determined $Q$ factors in the range of $10^6$ to $10^8$. We also note that our design for a target frequency of $\SI{400}{\kilo\hertz}$ shows similarity to recent work by Shi et al.~\cite{arXivSigmund24}, where the dilution factor of higher-order modes was optimized for frequencies above $\SI{1}{\mega\hertz}$ using a formulation similar to ours. For all designs, the final volume fraction is equal to the volume limit, i.e. $V_m = V_c =0.5$.

\pgfplotsset{
  table/search path={./figures/data},
}

%
%
\begin{figure}
	\centering
	\begin{tblr}{
			cells={valign=m,halign=c},
			row{1}={font=\bfseries,rowsep=0pt},
			colspec={QQQQQQQ},
			column{1}={valign=m},
		}
		iter. 10  & iter. 50 & iter. 150 & iter. 450& iter. 1000 &Final Design &  & \\
		\addresonator{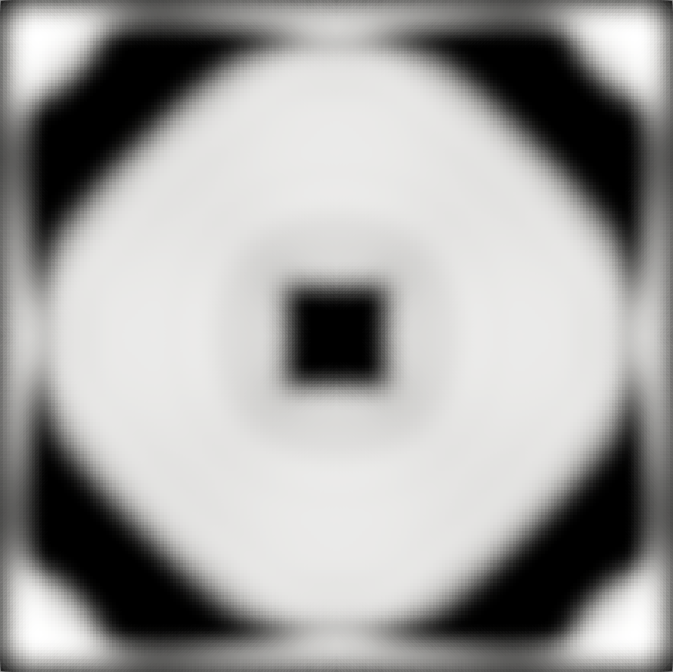} & \addresonator{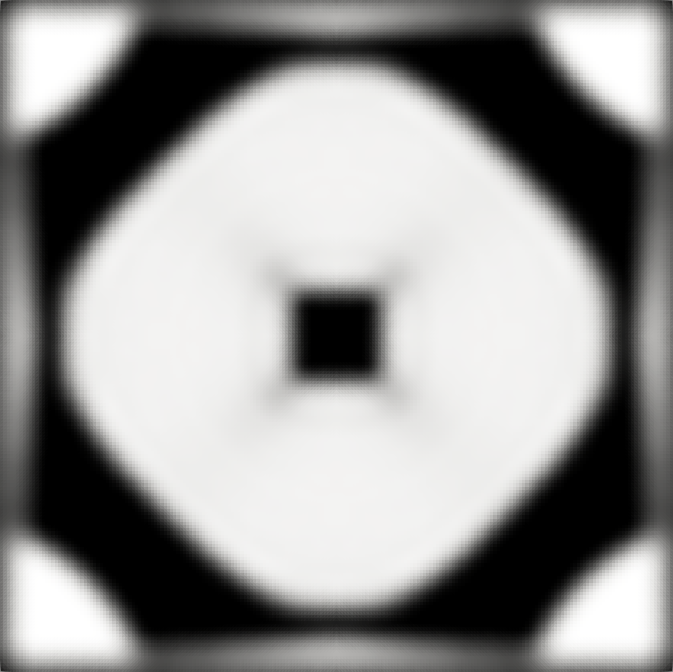} & \addresonator{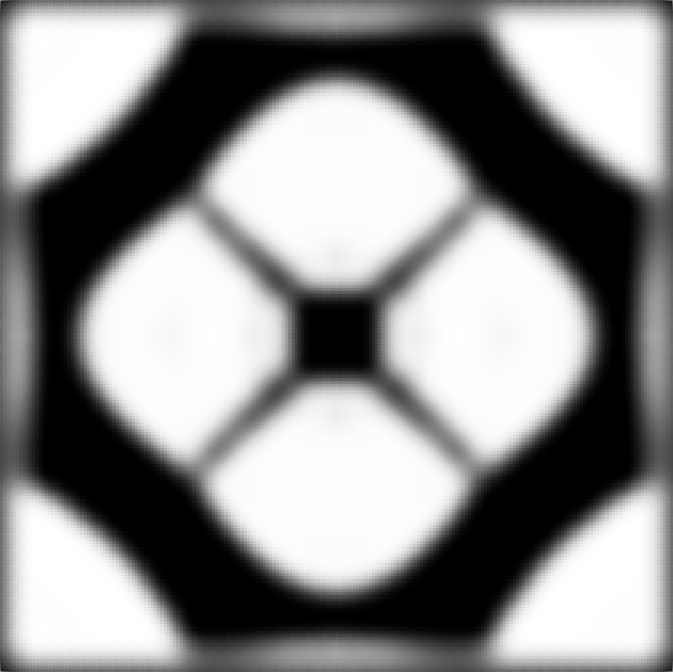} & \addresonator{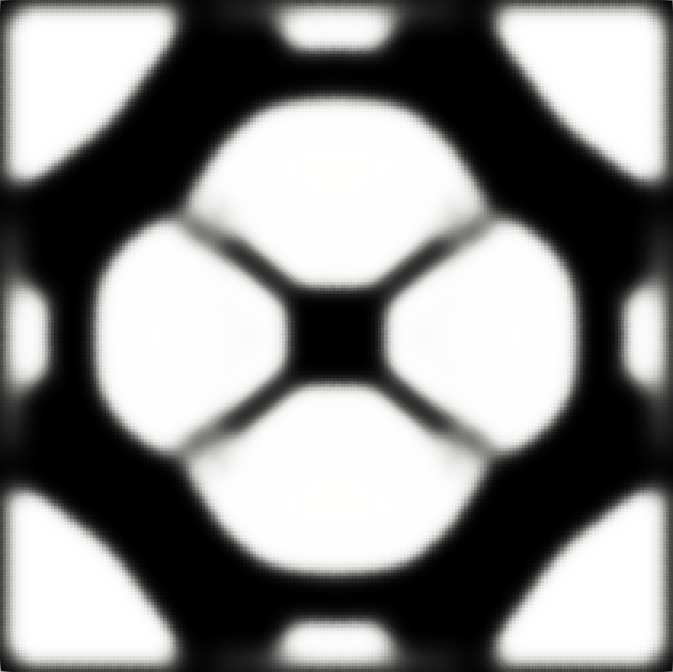} & \addresonator{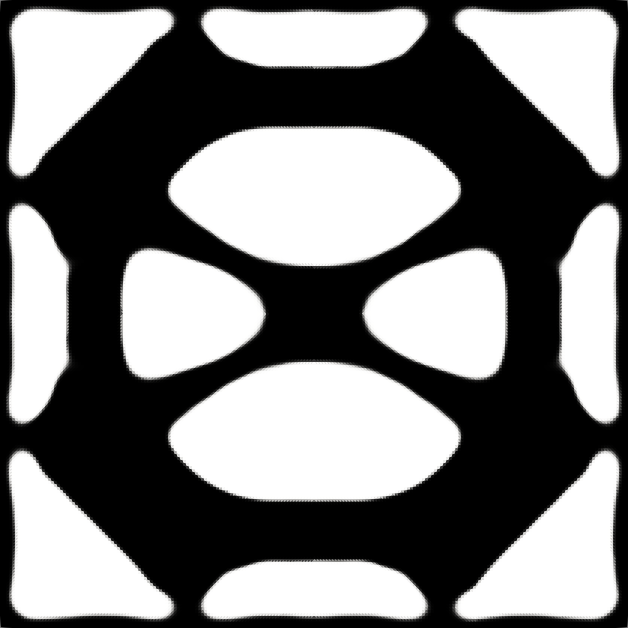} & \addresonator{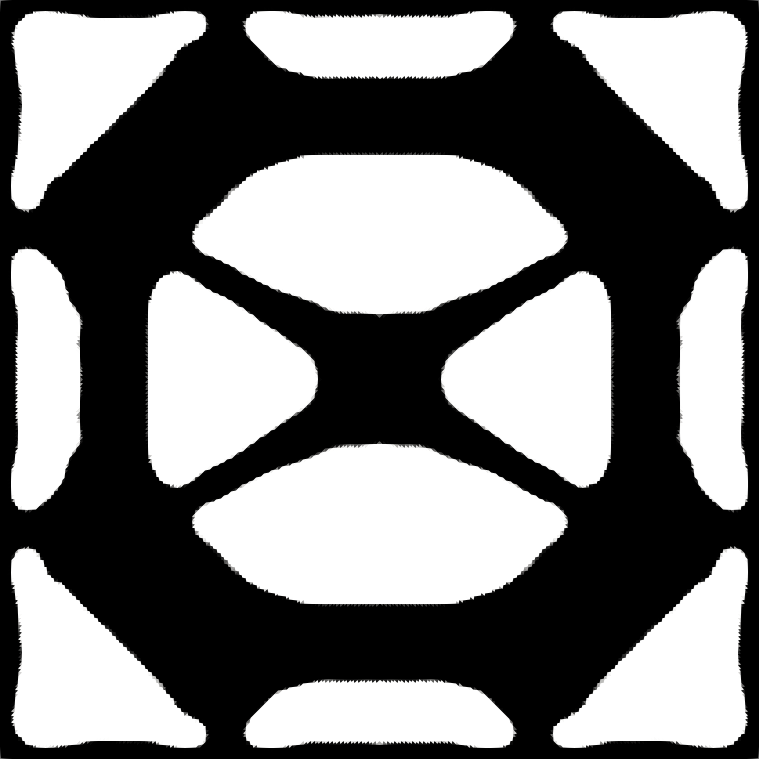} & \imagetop{\begin{tikzpicture} 		
				\fill[top color=black,bottom color=white]
				(-0.1,-0.72) rectangle (0.1,0.75); 
				\node [xshift=0.55cm] (0.0,0.0)  {$\bar{\rho}^{e}_I$};
				\node [xshift=0.4cm, yshift=0.6cm] (0.0,0.0)  {$1$};
				\node [xshift=0.4cm, yshift=-0.6cm] (0.0,0.0) {$0$};
		\end{tikzpicture}} \\
		\addresonator{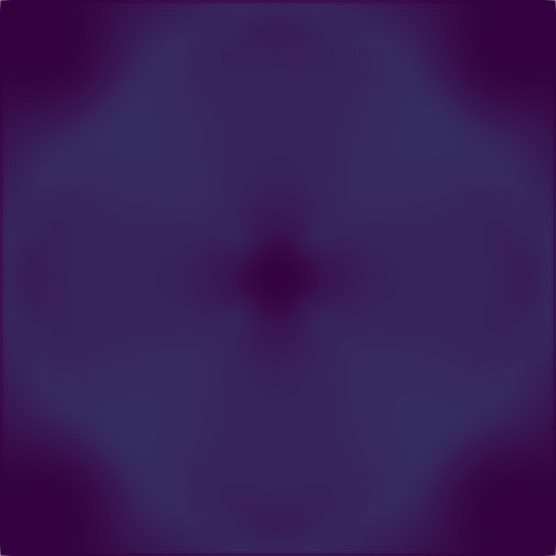} & \addresonator{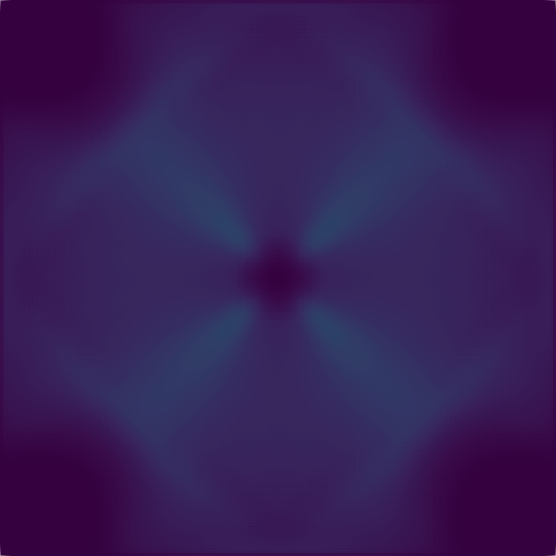} & \addresonator{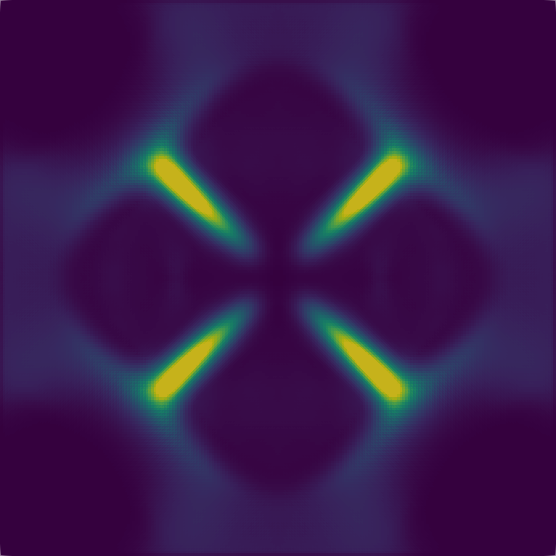} & \addresonator{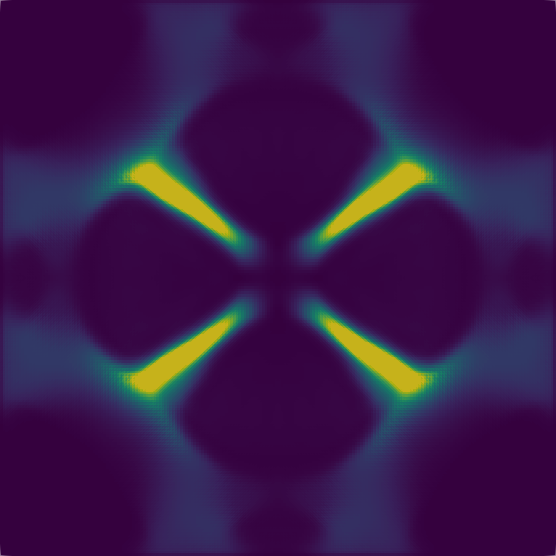} & \addresonator{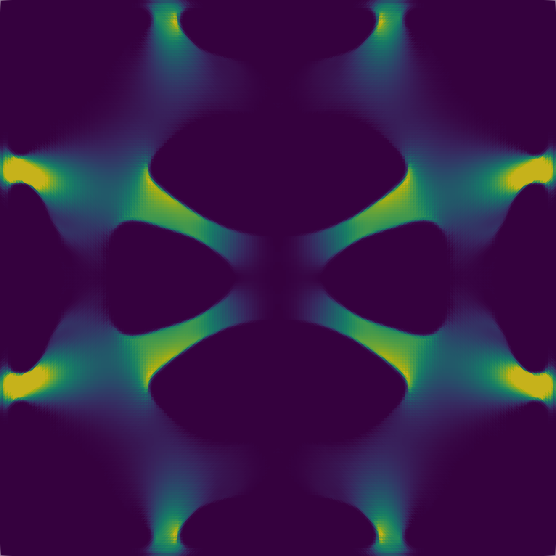}& \addresonator{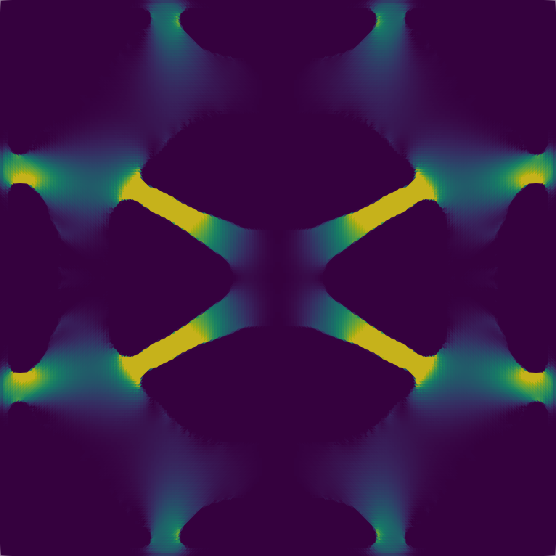} & \imagetop{\begin{tikzpicture} 		
				\node [xshift=0.5cm] (0.0,0.0)  {$ \widetilde{\kappa}_e $};
				\node [xshift=0.4cm, yshift=0.6cm] (0.0,0.0)  {$1$};
				\node [xshift=0.4cm, yshift=-0.6cm] (0.0,0.0) {$0$};
				\fill[top color=Dandelion,bottom color=BlueViolet, middle color = Green!25]
				(-0.1,-0.72) rectangle (0.1,0.75); 
		\end{tikzpicture}} \\
		\addresonator{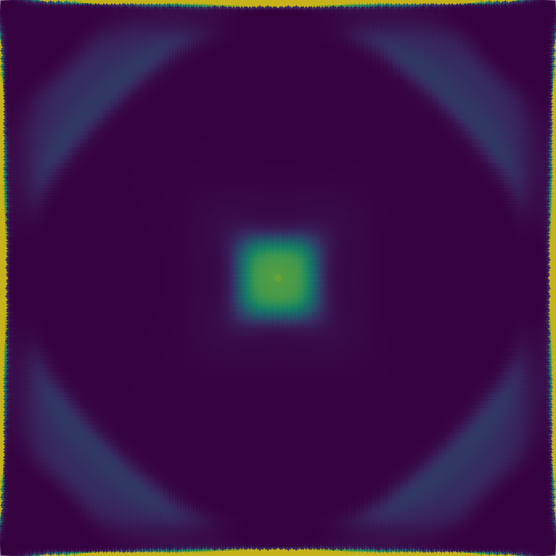} & \addresonator{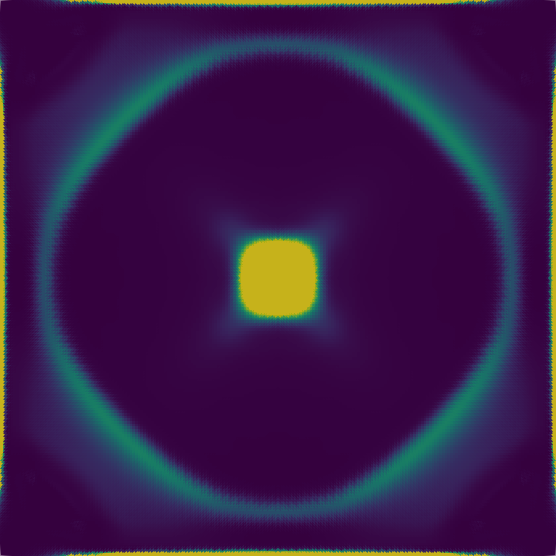} & \addresonator{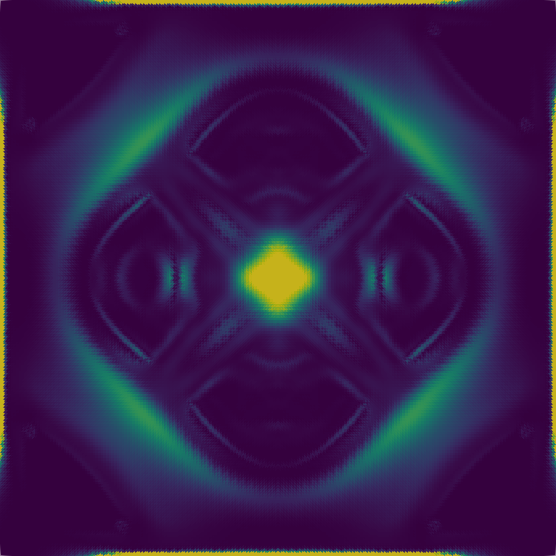} & \addresonator{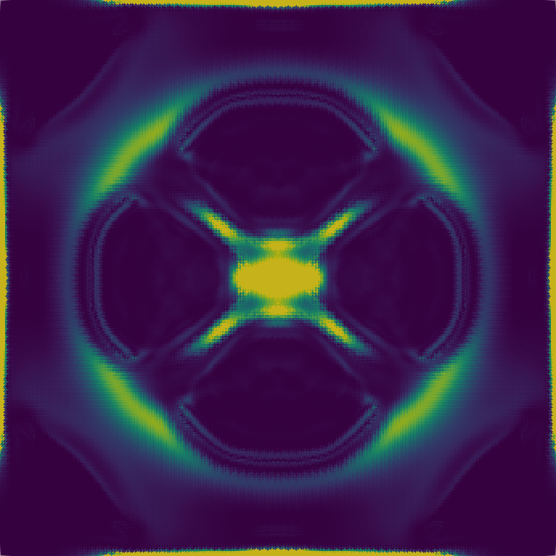} & \addresonator{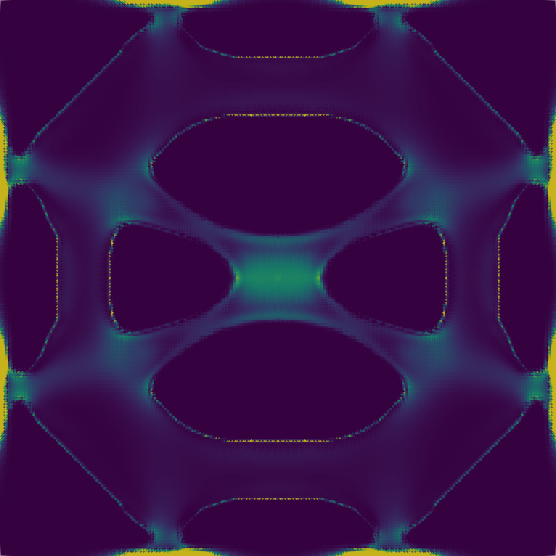}& \addresonator{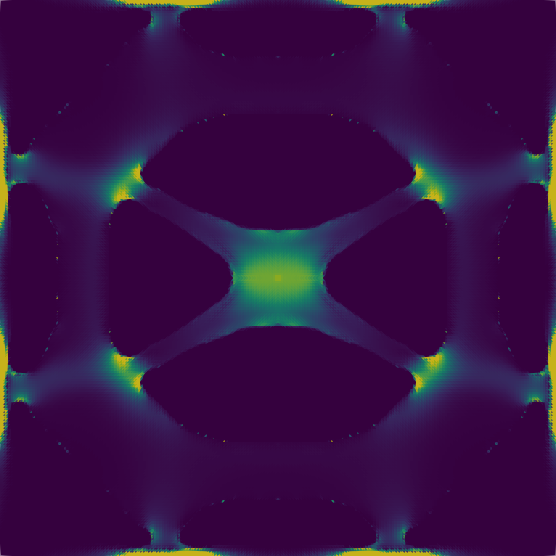} &\imagetop{\begin{tikzpicture} 		
				\fill[top color=Dandelion,bottom color=BlueViolet, middle color = Green!25]
				(-0.1,-0.72) rectangle (0.1,0.75); 
				\node [xshift=0.5cm] (0,0)  {$\overline{\kappa}_e$};
				\node [xshift=0.4cm, yshift=0.6cm] (0,0.0)  {$1$};
				\node [xshift=0.4cm, yshift=-0.6cm] (0,0.0) {$0$};
		\end{tikzpicture}} 
	\end{tblr}	

    \includegraphics{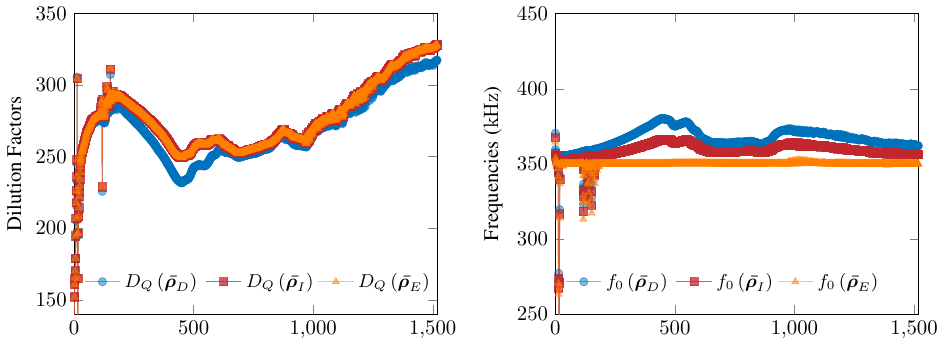}

	\hfill
	\caption{Optimization of a square resonator with a prescribed lower frequency $f_{\min}= \SI{350}{\kilo\hertz}$. The top row shows the evolution of the intermediate projected density, while the second and third rows depict, respectively, the element-level contributions to the geometrically nonlinear and linear modal stiffnesses; these are computed, respectively, as $\widetilde{\kappa}_e = \bs{\Phi}_e^\intercal \tilde{\bs{k}}_e \bs{\Phi}_e/2000$, and $\overline{\kappa}_e = \bs{\Phi}_e^\intercal \bar{\bs{k}}_e \bs{\Phi}_e / 2$ (they are also normalized). The color bar has been capped for both quantities to give a clearer depiction of these fields (as this limit was exceeded locally). The bottom two graphs show the evolution of the dilution factor (left) and fundamental resonance frequency (right) throughout the optimization for all three fields of the robust projection.}
	\label{tbl:f350}
\end{figure}
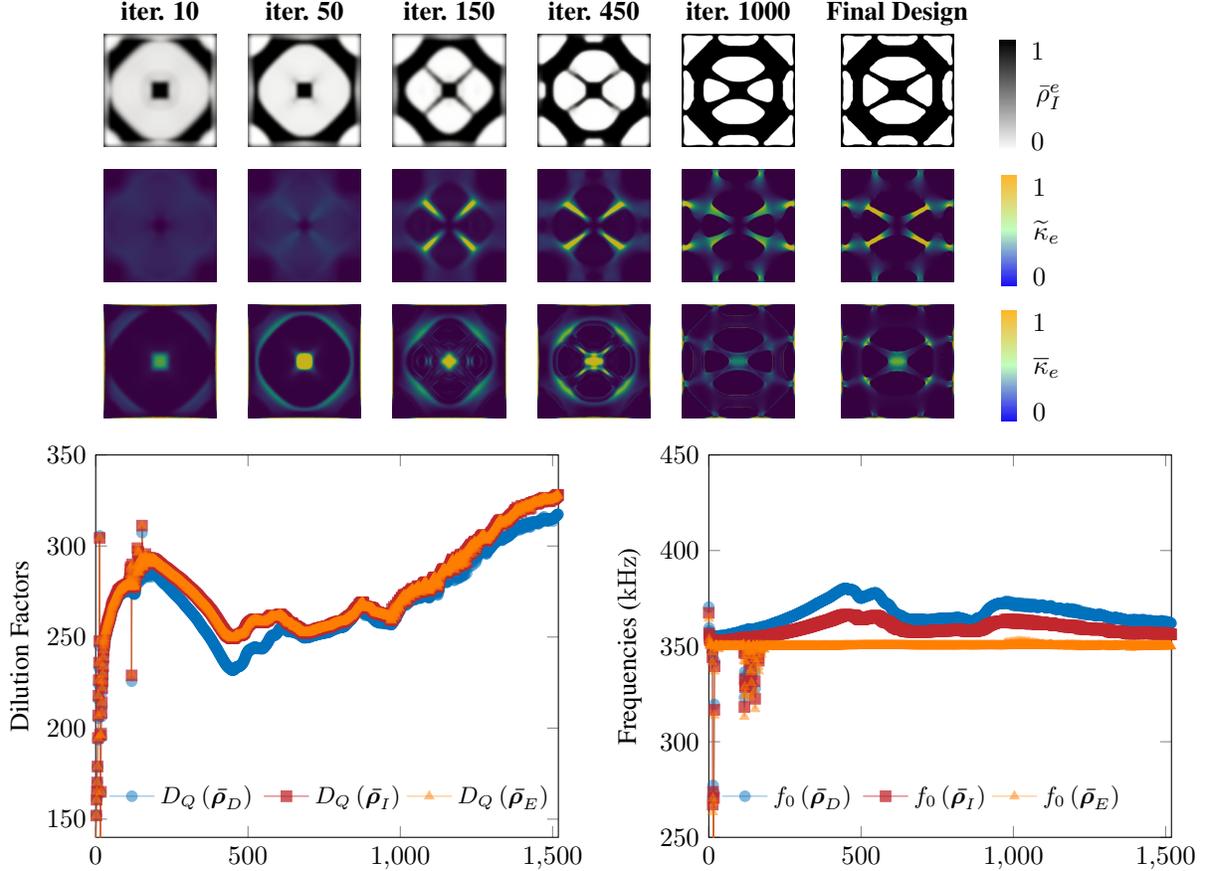

%
%
\begin{figure}
	\centering
	\begin{tblr}{
			cells={valign=m,halign=c},
			row{1}={font=\bfseries,rowsep=0pt},
			colspec={QQQQQQQ},
			column{1}={valign=m},
		}
		iter. 10  & iter. 50 & iter. 150 & iter. 450& iter. 1000 &Final Design &  & \\
		\addresonator{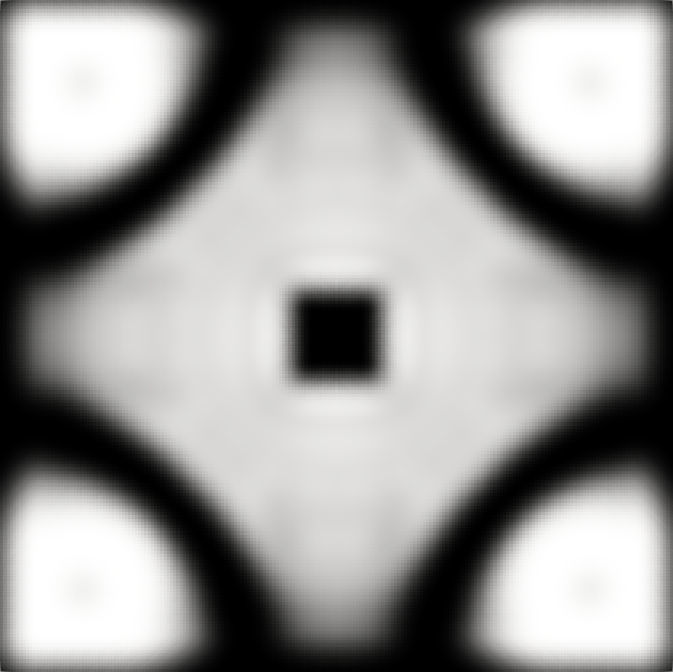} & \addresonator{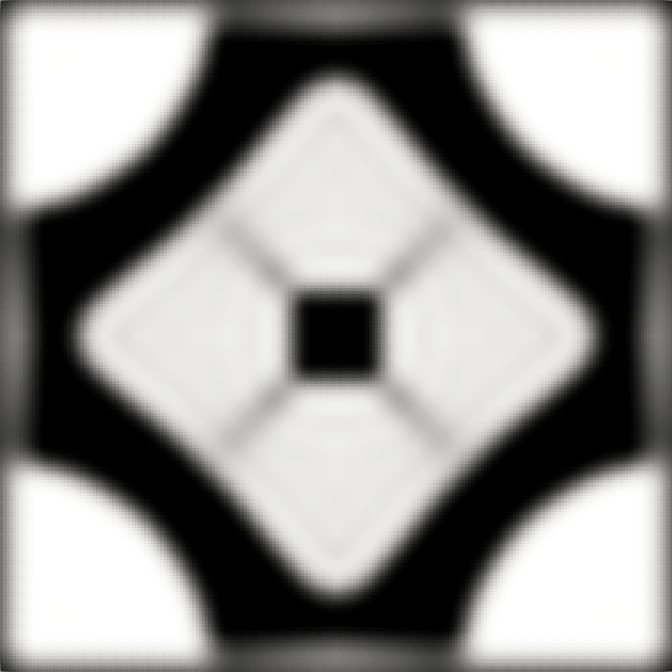} & \addresonator{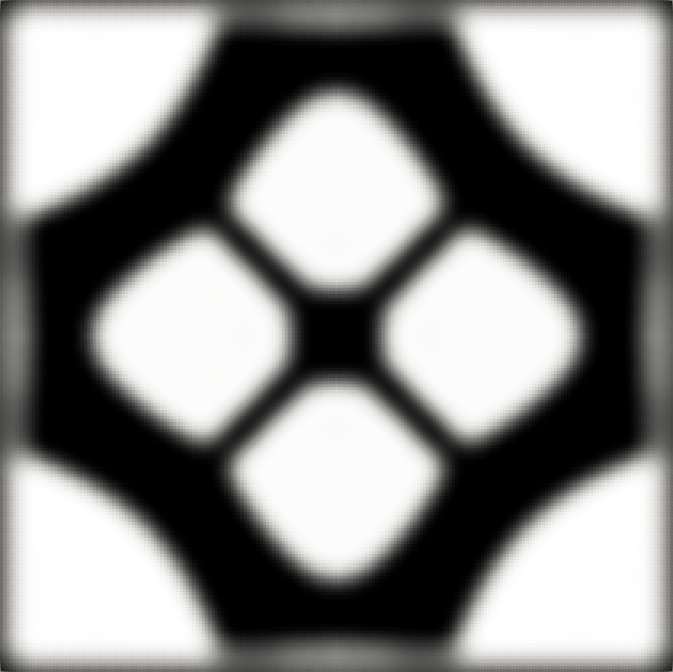} & \addresonator{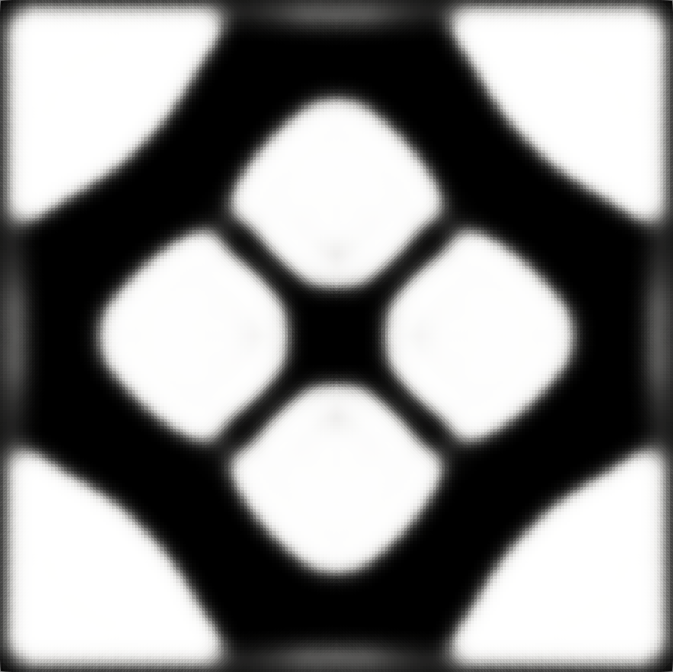} & \addresonator{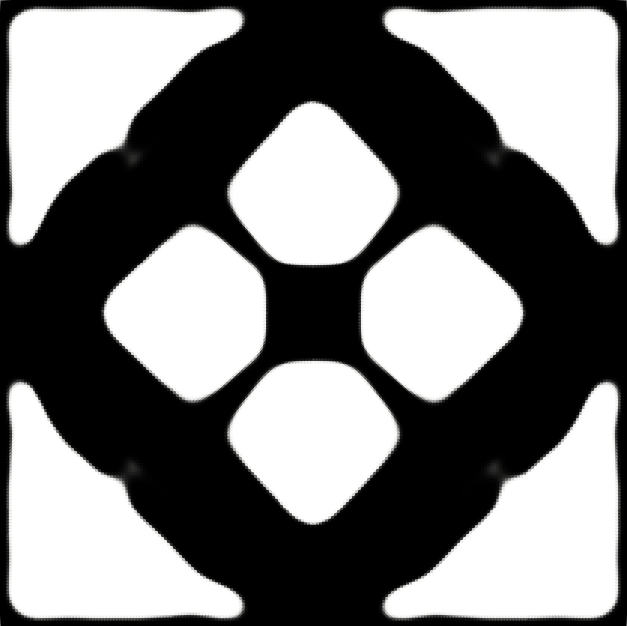} & \addresonator{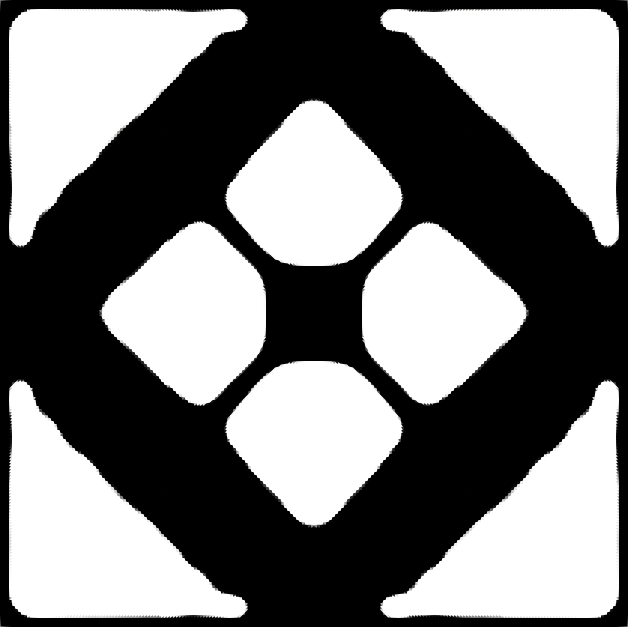} & \imagetop{\begin{tikzpicture} 		
				\node [xshift=0.55cm] (0.0,0.0)  { $\bar{\rho}^{e}_I$ };
				\node [xshift=0.4cm, yshift=0.6cm] (0.0,0.0)  {$1$};
				\node [xshift=0.4cm, yshift=-0.6cm] (0.0,0.0) {$0$};
				\fill[top color=black,bottom color=white]
				(-0.1,-0.72) rectangle (0.1,0.75); 
		\end{tikzpicture}} \\
		\addresonator{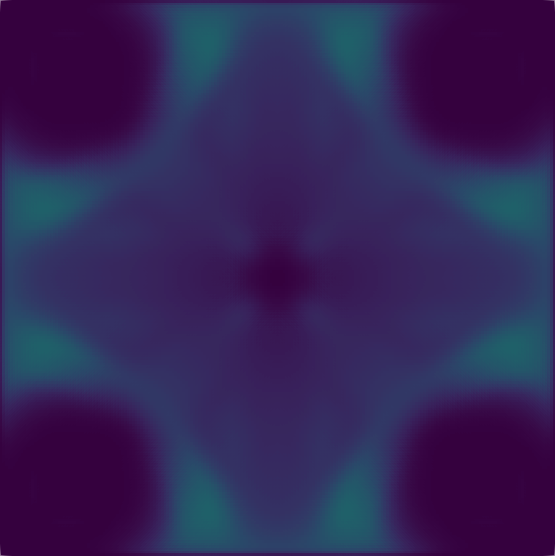} & \addresonator{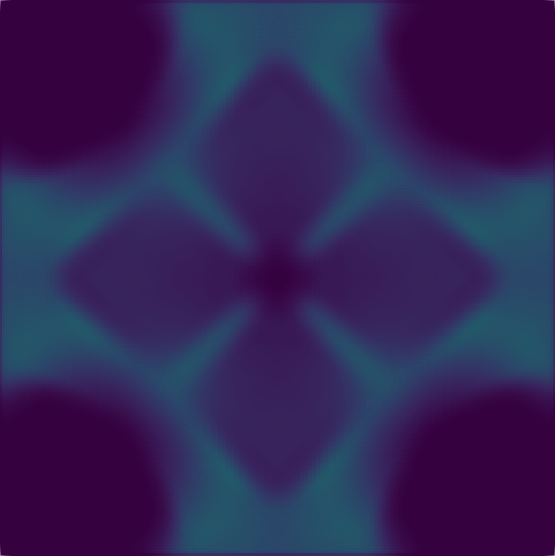} & \addresonator{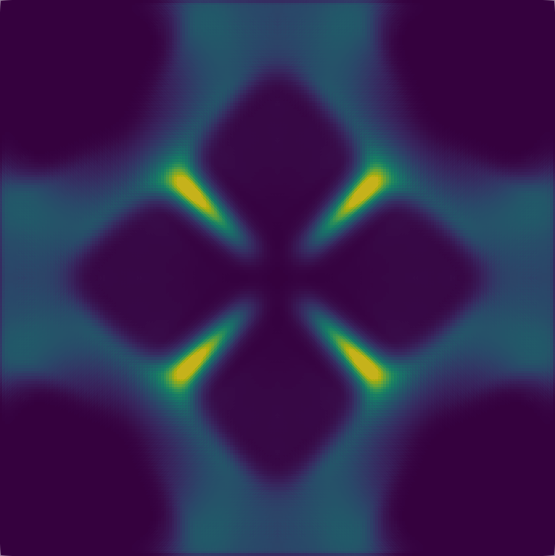} & \addresonator{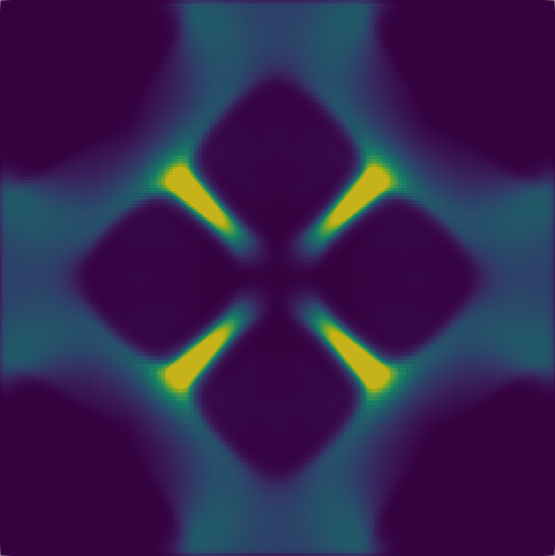} & \addresonator{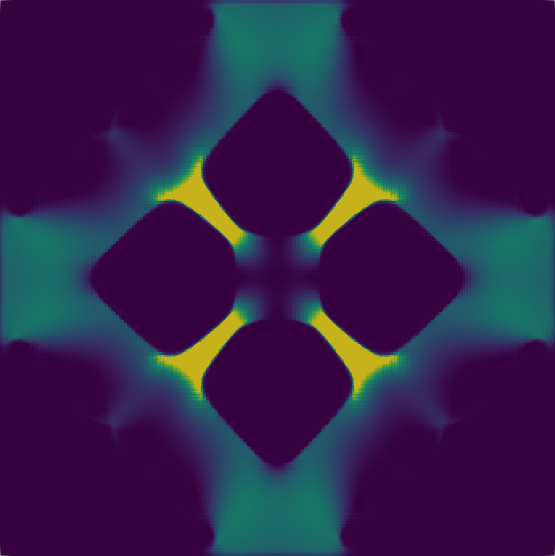}& \addresonator{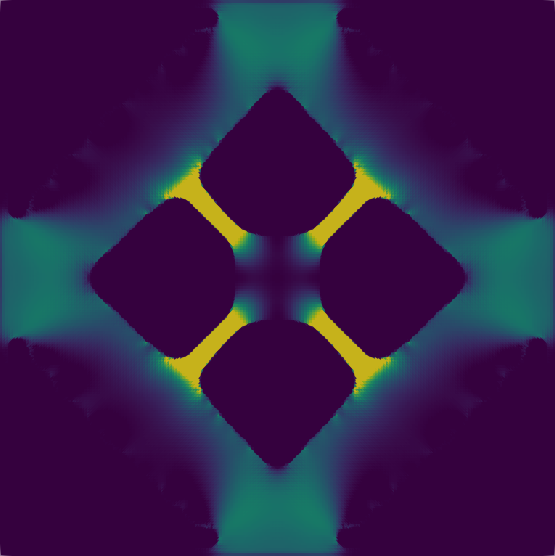} & \imagetop{\begin{tikzpicture} 		
				\node [xshift=0.5cm] (0.0,0.0)  { $ \widetilde{\kappa}_e$ };
				\node [xshift=0.4cm, yshift=0.6cm] (0.0,0.0)  {$1$};
				\node [xshift=0.4cm, yshift=-0.6cm] (0.0,0.0) {$0$};
				\fill[top color=Dandelion,bottom color=BlueViolet, middle color = Green!25]
				(-0.1,-0.72) rectangle (0.1,0.75); 
		\end{tikzpicture}} \\
		\addresonator{f400k10} & \addresonator{f400k50} & \addresonator{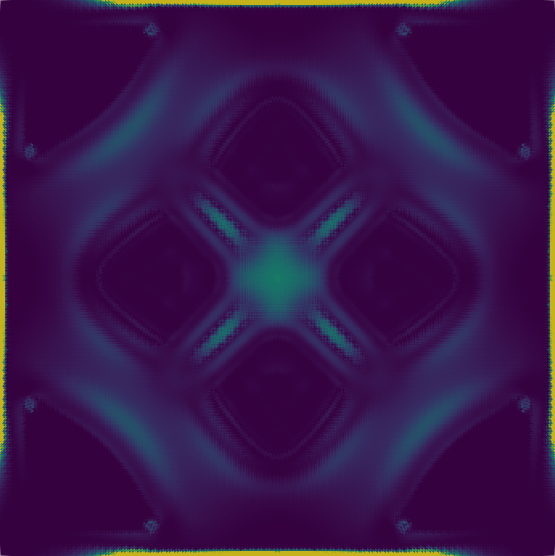} & \addresonator{f400k450} & \addresonator{f400k1000}& \addresonator{f400k1519} &\imagetop{\begin{tikzpicture} 		
				\fill[top color=Dandelion,bottom color=BlueViolet, middle color = Green!25]
				(-0.1,-0.72) rectangle (0.1,0.75); 
				\node [xshift=0.5cm] (0,0)  {$\overline{\kappa}_e$};
				\node [xshift=0.4cm, yshift=0.6cm] (0,0.0)  {$1$};
				\node [xshift=0.4cm, yshift=-0.6cm] (0,0.0) {$0$};
		\end{tikzpicture}} 
	\end{tblr}	
	
	\includegraphics{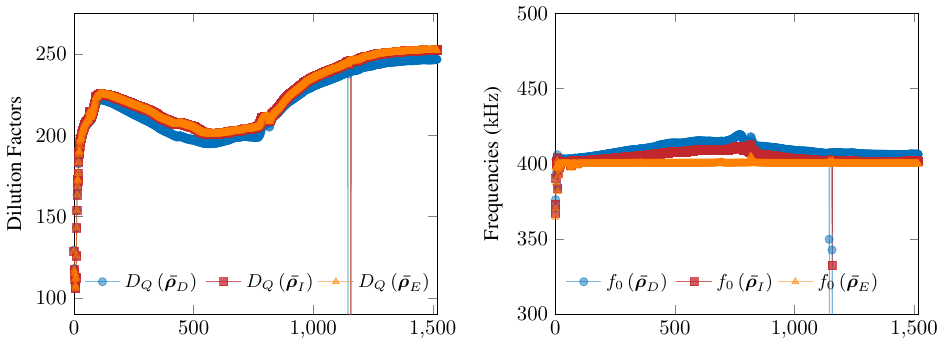}

	\hfill
	\caption{Optimization of a square resonator with a prescribed lower frequency $f_{\min}= \SI{400}{\kilo\hertz}$. The top row shows the evolution of the intermediate projected density, while the second and third rows depict, respectively, the element-level contributions to the geometrically nonlinear and linear modal stiffnesses; these are computed, respectively, as $\widetilde{\kappa}_e = \bs{\Phi}_e^\intercal \tilde{\bs{k}}_e \bs{\Phi}_e/2000$, and $\overline{\kappa}_e = \bs{\Phi}_e^\intercal \bar{\bs{k}}_e \bs{\Phi}_e / 2$ (they are also normalized). The color bar has been capped for both quantities to give a clearer depiction of these fields (as this limit was exceeded locally). The bottom two graphs show the evolution of the dilution factor (left) and fundamental resonance frequency (right) throughout the optimization for all three fields of the robust projection.}
	\label{tbl:f400}
\end{figure}
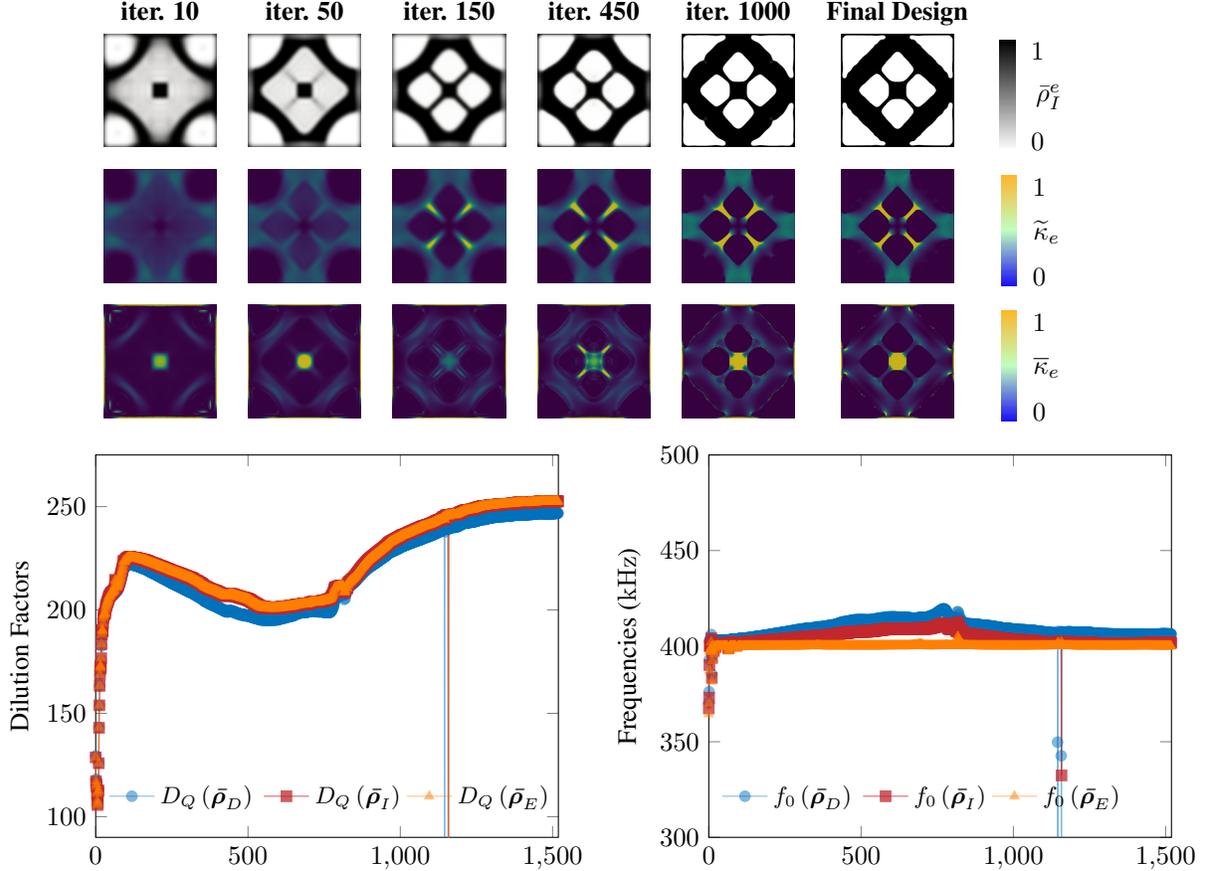

As apparent from the graphs in Figures \ref{tbl:f300},\ref{tbl:f350}, and \ref{tbl:f400}, the dilution factor initially increases sharply, after which it decreases during some iterations; this is then followed by another overall increase.
This behavior can be explained by the disappearance of intermediate-density elements as the projection slope is increased to force the optimizer converge to a fully black-and-white design.
Elements with intermediate densities can have high geometric stiffness and low bending stiffness, which makes them beneficial for the optimizer. As they disappear, the optimizer pushes the design towards a new viable point in the evolving design space (since the continuation scheme changes the optimization problem), thereby reducing the objective function.

The figures show that the regions of high geometrically nonlinear and linear modal stiffnesses (second and third rows, respectively) do not overlap. The linear modal stiffness---i.e., the denominator in \eqref{eq:DQ objective}---has high values close to the outermost border, as well as in the central region. Conversely, the geometrically nonlinear modal stiffness---i.e., the numerator in~\eqref{eq:DQ objective}---has the highest values along the tethers that connect the central pad to the outer boundary. This distribution of geometrically nonlinear and linear modal stiffnesses (which are proportional to energy stored in stretching and bending, respectively) is in accordance with the model of dissipation dilution discussed by Engelsen et al.~\cite{Engelsen2021proceeding}---i.e., the bending energy dominates the central and clamping regions of the mode shape, while the stretching energy dominates the region in which the highest curvature of the mode is obtained.

Another observation, specifically related to Figure~\ref{tbl:f400}, is that at some iterations (around iteration 1150) both the objective function and the fundamental frequency drop drastically. At these iterations, the updated design features regions with slight compression, which yielded these outlying values. However, the overall objective remained positive, which indicates that the fundamental mode of those systems remains \textit{physical}. Fortunately, this did not strongly influence the optimizer, as shown by the succeeding designs. This phenomenon, which was more commonly seen in our early optimizations, is discussed in more detail later in Section~\ref{sec:conclusion} and~\ref{app:compressive_void_regions}.

\newcommand{\addcomsol}[1]{\imagetop{\includegraphics[height=2cm]{figures/#1}}}
\newcommand{\addcomsolSize}[2]{\imagetop{\includegraphics[height=#2 cm]{figures/#1}}}
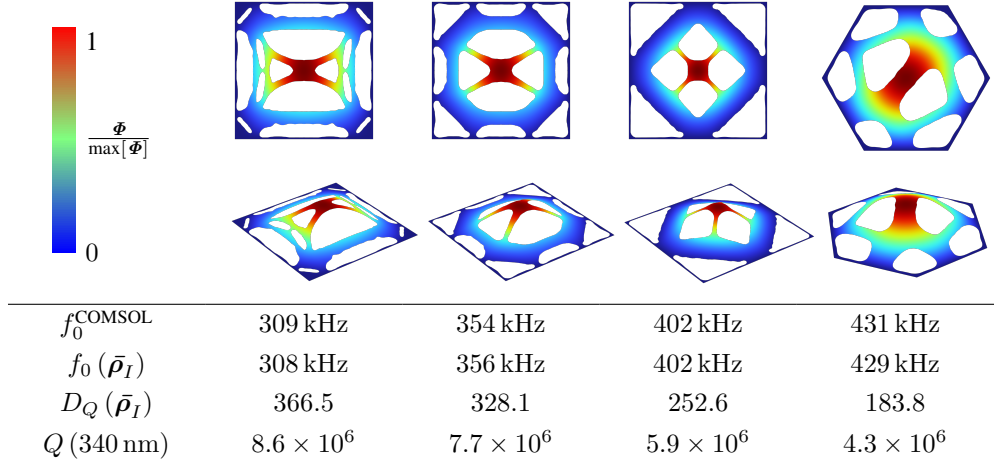
\begin{figure}[!ht]
	\centering
	\begin{tblr}{
					cells={valign=m,halign=c},
					colspec={QQQQQ},
					column{1}={valign=m},
				}
				\parbox[c]{2.2cm}{
				\centering
				\imagetop{\begin{tikzpicture} 		
						\fill[top color=red,bottom color=blue,middle color=green!50, ]
						(0,0) rectangle (0.3,3); 
						\node [xshift=25, yshift =42.5]  {$\frac{\bs{\Phi}}{ \text{max}\left[\bs{\Phi}\right]} $};
						\node [xshift=15, yshift =80]  {1};
						\node [xshift=15, yshift =0]  {$0$};
				\end{tikzpicture}}
				}

				& \parbox[c]{2.2cm}{%
					\centering
					\addcomsol{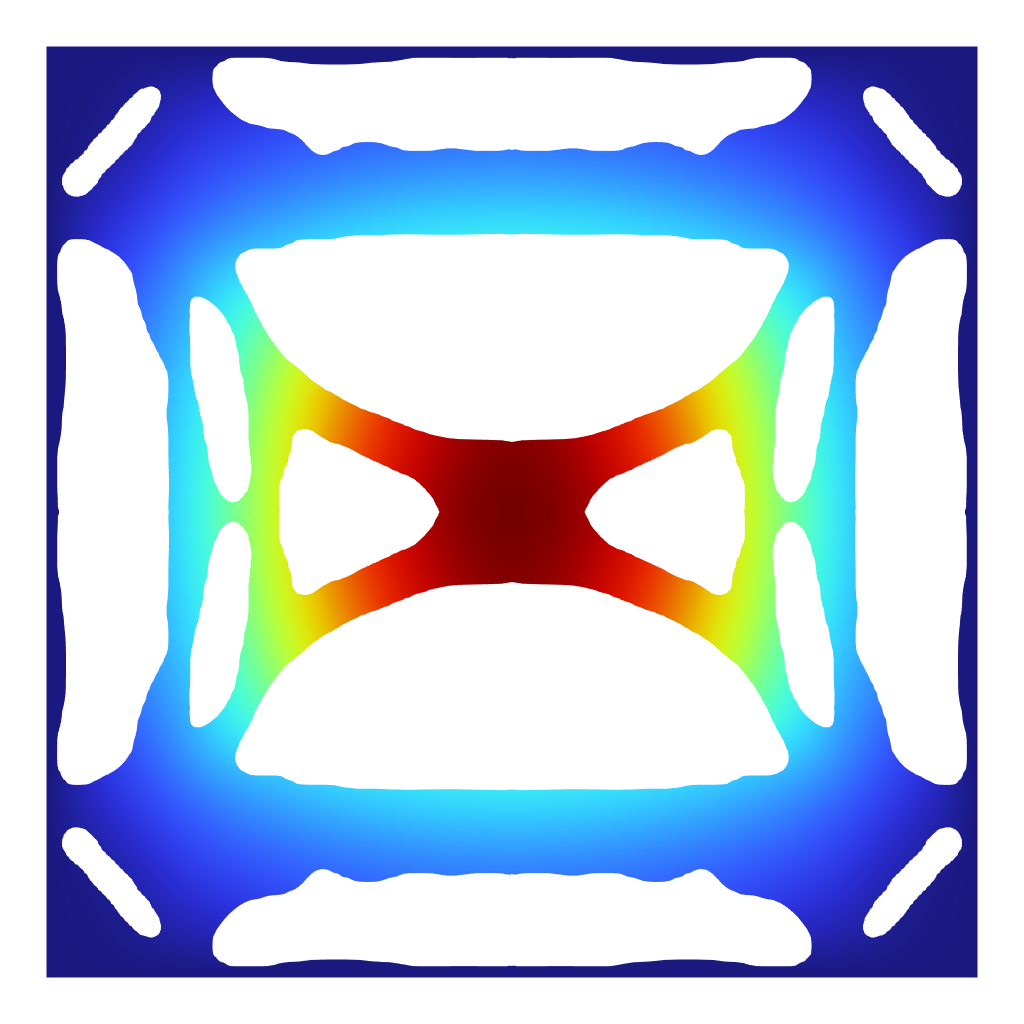}\\
					\addcomsol{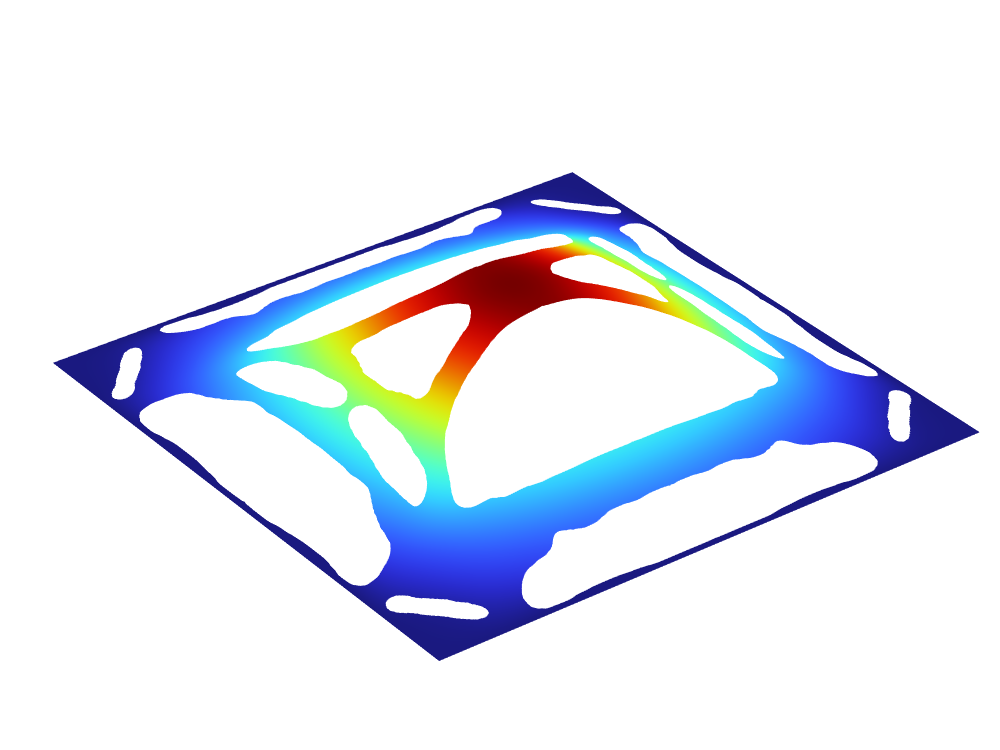}}%
				 & \parbox[c]{2.2cm}{%
				 	\centering
				\addcomsol{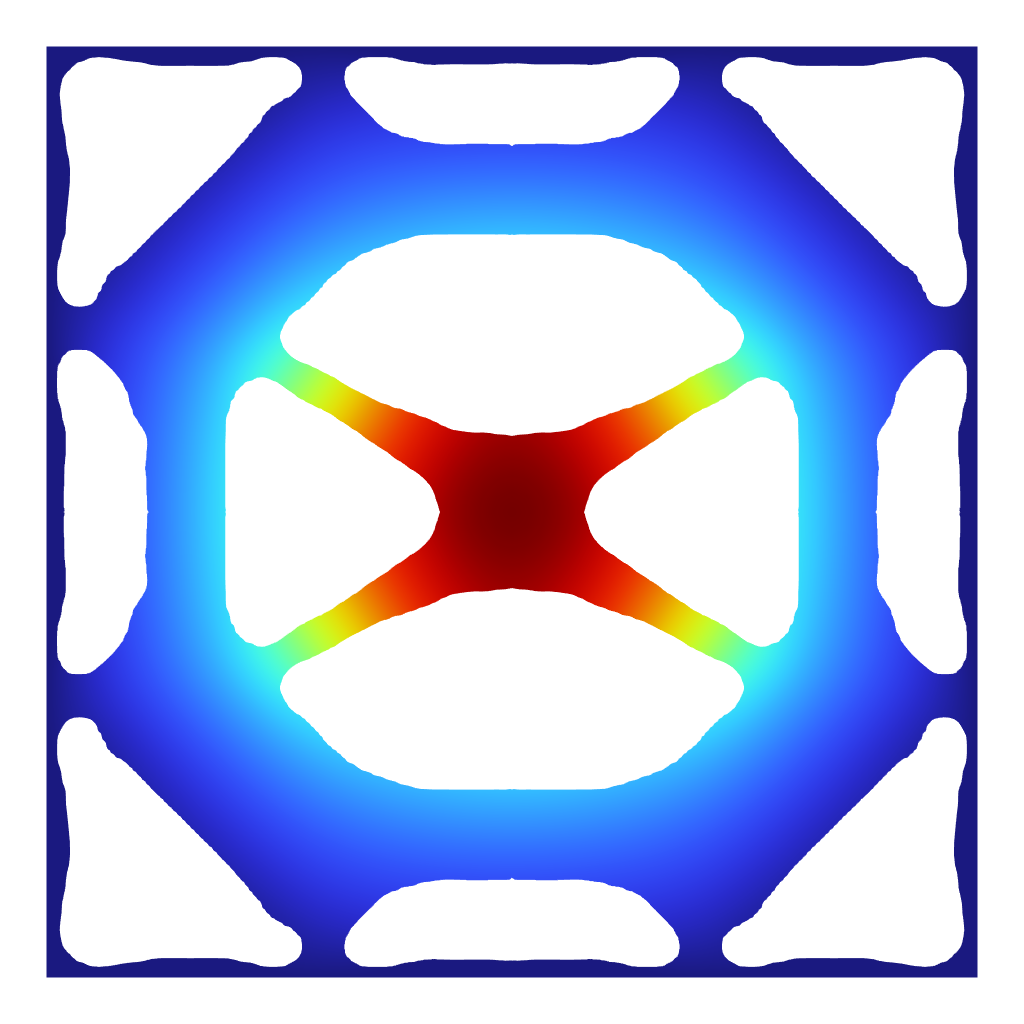}\\
				\addcomsol{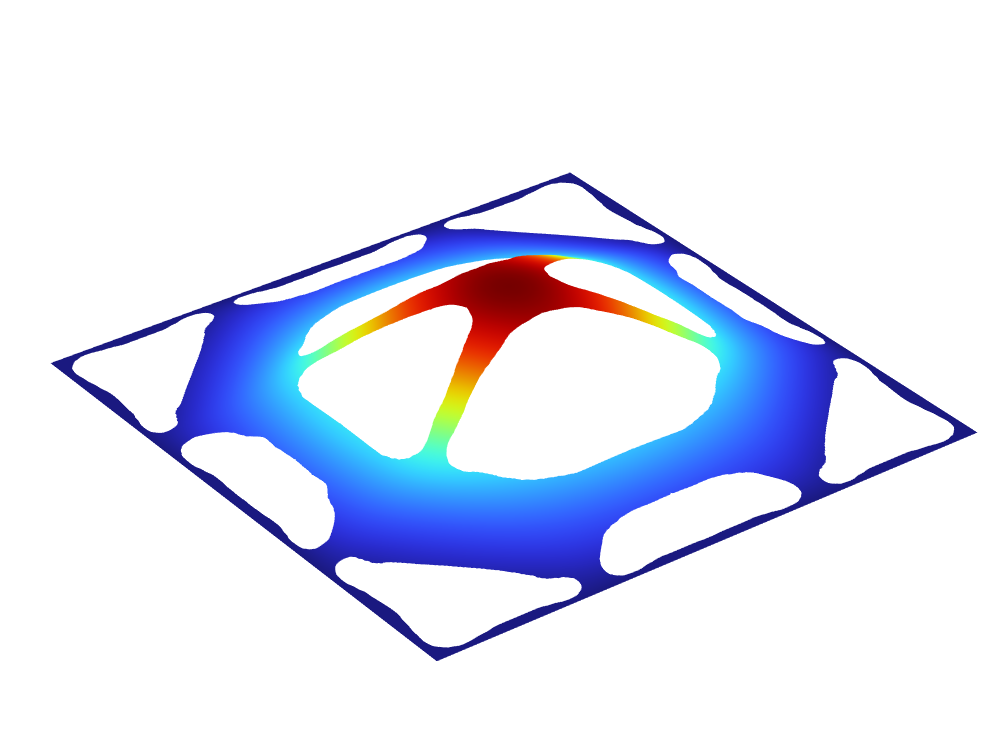}} & \parbox[c]{2.2cm}{%
				\centering
					\addcomsol{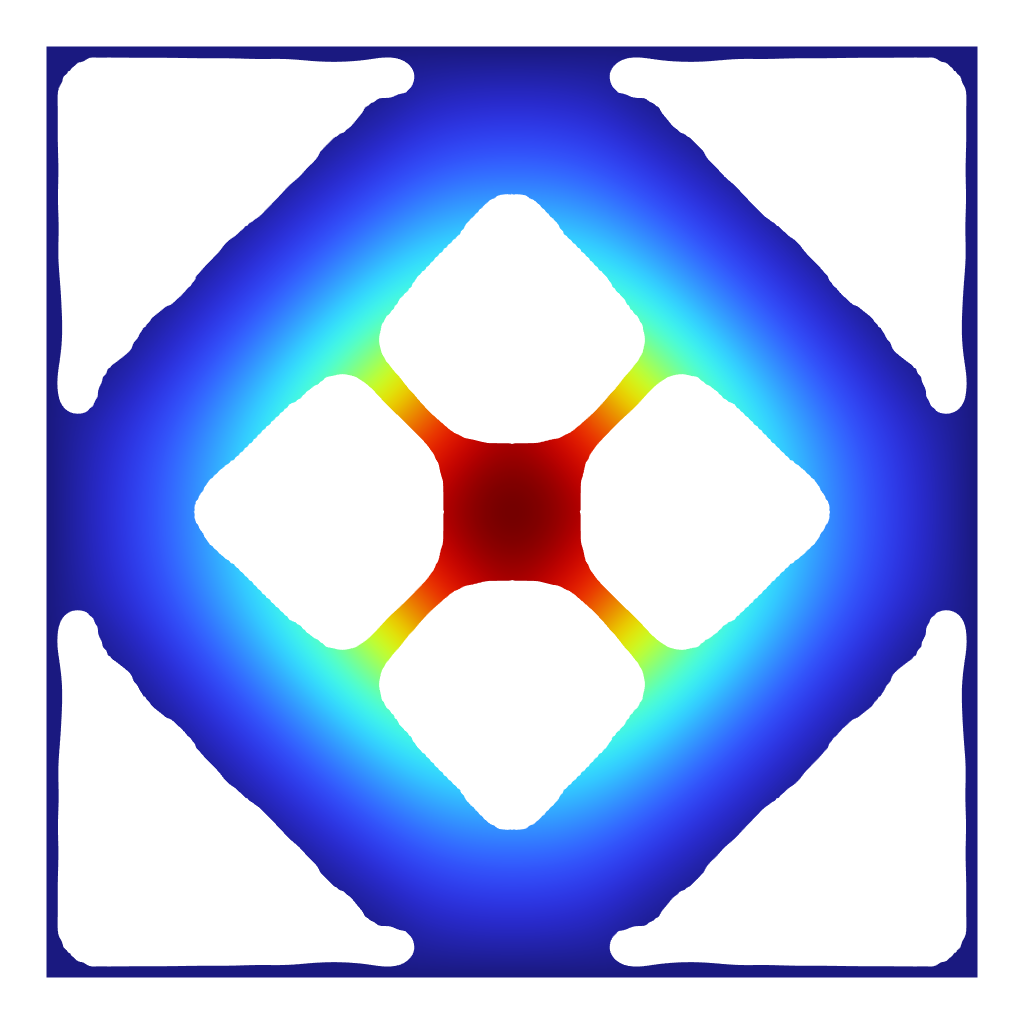}\\
					\addcomsol{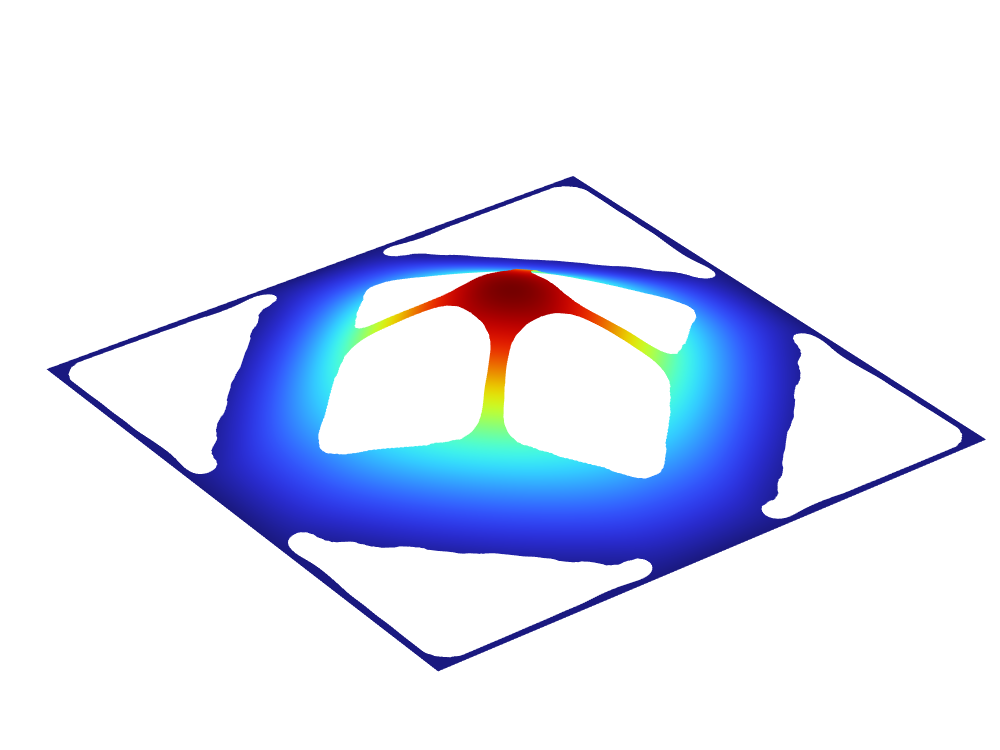}} & \parbox[c]{2.2cm}{%
					\centering
						\addcomsolSize{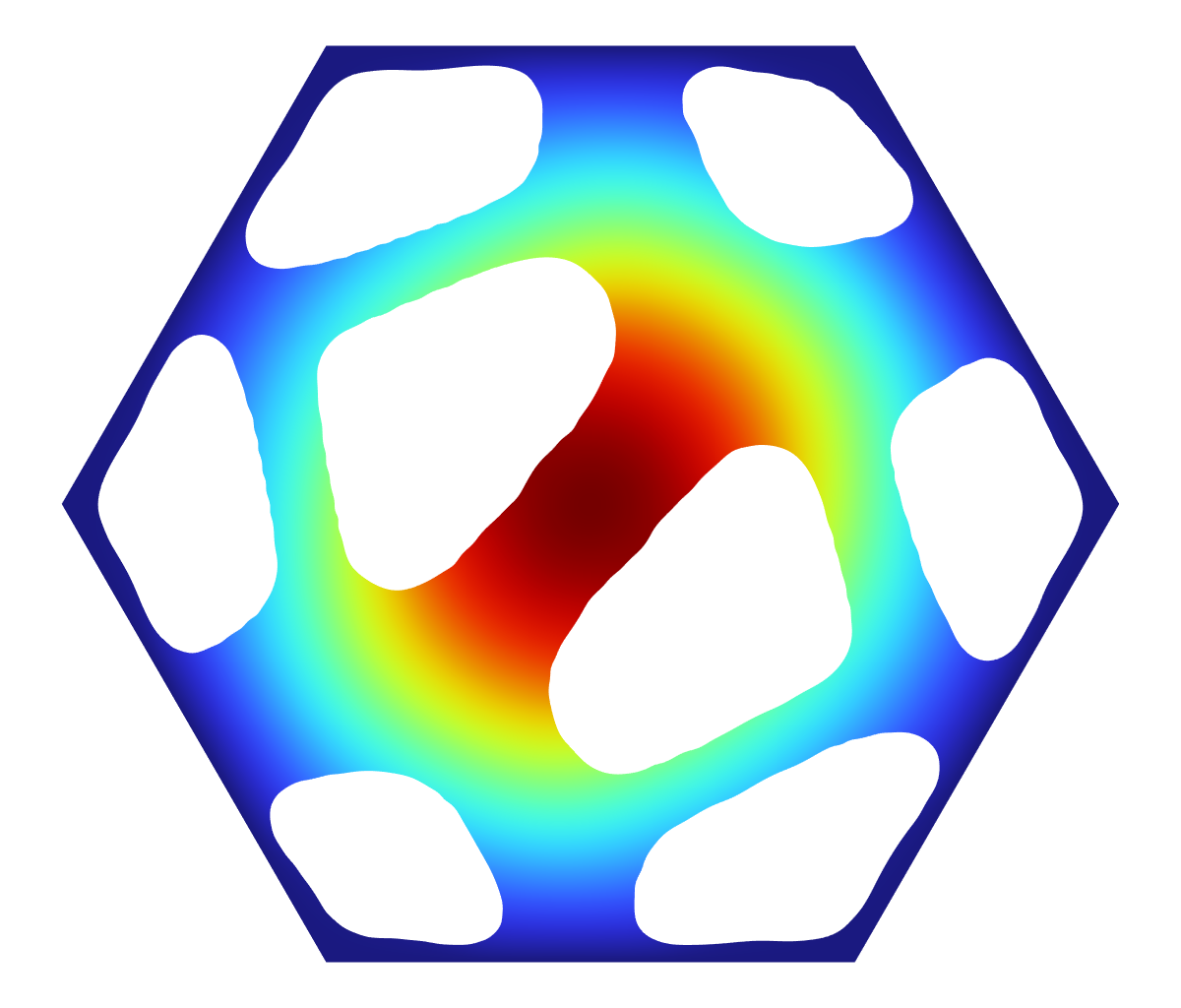}{2.12}\\
						\addcomsolSize{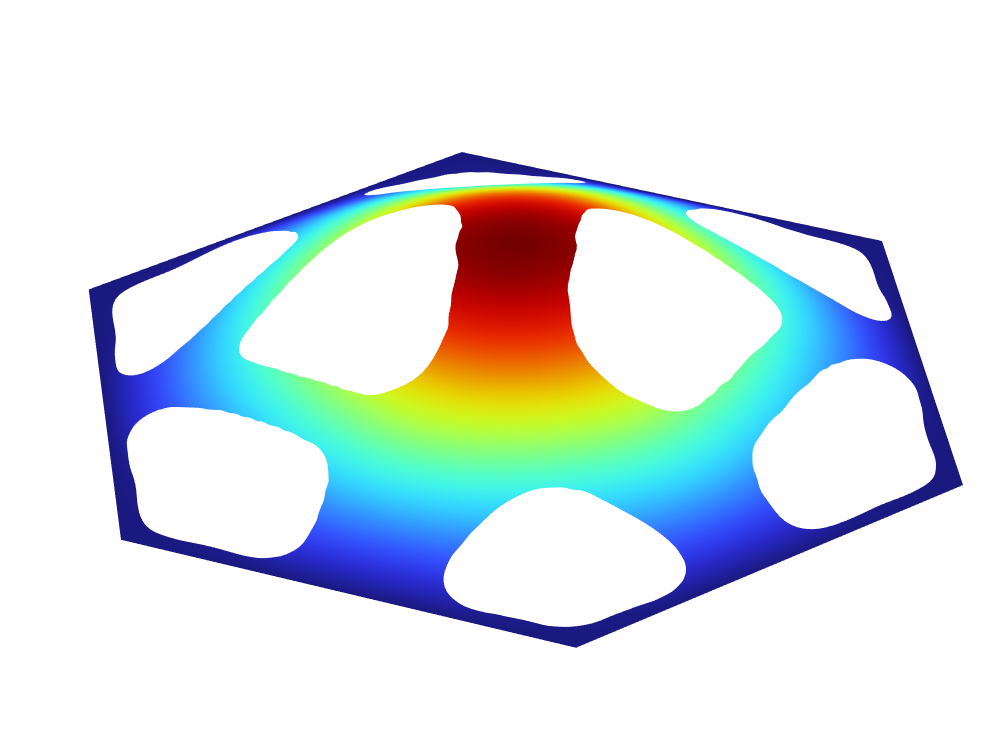}{1.8}} \\ \hline

			$f_0^{\text{COMSOL}}  $ & $\SI{309}{\kilo\hertz}$ & $\SI{354}{\kilo\hertz}$ &  $\SI{402}{\kilo\hertz}$ & $\SI{431}{\kilo\hertz}$ \\
			$f_0\left( \bs{\bar{\rho}}_I \right)$  & $\SI{308}{\kilo\hertz}$ & $\SI{356}{\kilo\hertz}$ &  $\SI{402}{\kilo\hertz}$ & $\SI{429}{\kilo\hertz}$ \\
			$D_Q\left( \bs{\bar{\rho}}_I \right)$ & $366.5$ & $328.1$ & $252.6$ & $183.8$ \\ 
			$Q\left(\SI{340}{\nano\metre}\right)$ & $8.6\times 10^{6}$ & $7.7\times10^{6}$ & $5.9\times10^6$ & $4.3\times10^6$ \\ 
\end{tblr}	

\caption{Overview of the performance all the final designs, taking $\bs{\bar{\rho}}_I$ as the design blue-print. COMSOL Multiphysics was used to verify the resonance frequencies, and to visualize the mode shapes. The color bar on the left indicates the out-of-plane displacement magnitude of each normalized (fundamental) mode shape. In the bottom row of the table, we add the computed $Q$ factor based on a membrane thickness of $\SI{340}{\nano\metre}$, for which we approximated the intrinsic $Q$ factor as $Q_0 \approx 6900\times h / \SI{100}{\nano\metre}$, as done by D. Shin et al. \cite{shin21}. For all designs, th final volume fraction is equal to the volume limit, i.e. $V_m = V_c =0.5$.}
\label{tbl:comsol}
\end{figure}

\subsection*{Non-convexity of the objective landscape}
An aspect that we now discuss is the non-convexity of our optimization problem. In fact, most problems in TO are non-convex, which means that their objective function landscape has multiple local optima~\cite{sigmund2004}. The objective function landscape is the result of the highly-dimensional parameterization and all other aspects that make up the optimization problem. Therefore, finding ``optimal'' designs in TO can be considered somewhat of an art as well as a science, as parameters such as the filter radius, initial material distribution, constraint values, convergence criteriaon, and the projection continuation scheme all influence the objective function landscape and the way in which the optimizer traverses it.

To showcase the non-convexity of the objective landscape for the problem at hand, we consider two identical optimizations where the only difference is the initial material distribution---i.e., a different starting point in our highly-dimensional design space. For our first optimization we again use a homogeneous distribution of material by setting the design density of all elements to $\rho_e = 0.5$. In the second optimization, we start with an \textit{X-shaped} resonator geometry inspired by the work of Norte et al.~\cite{norte16}, where the central pad is connected by tethers to the corners of the domain. For simplicity, we increased the projection slope in~\eqref{eq:projection} by 100\% every \num{150} iterations. We also set $f_{\min} = \SI{300}{\kilo\hertz}$, until it reached $\beta = 128$ (resulting in a total of \num{1200} iterations).

\begin{figure}[!ht]
	\centering
	\begin{tblr}{
			cells={valign=m,halign=c},
			row{1}={bg=lightgray,font=\bfseries,rowsep=0pt},
			colspec={QQQQQQQ},
			column{1}={valign=m},
		}
		Initial  Design &  & &  &  Final Design  & $D_Q\left( \bs{\bar{\rho}}_I \right)$ & $f_0 \left( \bs{\bar{\rho}}_I \right)$ \\
		\addresonator{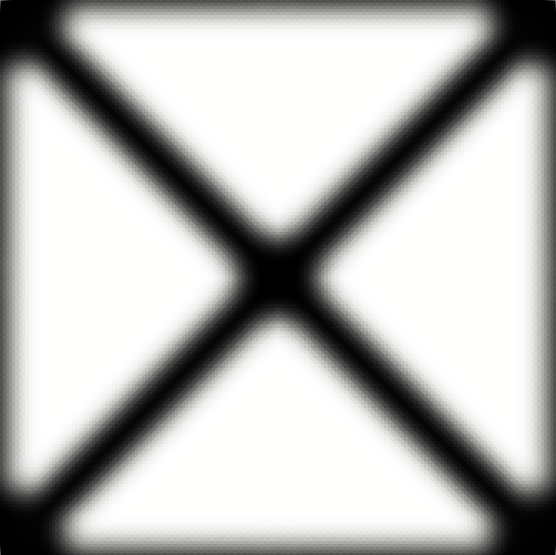} &\addresonator{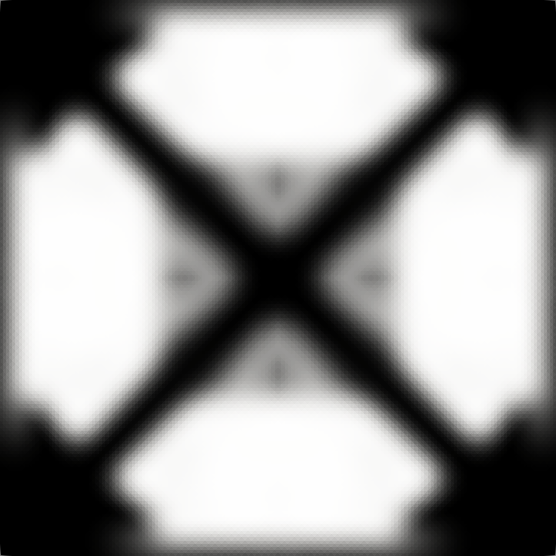} & \addresonator{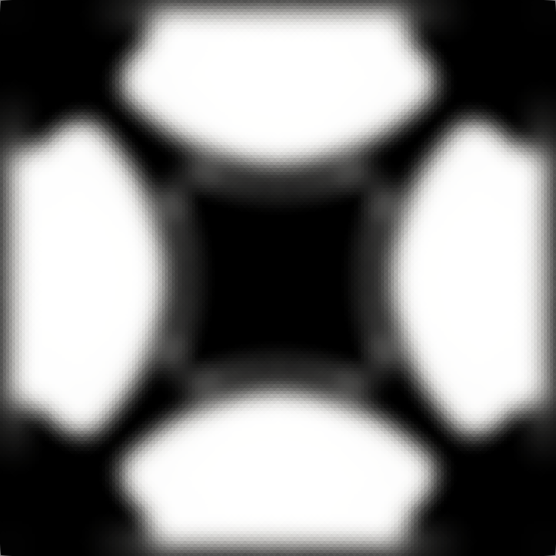}&\addresonator{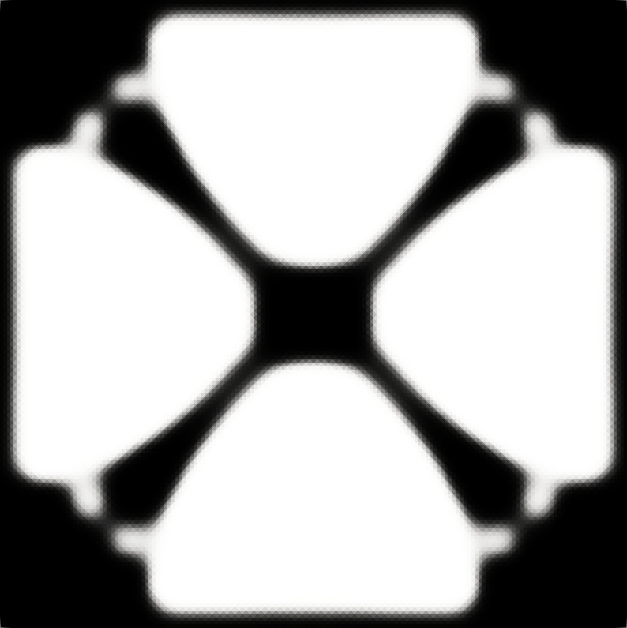}&\addresonator{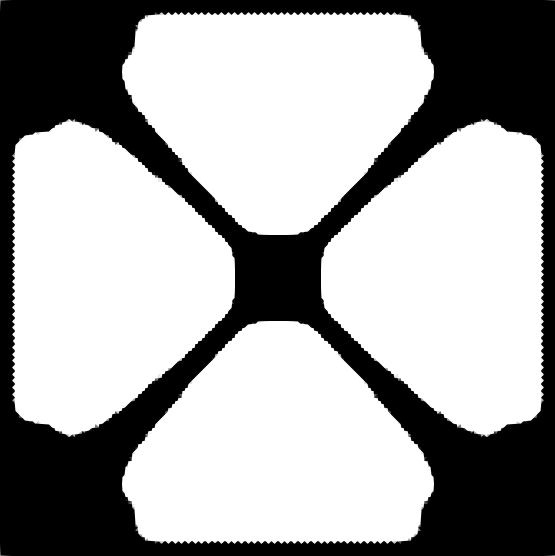} & 241.0 & $\SI{376.2}{\kilo\hertz}$  \\
		\addresonator{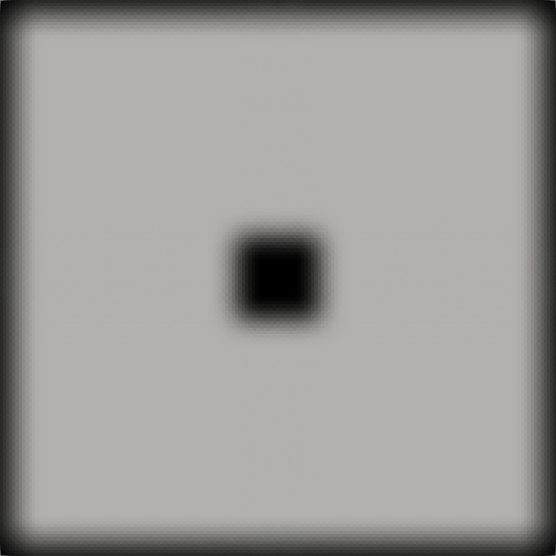}&\addresonator{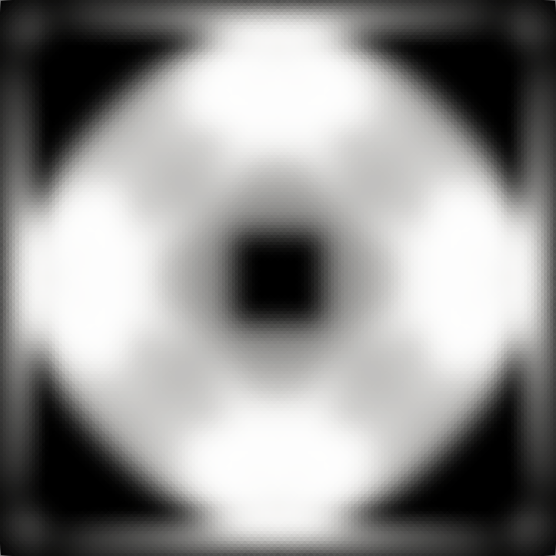} &\addresonator{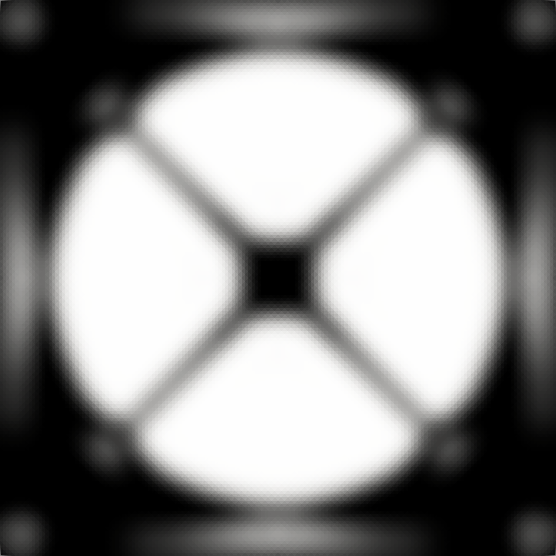} &\addresonator{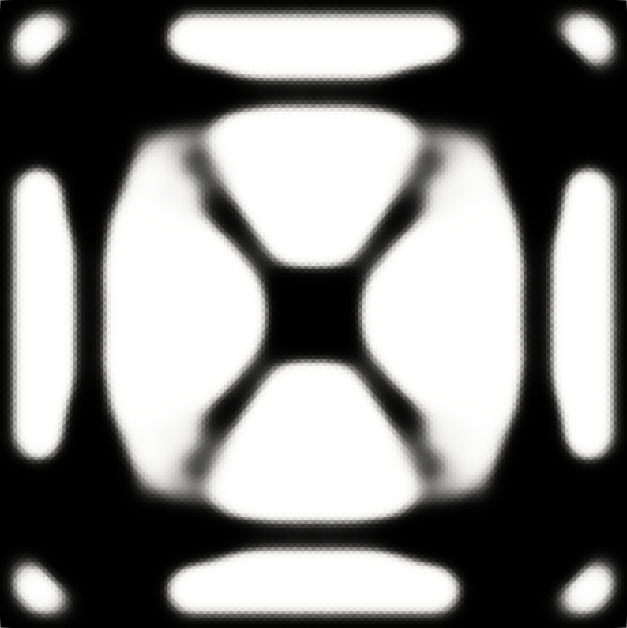}&\addresonator{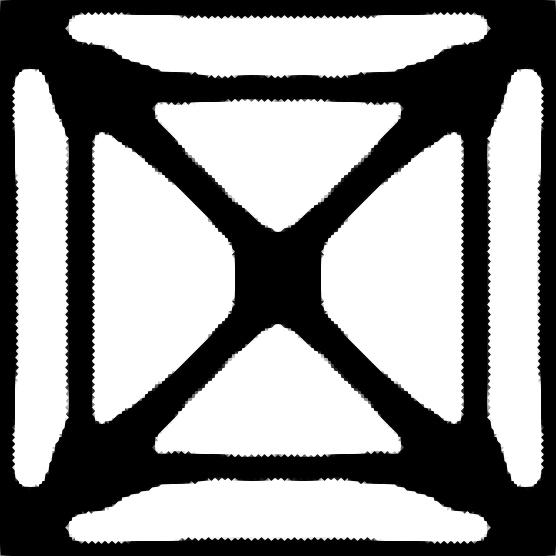} & 282.5 & $\SI{306.7}{\kilo\hertz}$ 
	\end{tblr}
	\caption{The design evolution of two optimizations that demonstrate the non-convexity of our objective function landscape. The starting material density distribution was different, i.e., we start our optimization from two different points in our highly-dimensional design space. All other parameters of the optimization were kept the same. The frequency lower bound was set to $f_{\min} = \SI{300}{\kilo\hertz}$, and a continuation scheme that doubled the projection slope $\beta$ every 150 iterations was used. The final designs have unique geometries, and show different fundamental frequencies and dilution factors.}
	\label{tbl:nonconvex}
\end{figure}

The design evolution and performance are summarized in Figure~\ref{tbl:nonconvex}. These results show that, depending on our starting design, the optimizer converges to two completely different local minima. 
It is therefore worth noting that the fact that our objective function is non-convex should always be kept in mind when viewing our results.

\subsection{Hexagonal computational domain}\label{sec:hexagon simulations}

\begin{figure}[t]
	\centering
    \includegraphics{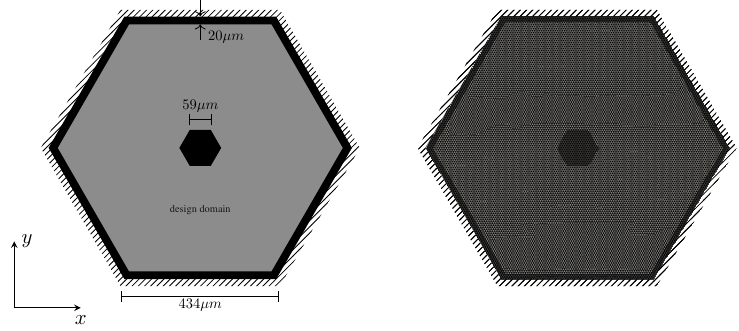}
	\caption{Left: The hexagonal design domain setup used in the numerical study. Right: The mesh used to model this domain. In the dark inner hexagon and outermost boundary, the design density was set as $\rho_e=1$. The dashed border indicates the homogeneous Dirichlet boundary conditions prescribed for the outermost elements, i.e. zero displacement and rotation.}
	\label{fig:domainHexagon}
\end{figure}

\begin{figure}
	\centering
	\begin{tblr}{
			cells={valign=m,halign=c},
			row{1}={font=\bfseries,rowsep=0pt},
			colspec={QQQQQQQ},
			column{1}={valign=m},
		}
		iter. 10  & iter. 50 & iter. 150 & iter. 590 & iter. 650 & Final Design &  & \\
		\addresonator{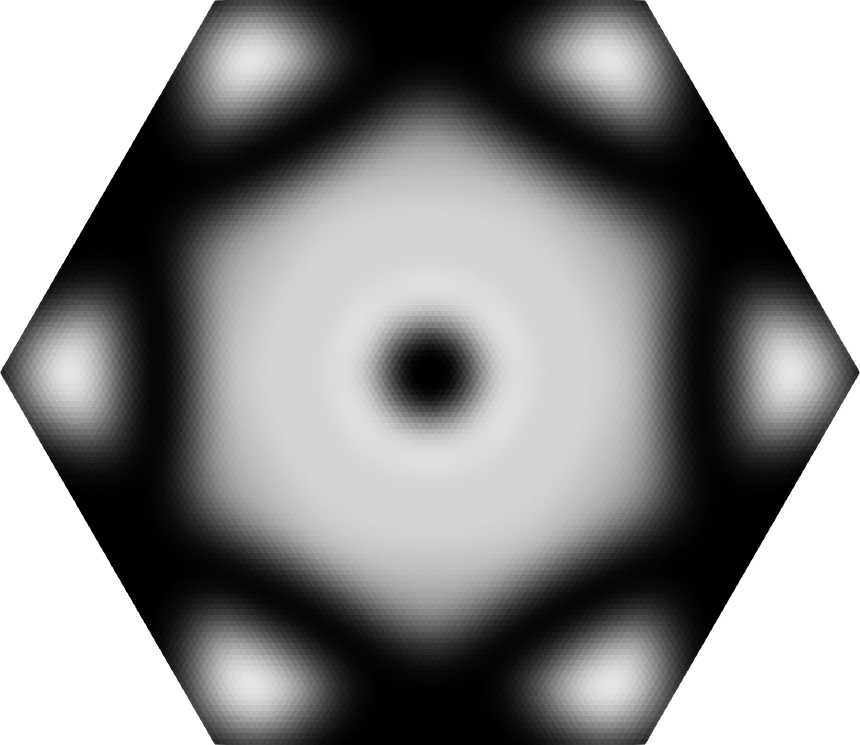} & \addresonator{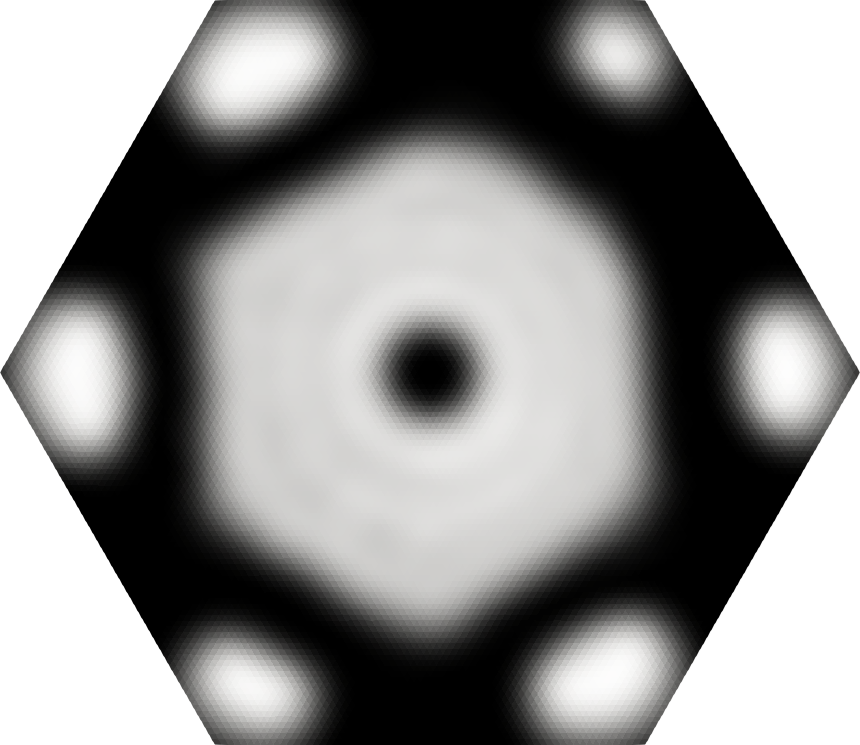} & \addresonator{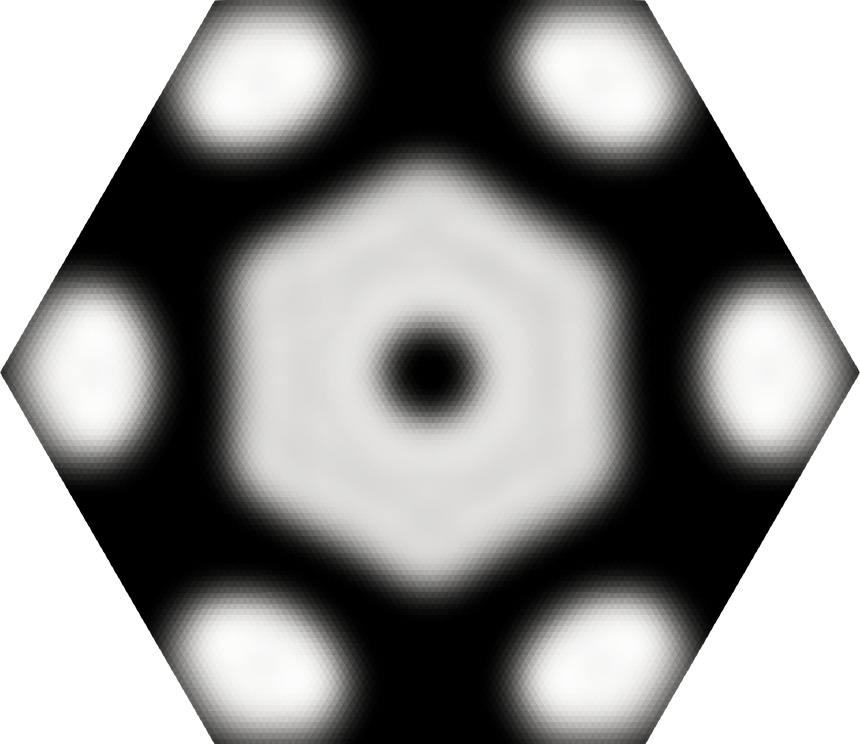} & \addresonator{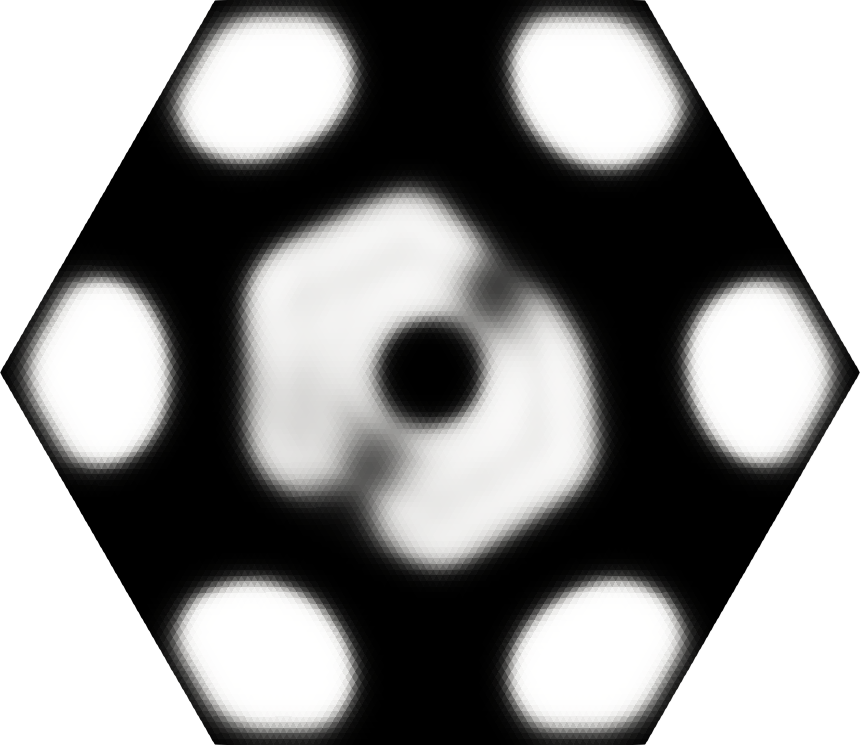} & \addresonator{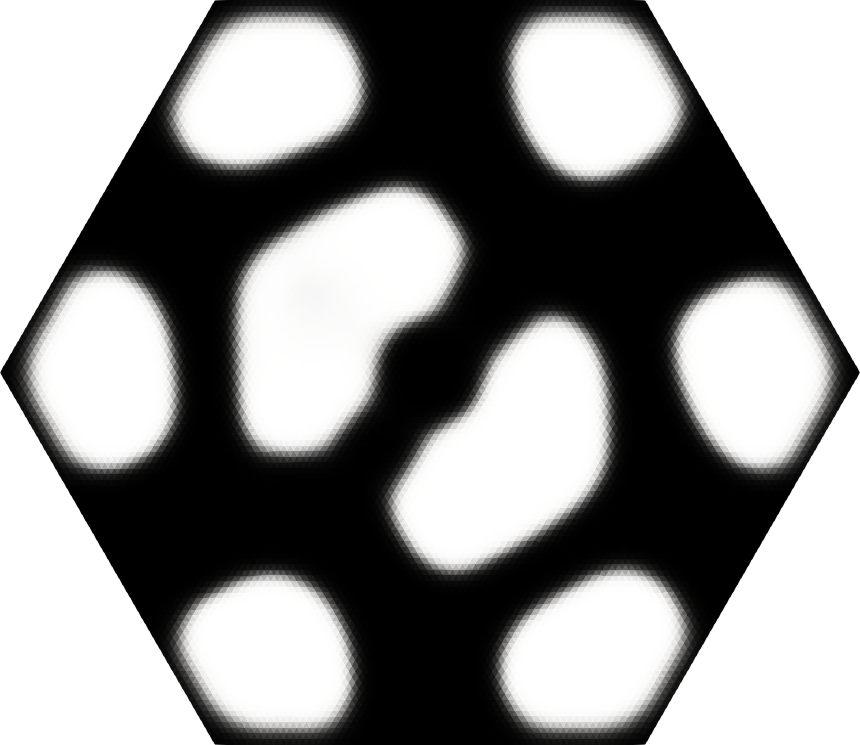} & \addresonator{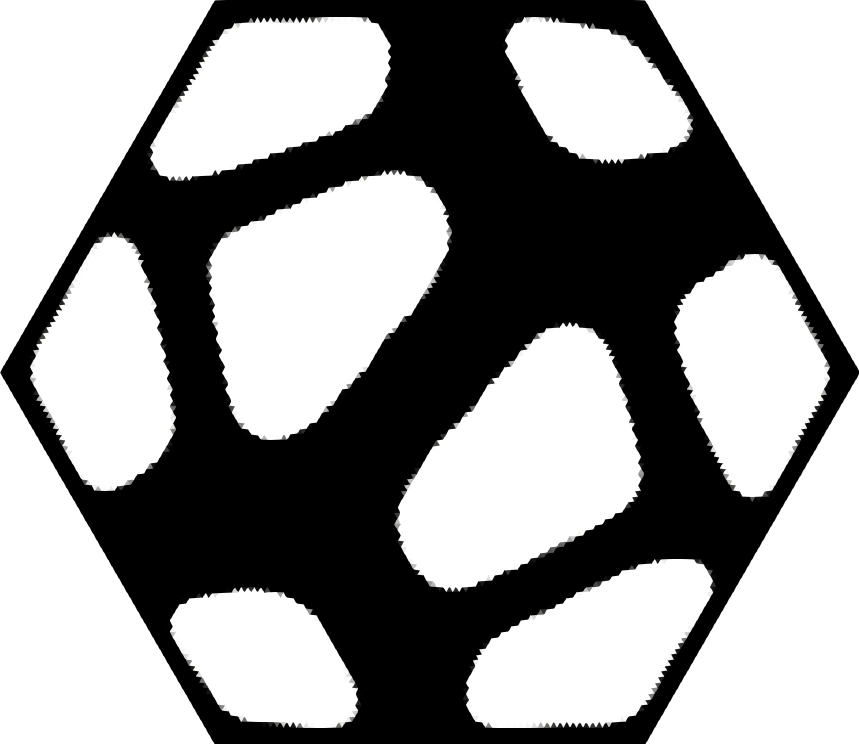} & \imagetop{\begin{tikzpicture} 		
				\node [xshift=0.55cm] (0.0,0.0)  { $\bar{\rho}^{e}_I$ };
				\node [xshift=0.4cm, yshift=0.6cm] (0.0,0.0)  {$1$};
				\node [xshift=0.4cm, yshift=-0.6cm] (0.0,0.0) {$0$};
				\fill[top color=black,bottom color=white]
				(-0.1,-0.72) rectangle (0.1,0.75); 
		\end{tikzpicture}} \\
		\addresonator{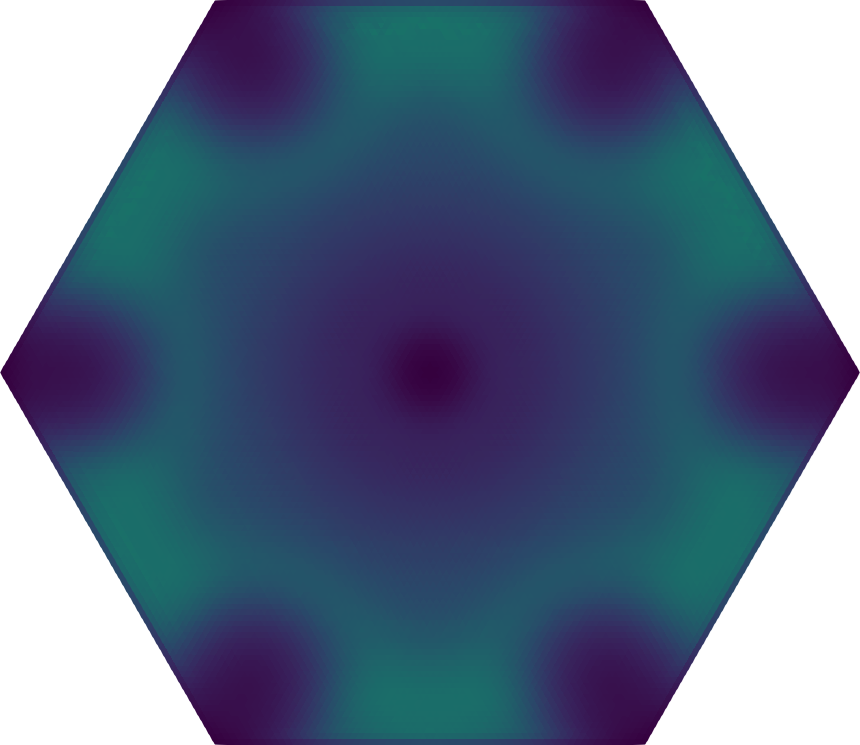} & \addresonator{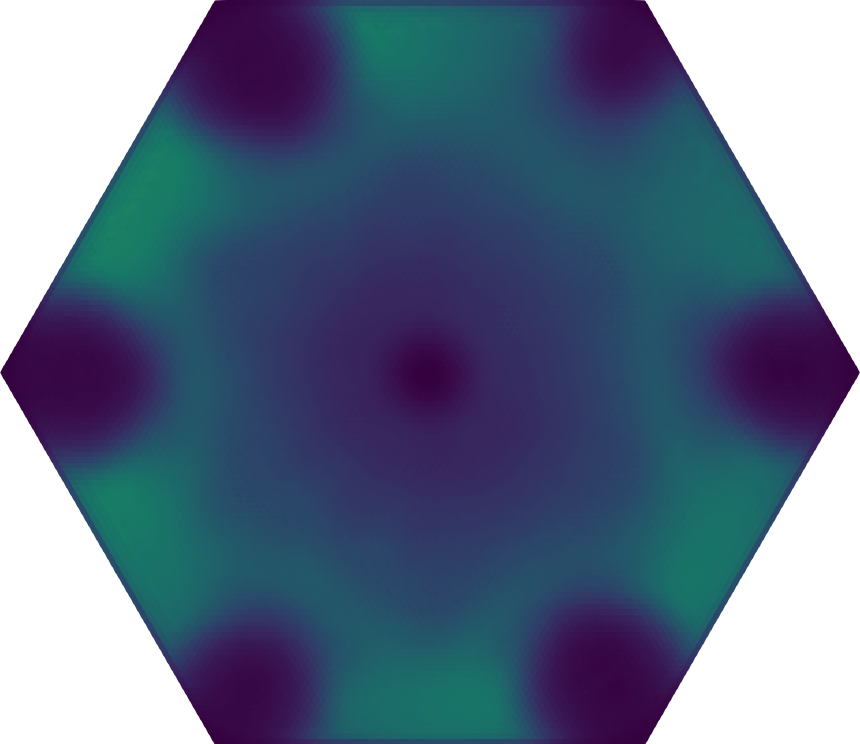} & \addresonator{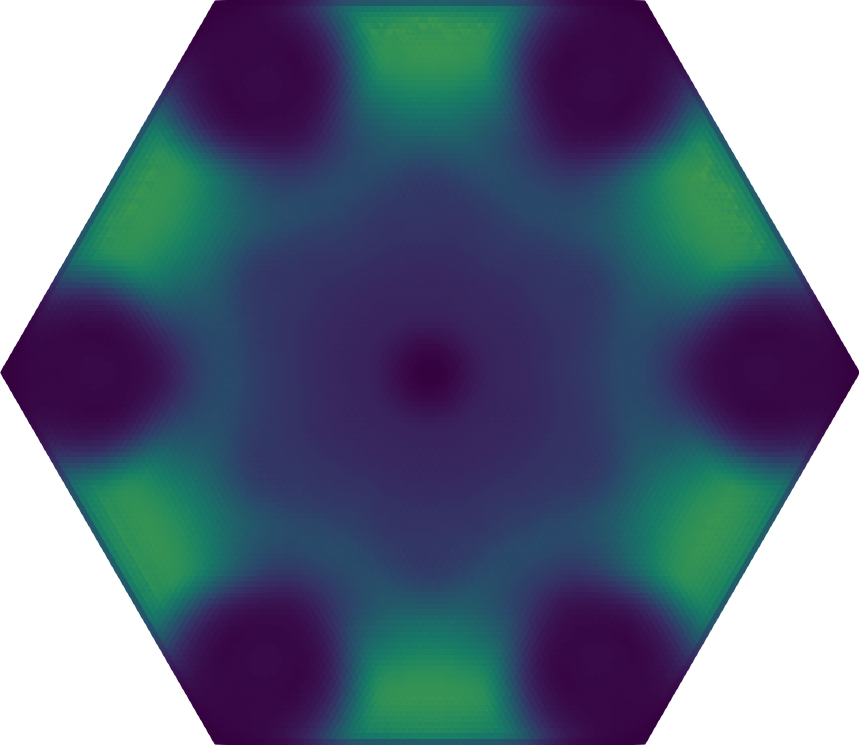} & \addresonator{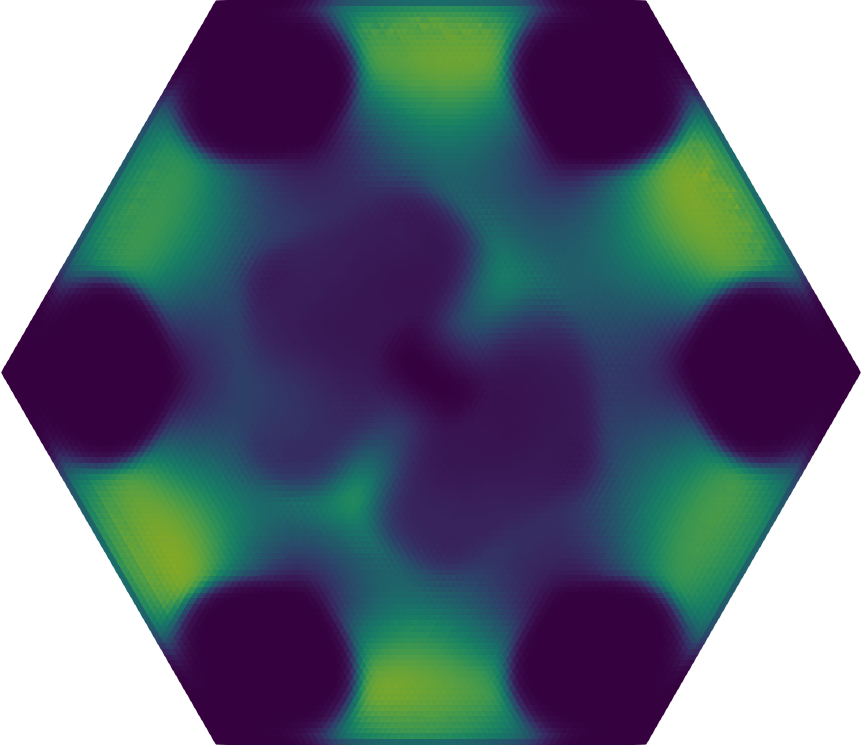} & \addresonator{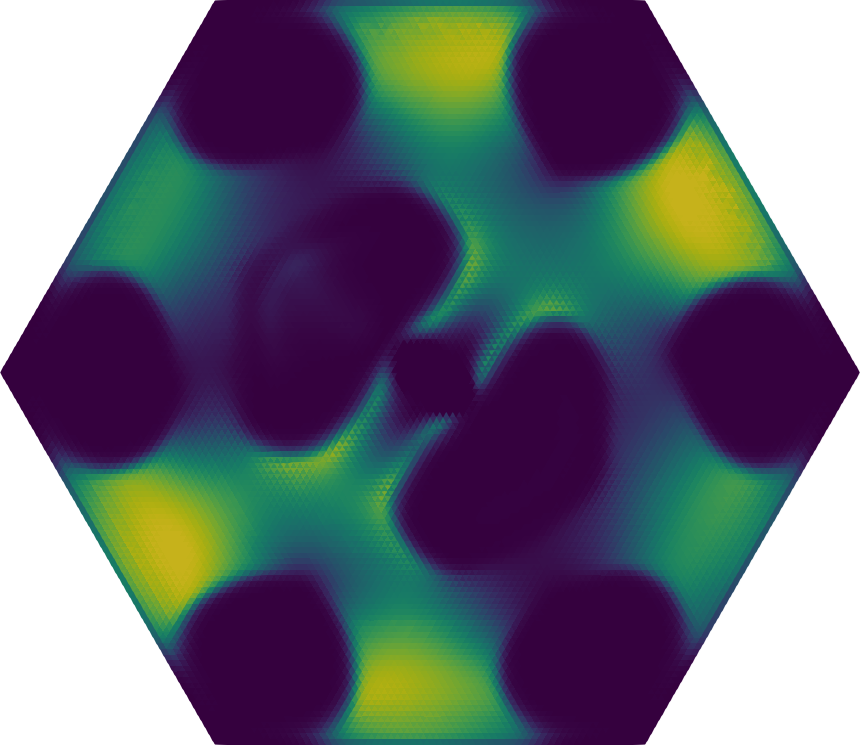}& \addresonator{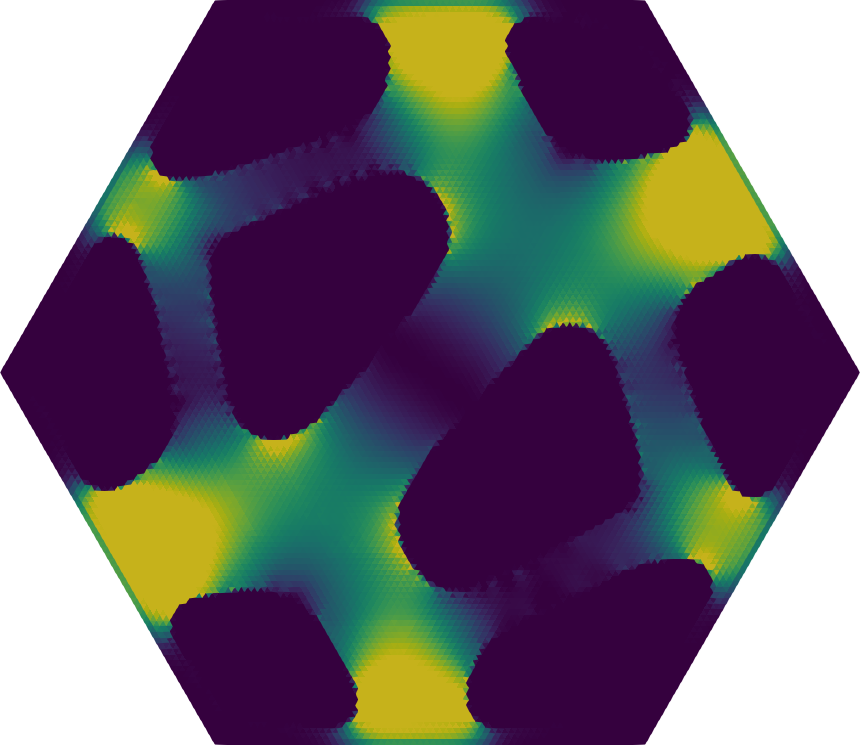} & \imagetop{\begin{tikzpicture} 		
				\node [xshift=0.5cm] (0.0,0.0)  { $\widetilde{\kappa}^e$ };
				\node [xshift=0.4cm, yshift=0.6cm] (0.0,0.0)  {$1$};
				\node [xshift=0.4cm, yshift=-0.6cm] (0.0,0.0) {$0$};
				\fill[top color=Dandelion,bottom color=BlueViolet, middle color = Green!25]
				(-0.1,-0.72) rectangle (0.1,0.75); 
		\end{tikzpicture}} \\
		\addresonator{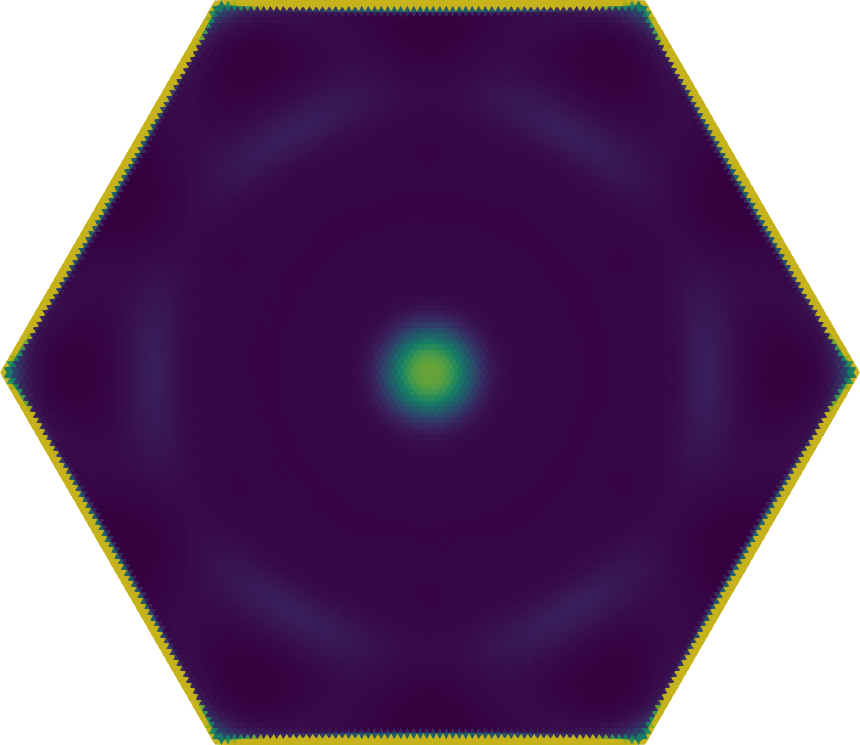} & \addresonator{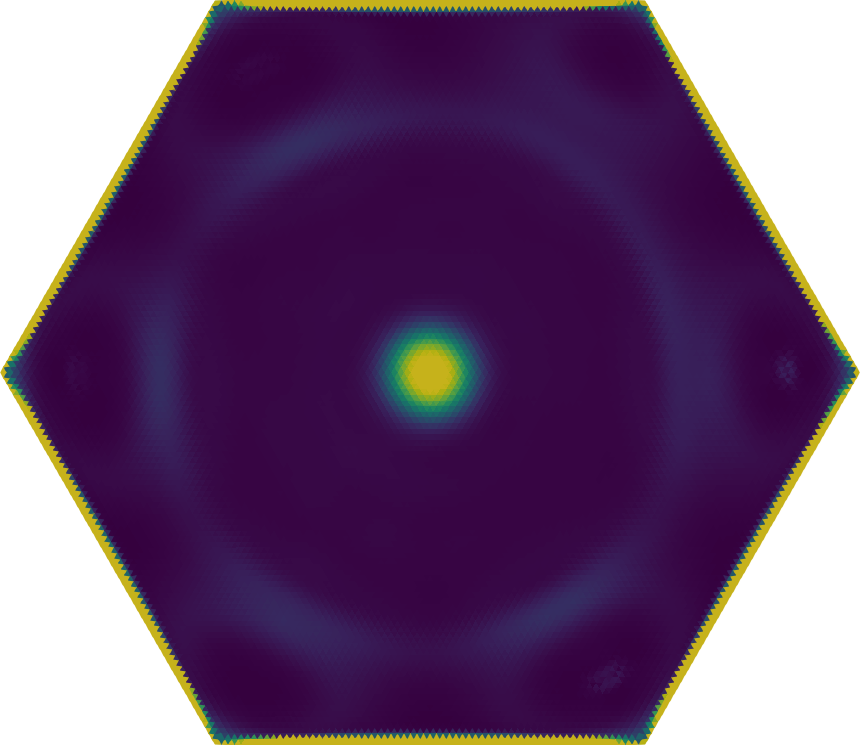} & \addresonator{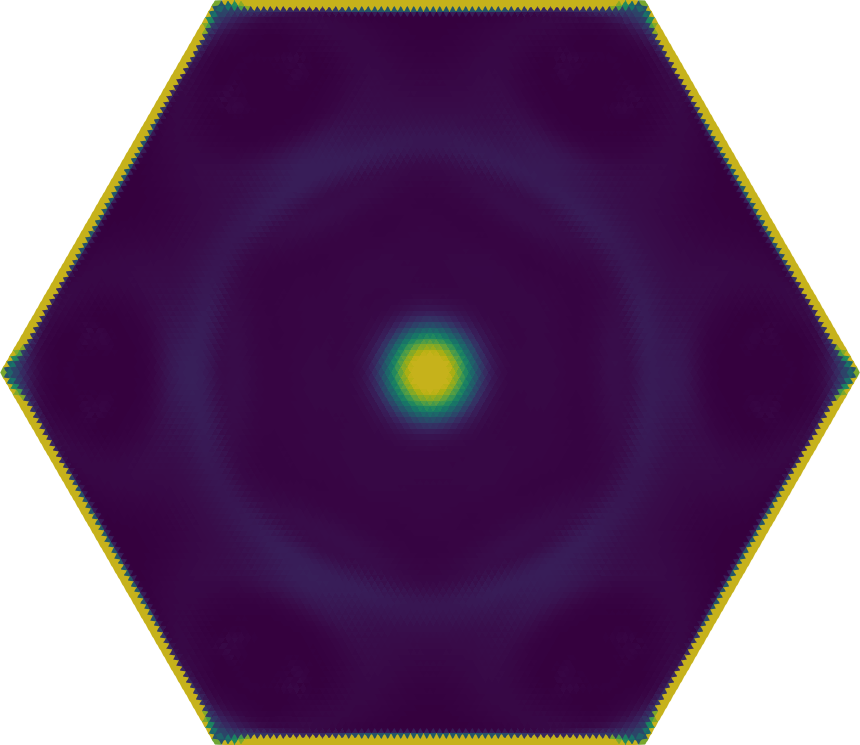} & \addresonator{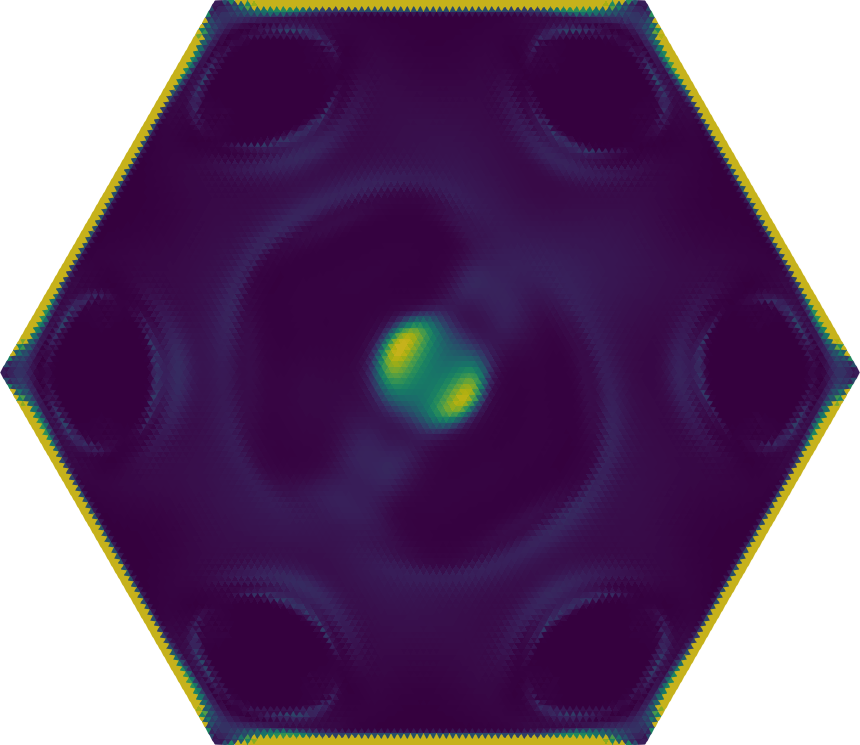} & \addresonator{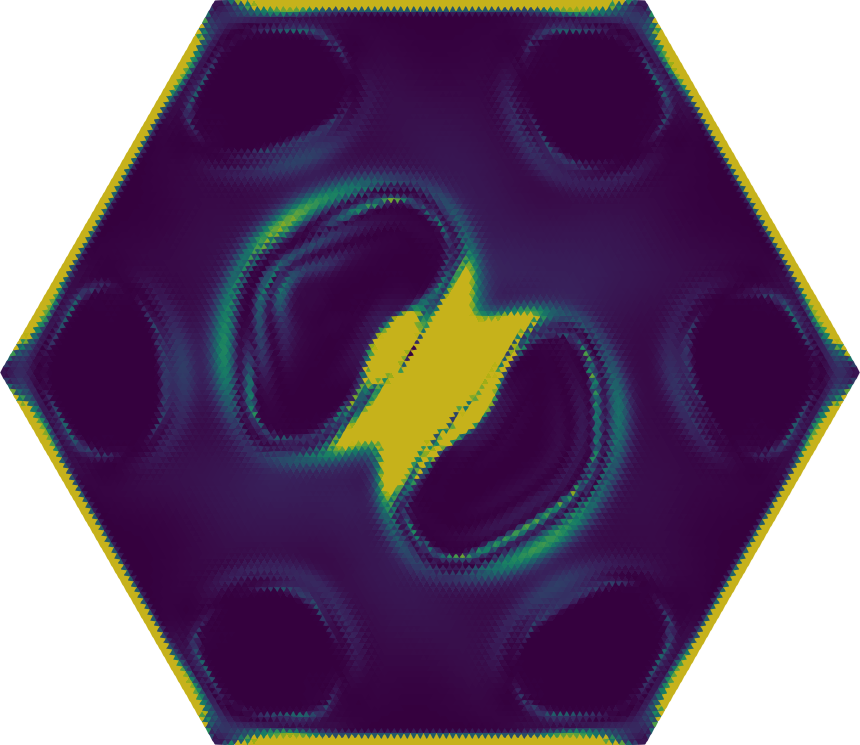}& \addresonator{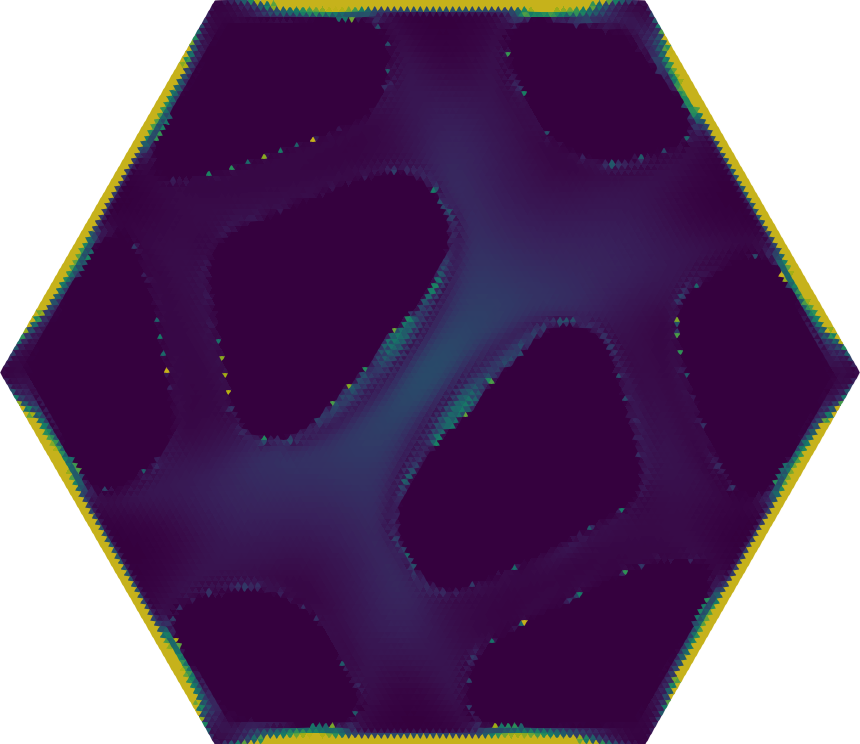} &\imagetop{\begin{tikzpicture} 		
				\fill[top color=Dandelion,bottom color=BlueViolet, middle color = Green!25]
				(-0.1,-0.72) rectangle (0.1,0.75); 
				\node [xshift=0.5cm] (0,0)  { $\overline{\kappa}^e$ };
				\node [xshift=0.4cm, yshift=0.6cm] (0,0.0)  {$1$};
				\node [xshift=0.4cm, yshift=-0.6cm] (0,0.0) {$0$};
		\end{tikzpicture}} 
	\end{tblr}

	\includegraphics{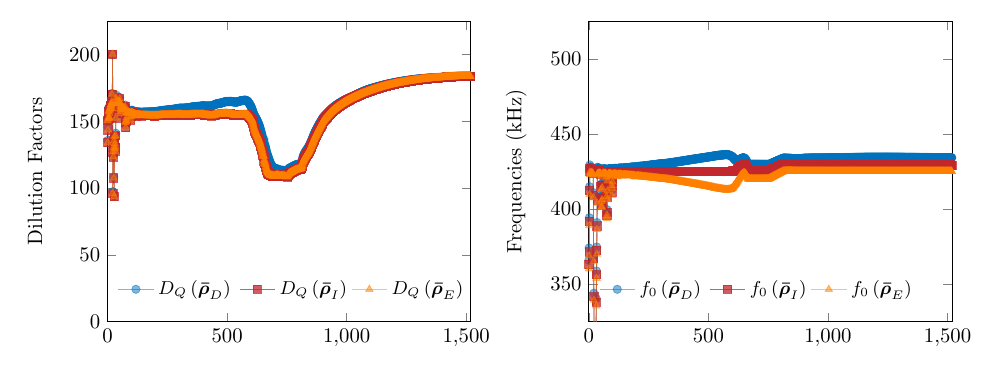}

	\hfill
	\caption{Optimization of a hexagonal resonator with a prescribed lower frequency $f_{\min}= \SI{400}{\kilo\hertz}$. The top row shows the evolution of the intermediate projected density, while the second and third rows depict, respectively, the element-level contributions to the geometrically nonlinear and linear modal stiffnesses; these are computed, respectively, as $\widetilde{\kappa}_e = \bs{\Phi}_e^\intercal \tilde{\bs{k}}_e \bs{\Phi}_e/1000$, and $\overline{\kappa}_e = \bs{\Phi}_e^\intercal \bar{\bs{k}}_e \bs{\Phi}_e $ (they are also normalized, but differently from their counterparts in the square domains). The color bar has been capped for both quantities to give a clearer depiction of these fields (as this limit was exceeded locally). At iteration 650, the linear modal stiffness was much higher due to the local curvature of the vibration mode at the center of the domain. This strongly increased the linear modal stiffness, and a figure showcasing this has been included in Appendix~\ref{ap:supporting_hex}. The bottom two graphs show the evolution of the dilution factor (left) and fundamental resonance frequency (right) throughout the optimization for all three fields of the robust projection.}
	\label{tbl:hex}
\end{figure}
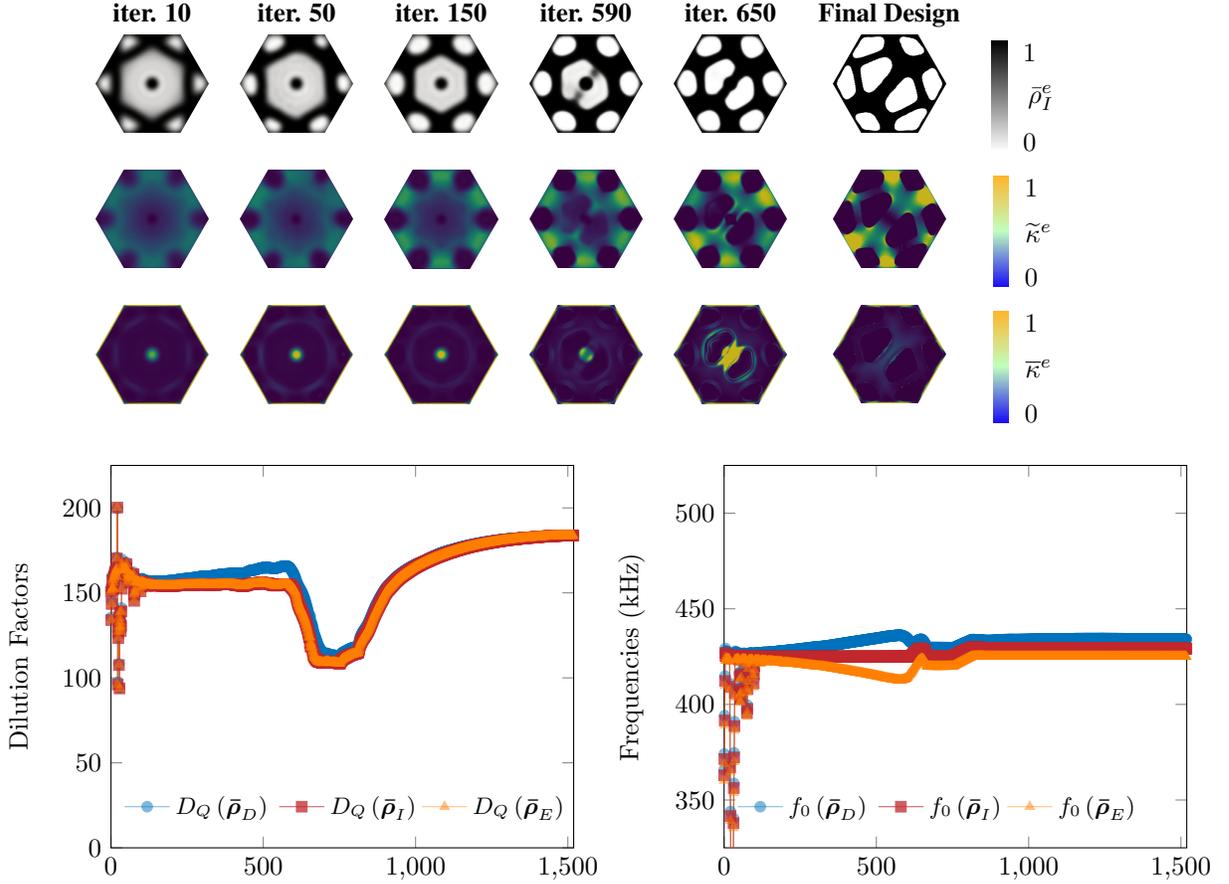

Herein we apply our methodology to the optimization of a resonator on a hexagonal computational domain. Consider the geometry depicted in Figure \ref{fig:domainHexagon}, which displays a hexagon with outer vertices dimensioned at $\SI{434}{\micro\metre}$. These lengths were chosen such that the total surface area is the same of the square domain studied in Section~\ref{sec:square simulations}. Again, clamped boundary conditions (i.e., zero displacement and zero rotation) are prescribed on the outer boundary of the domain. We also use the same continuation scheme that increases the projection slope $\beta$ by 1\% every 3 iterations, starting at $\beta = 1$ and ending at $\beta = 110$. We allow 50 iterations for the first and last slope value, resulting in a total of \num{1519} iterations. Noteworthy, we do not use symmetry boundary conditions but we optimize the entirety of the domain, allowing the optimizer to potentially discover asymmetric designs. We choose a filter radius $r_f = 48.8 $ (approximately 6 elements). As we model the entire domain, this filter radius is considerably larger than that used earlier for the square domain (although they both span the same number of elements). We design a hexagonal resonator for a target fundamental frequency $f_{\min} = \SI{425}{\kilo\hertz}$, for which the results have been summarized in Figure \ref{tbl:hex}. The fundamental mode shape and frequency were evaluated using COMSOL Multiphysics, and have been included in Figure \ref{tbl:comsol}.

In the figure, we see a similar behavior as compared to the optimizations for square resonators. We note that the dilution factor decreases halfway throughout the optimization, after which we have an overall increase. We also saw that there were two main structures during this optimization procedure. Initially, the outer region forms a ring around the central (fixed) region, connected by intermediate-density elements. As the projection slope is increased, these elements start to disappear and the optimizer is forced to tether the central pad to the outside region. To showcase this transition, we illustrate the structure at different iteration counts compared to the square domains. 
	
In our simulation, a relatively high frequency constraint was used for the hexagonal design. Optimizations targeting a fundamental frequency $f_{\min} \leq \SI{400}{\kilo\hertz}$ were commonly seen to result in asymmetric designs. A higher value of $f_{\min}$ results in more ``connectedness'' to the sides of the domain, and stabilizes the optimization procedure by restraining the design space (which was also discussed in \cite{gao20}). A limiting factor to the design of the structure we found here is the filter radius used, of which the absolute value is much higher than the one used for the square domains. This limited the feature size available, and therefore restrained the effects of dissipation dilution. This is because thinner features allow for a higher stress (and therefore geometrically nonlinear stiffness) in the tether regions, which enhances the effects of dissipation dilution, noted by M. Bereyhi et al. \cite{bereyhi19}. Although the resolution is rather course, we nevertheless obtained an interesting hexagonal structure, and demonstrate the general applicability of our optimization approach with these results.

\section{Discussion and conclusions}\label{sec:conclusion}
Dissipation dilution is a phenomenon that describes an increase in the quality factor of nanomechanical resonators under pretension. We expressed this effect as a function based on the ratio of geometrically nonlinear and linear modal stiffnesses obtained via finite element analysis. Although our function is equivalent to a function known from the literature that describes the same effect, it is more appealing for optimization since its sensitivity analysis formulation is straightforward. With this equation as an objective function we thereafter proposed a topology optimization approach that can effectively be used to design resonators with high $Q$ factor.  Furthermore, we provided adjoint (exact) sensitivity formulation, which was verified by comparing it to the sensitivities obtained through finite differences. Because the gradients of the newly defined objective function can be computed exactly, the proposed technique is suitable for high-dimensional design spaces. Consequently, the proposed technique has more design freedom than low-dimensional optimization techniques (e.g., the work of Shin et al.~\cite{shin21}), thereby improving the search for optimized resonator geometries. Additionally, our technique also does not rely on empirical models based on experimental data (e.g., the work of Gao et al.~\cite{gao20}).

In our numerical study we demonstrate the conventional trade-off between the $Q$ factor of a resonator and its fundamental mode frequency, and how this is reflected in resonator topology. We observe that for all our final designs, the element contributions to the geometrically nonlinear modal stiffness and the linear modal stiffness do not overlap; while the former is mainly confined to the tethers connected to the central region, the latter is mainly present at the boundaries and within the central pad itself. This is in agreement with known models of dissipation dilution~\cite{Engelsen2021proceeding}. The $Q$ factor of our final designs lies in the range of $\left[ 4, 9 \right] \times 10^6$; these $Q$ factors are similar to, but not higher than values found in the literature. We note, however, that the highest values found in literature are commonly obtained using thinner membranes and thinner features, which can in turn result in a higher dissipation dilution~\cite{bereyhi19}.

We showcased the generality of our methodology by optimizing a resonator on a hexagonal domain using the full computational domain without assuming any symmetry. We noticed that the behavior of the optimization procedure was different as compared to the optimization on square domains. Specifically, the central pad remained completely disconnected for the first half of the optimization, where the optimizer could not design tethers to connect to the central pad. As we increase the projections slope $\beta$ towards a black-and-white structure, a two-axis symmetric, two-tethered structure appeared, which remained almost unchanged for the remainder of the optimization. We observed that our full-domain hexagonal optimizations were more prone to becoming completely asymmetric. To counter this, we not only used a completely symmetric triangular finite element mesh (with six axes of symmetry), the design space was constrained by using a relatively high value of the frequency constraint $f_{\min} = \SI{425}{\kilo\hertz}$. Furthermore, the feature size of our hexagonal structure was constrained by the relatively large filter radius (resulting from modeling the entire domain, which necessitated a larger mesh size). As thinner features are known to enhance dissipation dilution~\cite{bereyhi19}, having a relatively coarse mesh size and large filter radius limited the overall objective of our structure significantly.


We discuss challenges related to our optimization problem, starting with the fact that our objective function landscape is non-convex. This was shown by running two nearly identical optimizations, with the only difference being the initial material distribution, yielding two different designs with different dilution factors. We argue that although almost all problems in TO are non-convex, the level of non-convexity in our problem is particularly high due to the high level of nonlinearity between the design variables and the objective. 
We therefore make no claims regarding attaining \textit{global optima}. It is therefore plausible that better-performing geometries exist, which could be found by changing the problem parameterization and/or by tuning the settings of the optimization procedure. We note a recent work in the field of TO that has been done to automate this process, e.g. the work of Ha and Carstensen~\cite{carstensen24} who used machine learning to determine these hyperparameters.

Another issue we encountered, is one that mainly concerned some of our initial optimizations, in which the objective function dropped to a negative (or significantly lower) value for some iterations. These downward spikes were caused by void regions subjected to compressive in-plane forces in the membrane, which negatively contribute to the geometrically nonlinear stiffness $\bs{\widetilde{K}}$. In some cases, this resulted in \textit{negative} objective values, as a result of instabilities in the eigenvalue analysis. These values were associated with spurious modes in the void region. In our topology optimization formulation we included measures to avoid compressive effects by interpolating the static displacement in Eq.~\eqref{eq:static_projection}. However, this phenomenon was not entirely mitigated, as can be seen in the bottom figures of Figure~\ref{tbl:f400}, where a downward spike in the objective function value indicates that part of the domain is under compression. One of these optimizations is discussed in detail in Appendix~\ref{app:compressive_void_regions}.
Overall, we found that meshes that were not symmetric with respect to the $\SI{45}{\degree}$ line spanning from the origin to the center of the domain were more prone to compressive regions. Another factor of influence was the frequency constraint, of which a higher value generally reduces the compressive regions, which can be explained by the fact that a higher fundamental frequency requires a higher level of stiffness in the domain.

When preparing our manuscript we learned about another work on TO of nanomechanical resonators where the authors present a similar formulation, but focus on higher-order modes in a square design domain~\cite{arXivSigmund24}. Our work expands on the MSc thesis of the first author~\cite{thesishendrik}, where the objective function and its corresponding sensitivity formulation were introduced for the first time. Here, however, we further discuss the non-convexity of our objective function landscape and the role of different parameters affecting the design optimization, such as the role of frequency constraints and the distribution of geometrically nonlinear and liner modal stiffnesses. Furthermore, for the first time we obtain a symmetric resonator design on a hexagonal domain without imposing symmetry a priori.

An interesting path for future research---as already demonstrated by the work of Shi et al.~\cite{arXivSigmund24}---would be to optimize the dilution factors of higher-order vibration modes, as they have higher frequencies by definition, and could be used (in combination with thinner membranes and higher pretension levels) to design resonators with a $Q f$ product beyond $10^{13}$---which would facilitate room-temperature experiments on quantum technologies. An obvious challenge associated with this direction is dealing with spurious modes in void regions, which do not constitute part of the resonator anyways. Another aspect that could be looked into is the addition of stress constraints to help ensure that designs can be realized by mature manufacturing techniques, which would be required for experimental validation of the methodology.

\section*{Acknowledgements}
The authors acknowledge fruitful discussions with Dr.~Richard Norte and Prof.~Peter G. Steeneken from the Mechanical Engineering faculty at Delft University of Technology, The Netherlands.
Dr.~F.~Alijani acknowledges financial support from the  European Union (ERC Consolidator, NCANTO, 101125458). Views and opinions expressed are however those of the author(s) only and do not necessarily reflect those of the European Union or the European Research Council. Neither the European Union nor the granting authority can be held responsible for them. Z.~Li acknowledges financial support from China Scholarship Council.

\appendix


\section{Material properties of \ce{Si3N4}}\label{ap:material properties}
The material values listed in COMSOL for \ce{Si3N4} have been listed in the table below.
\begin{table}[H]
    \centering
    \begin{tabular}{l|ccc} \toprule 
         \textbf{Property}&   \textbf{Symbol}&\textbf{Value}& \textbf{Unit}\\ \midrule
         Young's modulus&   $E$&$250$& $\unit{\giga\pascal}$\\ \hline 
         Poisson's ratio&   $\nu$&0.23& -\\ \hline 
         Mass density&  $\rho$&3100 & $\unit[per-mode=symbol]{\kilo\gram\per\metre\cubed}$\\\bottomrule
    \end{tabular}
\end{table}


\section{Sensitivity analysis}\label{ap:sensitivity analyses}

\subsection{Differentiating the geometric stiffness matrix}\label{ap:Gmatrix}
The geometric stiffness matrix $\widetilde{\bs{K}}$ has both an explicit and implicit dependence on $\bs{\rho}$, which must be taken into account when computing its sensitivity. The explicit dependence is a result of its direct dependence on each element's penalized Young's modulus, (which linearly scales the elasticity matrix $\bs{S}$). Its implicit dependence stems from the dependence of $\widetilde{\bs{K}}$ on $\bs{U}_{\sigma_0}(\bs{\rho})$. To incorporate both dependencies, the chain rule is applied in the computation of $\frac{d\widetilde{\bs{K}}}{d\rho}$. This is done as follows:
\begin{equation}\label{eq:derivative of G}
     \frac{d\widetilde{\bs{K}} }{d\rho_e}   = \frac{\partial \widetilde{\bs{K}} }{\partial \rho_e}  + \frac{\partial \widetilde{\bs{K}} }{\partial \bs{U}_{\sigma_0}}   \frac{d \bs{U}_{\sigma_0}}{d\rho_e}
\end{equation}
The partial derivative of $\widetilde{\bs{K}}$ with respect to $\bs{\rho}$ in (\ref{eq:derivative of G}) can be computed by differentiating the material interpolation function (Eq. \eqref{eq:ramp}), which acts on the local Young's modulus. Taking the derivative of $\widetilde{\bs{K}}$ with respect to $\bs{U}_{\sigma_0}$ results in a three-dimensional tensor for each element. This tensor can be computed as

\begin{equation}\label{eq:dGdu definition}
    \frac{\partial \widetilde{\bs{K}} }{\partial \bs{U}_{\sigma_0}} =  \frac{\partial \widetilde{\bs{K}}}{\partial \widehat{\bs{\sigma}} }\frac{\partial \widehat{\bs{\sigma}} }{\partial \bs{U}_{\sigma_0}}  \ \Rightarrow \ \frac{\partial \widetilde{K}_{ij} }{\partial u_a} = \frac{\partial \widetilde{K}_{ij}}{\partial \widehat{\sigma}_k}\frac{\partial \widehat{\sigma}_k}{\partial u_a} = \frac{\partial ^2 \varepsilon_k}{\partial u_i \partial u_j} \left[\bs{S}\overline{\bs{B}}\right]_{ka},
\end{equation}
where all terms on the right-hand side of the arrow sign are at element level, and the result represents the derivative of $\widetilde{{K}}_{ij}^{(e)}$ with respect to th $a$-th component of the (local) displacement vector.

An projection was applied to the displacements before the equilibrium stress configuration $\widetilde{\bs{\sigma}}$ was computed, as described by Equation~\eqref{eq:static_projection}. This must be taken into account in the sensitivity analysis. 

\begin{equation}\label{eq:derivative of G with displacement interpolation}
     \frac{d\widetilde{\bs{K}} }{d\rho}  = \frac{\partial \widetilde{\bs{K}} }{\partial \rho}  + \frac{\partial \widetilde{\bs{K}} }{\partial \tilde{\bs{U}}}  \left( \bs{U}_{\sigma_0} \circ \frac{d\bs{\alpha}}{d\rho} +  \bs{\alpha}\circ \frac{d \tilde{\bs{U}}_{\sigma_0}}{d \rho}\right),
\end{equation}
\noindent
where $\circ$ represents the Hadamard product (element-wise multiplication). As $\bs{\alpha}$ only depends on the element of interest, the second right-hand-side term will be zero for all elements, except for the element for which $\frac{d\widetilde{\bs{K}} }{d\rho}$  is computed.

\subsection{Detailed objective sensitivity analysis}
The Lagrangian is constructed by augmenting the objective (Eq.~\eqref{eq:DQ objective}) with the null form of the static and modal analyses. Here, the mass-normalization was added for completeness, and because this was done in~\cite{tsai13}. This results in the following expression for the Lagrangian 

\begin{align}\label{eq:full lagrangian}
    \begin{split}
            L = \frac{\bs{\Phi} ^{\intercal} \widetilde{\bs{K}} \bs{\Phi} }{\bs{\Phi}^{\intercal} \overline{\bs{K}} \bs{\Phi}} + \bs{\lambda}_1^{\intercal} \left(\bs{F} - \overline{\bs{K}} \bs{U}_{\sigma_0} \right) &+ \bs{\lambda}_2^{\intercal} \left( \overline{\bs{K}}+\widetilde{\bs{K}}-\omega^2 \bs{M}  \right) \bs{\Phi} + \lambda_3 \left( \bs{\Phi} ^{\intercal} \bs{M} \bs{\Phi} -1 \right).
    \end{split}
\end{align}
Differentiating (\ref{eq:full lagrangian}) with respect to a design density $\rho_e$ results in

\begin{align}  \label{eq: lagrangian with adjoint problem}
    \begin{split}
        \frac{d L}{d \rho_e} =&\frac{\bs{\Phi}^{\intercal} \frac{\partial \widetilde{\bs{K}} }{\partial \rho_e}\bs{\Phi}}{\bs{\Phi}^{\intercal} \overline{\bs{K}} \bs{\Phi}} - \frac{ \bs{\Phi}^{\intercal} \widetilde{\bs{K}} \bs{\Phi} \cdot  \bs{\Phi}^{\intercal} \frac{d \overline{\bs{K}} }{d \rho_e}\bs{\Phi} }{\left( \bs{\Phi}^{\intercal} \overline{\bs{K}} \bs{\Phi} \right)^2} + \bs{\lambda}_1^{\intercal} \left( \frac{d \bs{F}}{d \rho_e} -  \frac{d \overline{\bs{K}}}{d \rho_e}\bs{U}_{\sigma_0}\right)  \\&+\bs{\lambda}_2^{\intercal} \left( \frac{d\overline{\bs{K}}}{d \rho_e} + \frac{\partial \widetilde{\bs{K}} }{\partial \rho_e}   - \frac{d\omega^2}{d \rho_e}\bs{M} - \omega^2 \frac{d\bs{M}}{d \rho_e} \right) \bs{\Phi}  + \lambda_3 \left( \bs{\Phi} ^ T \frac{d \bs{M}}{d \rho_e} \bs{\Phi} \right) 
        \\ &  + \left[ \frac{2 \bs{\Phi}^{\intercal} \widetilde{\bs{K}} }{ \bs{\Phi}^{\intercal} \overline{\bs{K}} \bs{\Phi}}- \frac{\bs{\Phi}^{\intercal} \widetilde{\bs{K}} \bs{\Phi} \cdot 2 \bs{\Phi}^{\intercal} \overline{\bs{K}} }{\left( \bs{\Phi}^{\intercal} \overline{\bs{K}} \bs{\Phi}\right)^2}+ \bs{\lambda}_2^{\intercal}  \left( \overline{\bs{K}}+\widetilde{\bs{K}}-\omega^2 \bs{M}  \right) + \lambda_3 \cdot 2\bs{\Phi}^{\intercal} \bs{M} \right] \frac{d\bs{\Phi}}{d \rho_e} 
        \\&   +  \left[ \frac{\bs{\Phi}^{\intercal} \frac{\partial \widetilde{\bs{K}} }{\partial \bs{U}_{\sigma_0}} \bs{\Phi} } { \bs{\Phi}^{\intercal} \overline{\bs{K}} \bs{\Phi}} - \bs{\lambda}_1 ^{\intercal} \overline{\bs{K}} + \bs{\lambda}_2^{\intercal}  \frac{\partial \widetilde{\bs{K}} }{\partial \bs{U}_{\sigma_0}} \bs{\Phi} \right]  \frac{d \bs{U}_{\sigma_0}}{d \rho_e},
    \end{split}
\end{align}
where all terms that are multiplied with derivatives of $\bs{\Phi}$ and $\bs{U}_{\sigma_0}$ have been grouped. Calculating $ \frac{d \bs{\Phi}}{d \rho_e}$ and $ \frac{d \bs{U}_{\sigma_0}}{d \rho_e}$ is very expensive from a computational point of view, as the mode shape or static displacement in every point of the domain depends on the design parameters in every element. In order to omit them from the analysis, the expressions in the square brackets of (\ref{eq: lagrangian with adjoint problem}) are equated to zero. This results in the two adjoint problems for the Lagrange multipliers, namely

\begin{equation}\label{eq: adjoint phi}
     \left( \overline{\bs{K}}+\widetilde{\bs{K}}-\omega^2 \bs{M}  \right)\bs{\lambda}_2  = \left(- \frac{2   }{ \bs{\Phi}^{\intercal} \overline{\bs{K}} \bs{\Phi}} \widetilde{\bs{K}} + \frac{2\bs{\Phi}^{\intercal} \widetilde{\bs{K}} \bs{\Phi}  }{\left( \bs{\Phi}^{\intercal} \overline{\bs{K}} \bs{\Phi}\right)^2} \overline{\bs{K}} + 2\lambda_3 \bs{M}\right) \bs{\Phi},
\end{equation}

\begin{equation}\label{eq: adjoint u}
       \overline{\bs{K}}\bs{\lambda}_1  = \left( \frac{\bs{\Phi}^{\intercal} }{ \bs{\Phi}^{\intercal} \overline{\bs{K}} \bs{\Phi}}  + \bs{\lambda}_2^{\intercal}  \right)  \frac{\partial \widetilde{\bs{K}} }{\partial \bs{U}_{\sigma_0}}\bs{\Phi}.
\end{equation}
In order to find a solution for $\bs{\lambda}_2$ in equation (\ref{eq: adjoint phi}), the approach presented by Tsai and Cheng~\cite{tsai13}) was used. This involves expressing $\bs{\lambda}_2$ as a linear combination of the mode shape $\bs{\Phi}$ (multiplied by an arbitrary scalar $c$) and an unknown vector denoted as $\bs{\psi}$, 

\begin{equation}\label{eq:lam2 expression}
    \bs{\lambda}_2 \equiv c\bs{\Phi} + \bs{\psi}.
\end{equation}
The solution to $\bs{\lambda}_2$ exists if and only if the right-hand term of (\ref{eq: adjoint phi}) is orthogonal to $\bs{\Phi}$. Imposing this condition results in the following expression for $\lambda_3$,

\begin{equation}\label{eq:lam3}
    \left(- \frac{2   }{ \bs{\Phi}^{\intercal} \overline{\bs{K}} \bs{\Phi}} \widetilde{\bs{K}} + \frac{2\bs{\Phi}^{\intercal} \widetilde{\bs{K}} \bs{\Phi}  }{\left( \bs{\Phi}^{\intercal} \overline{\bs{K}} \bs{\Phi}\right)^2} \overline{\bs{K}} + 2\lambda_3 \bs{M}\right) \bs{\Phi} = \bs{0}.
\end{equation}
When pre-multiplying the system in (\ref{eq:lam3}) with $\bs{\Phi}$, the first two fractions cancel each other\footnote{Due to the equality of $\frac{2 \bs{\Phi}^{\intercal} \widetilde{\bs{K}} \bs{\Phi}}{ \bs{\Phi}^{\intercal} \overline{\bs{K}} \bs{\Phi}} = \frac{\bs{\Phi}^{\intercal} \widetilde{\bs{K}} \bs{\Phi} \cdot 2 \bs{\Phi}^{\intercal} \overline{\bs{K}} \bs{\Phi} }{\left( \bs{\Phi}^{\intercal} \overline{\bs{K}} \bs{\Phi}\right)^2}$}, resulting in $2\lambda_3 \bs{\Phi}^\intercal\bs{M}\bs{\Phi}=0$. From this it follows that $\lambda_3 = 0$. In principle, the mass-normalization constraint could have been omitted from the initial Lagrangian entirely, but it was included in the detailed analysis because it was non-zero in the original work by Tsai and Cheng~\cite{tsai13}. The following system has to be solved for the particular solution $\bs{\psi}$,

\begin{equation}\label{eq:system psi}
     \left( \overline{\bs{K}}+\widetilde{\bs{K}}-\omega^2 \bs{M}  \right)\bs{\psi}  = \left(- \frac{2   }{ \bs{\Phi}^{\intercal} \overline{\bs{K}} \bs{\Phi}} \widetilde{\bs{K}} + \frac{2\bs{\Phi}^{\intercal} \widetilde{\bs{K}} \bs{\Phi}  }{\left( \bs{\Phi}^{\intercal} \overline{\bs{K}} \bs{\Phi}\right)^2} \overline{\bs{K}}\right) \bs{\Phi}.
\end{equation}
The system described here is singular, which means that infinitely many solutions exist. At this point, we choose one of the solutions for $\bs{\psi}$ and compute the scalar value $c$ based on our chosen solution. This value was chosen to satisfy

\begin{equation}
    \bs{\lambda}_2^\intercal \bs{M} \bs{\Phi} = 0 \ \ \Rightarrow \ \ c = - \bs{\psi}^\intercal \bs{M} \bs{\Phi},
\end{equation}
omitting the computation of $\frac{d \omega^2}{d\rho}$ in  \eqref{eq: lagrangian with adjoint problem}, which is computationally favorable.

\subsection{Finite difference check}\label{ap:FDcheck}

\begin{figure}[!ht]
  \savebox{\imagebox}{\scalebox{0.34}{
    \includegraphics{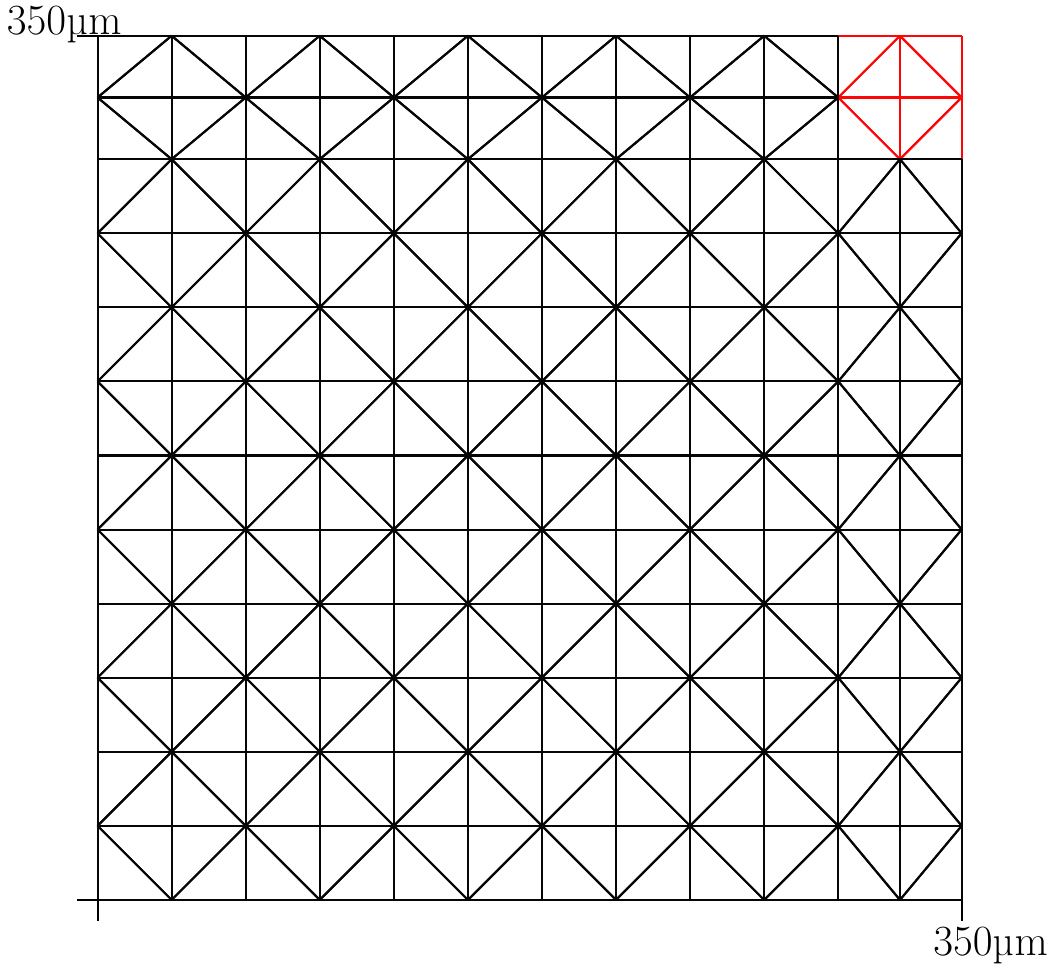}
  }}%
  \begin{subfigure}[t]{.4\linewidth}
    \centering\usebox{\imagebox} 
    \caption{}
    \label{fig:fdappendix}
  \end{subfigure} \qquad
  \begin{subfigure}[t]{.45\linewidth}
    \centering\raisebox{\dimexpr.5\ht\imagebox-.5\height}{
    \includegraphics{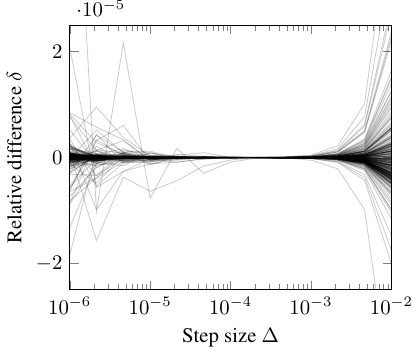}
    }
    \caption{}
    \label{fig:fdcheck_hom}
  \end{subfigure}
  
  \caption{(a) Mesh used for the finite difference verification of the objective function sensitivity given by Eq.~\eqref{eq:DQ sens main expression}. (b) Relative difference between the analytical sensitivity \eqref{eq:DQ sens main expression} and the central difference value as a function of step size (change in element density).}
    \label{fig:FD_check}
\end{figure}

We verified the accuracy of the analytical expression for the sensitivity given by~\eqref{eq:DQ sens main expression} with the finite difference (FD) method using a central difference scheme. To that end, we used the finite element discretization on a square of size $\SI{350}{\micro\metre} \times \SI{350}{\micro\metre}$ shown in Figure~\ref{fig:fdappendix}; symmetry conditions were prescribed along top and right edges, while nodes on the bottom and left edges were fully clamped. A fixed density $\rho_e = 1$ was set to all elements in the red region, as well as all elements in direct contact with bottom and left edges. A pseudo-random density value between 0 and 1 was set in all other elements. All designs were evaluated with a thickness of $\SI{340}{\nano\metre}$ and a pretension of $\SI{1.1}{\giga\pascal}$, imitating the values of \ce{Si3N4} resonators used by Li et al.~\cite{zichao23} (see also Appendix~\ref{ap:material properties} for the material properties).
With the model set, the density of individual elements was perturbed by a small value (i.e., the step size), and the objective values of the perturbed designs were computed. This was done for every element, as well as for different step sizes.

The results are summarized in Figure \ref{fig:fdcheck_hom}, which shows the relative difference between the analytical sensitivity and the FD computed one as a function of the step size.
Noteworthy, while too small a value induces numerical error, a step size that is too large results in an inaccurate FD approximation. This results in divergence  towards the left and right sides of the graph, respectively. The smallest relative difference was found for a step size of $\Delta = 2 \cdot 10^{-4}$, where it converged to $ \delta = 8 \cdot 10^{-9}$.
The convergence of the relative difference to a very low value at small step sizes indicates that the adjoint formulation is correct and has been implemented correctly.

\pgfplotsset{
  table/search path={./figures/data},
}

\section{Spurious modes in void regions with compressive in-plane forces}\label{app:compressive_void_regions}

\begin{figure}[!ht]
	\centering
    \includegraphics{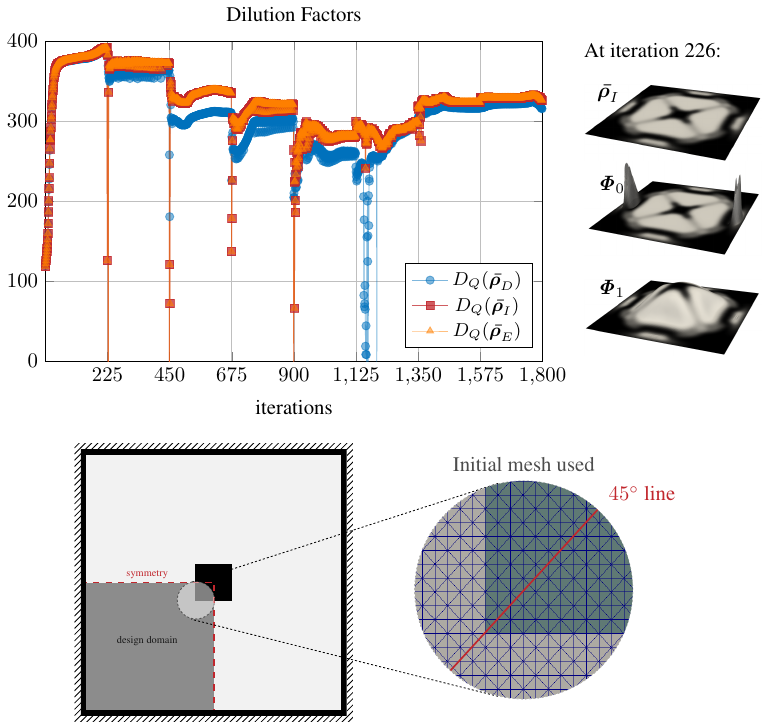}
	\caption{One of our early optimizations for a square resonator, in which spurious modes were commonly seen. These instabilities, which were caused by void regions of the design with compressive in-plane forces, manifested as sharp decreases in the dilution factor objective. These instabilities were more commonly encountered when using a mesh that was not symmetric with respect to the $\SI{45}{\degree}$ line from the origin to the center of the domain, and when updating abruptly the projection slope $\beta$ of our continuations scheme.}
	\label{fig:instability}
\end{figure}

In our earlier optimizations we frequently observed the appearance of spurious modes---which at times even dominated the fundamental mode---in void regions with compressive in-plane forces. Figure~\ref{fig:instability} shows one such optimization, which targeted a fundamental frequency $f_{\min}=\SI{300}{\kilo\hertz}$ and where we used a continuation scheme that doubled the projection slope $\beta$ every 225 iterations, finalizing with a slope $\beta = 128$ at a total of \num{1800} iterations.

As apparent from the figure, we encountered iterations in which the dilution factor objective dropped drastically. These commonly occurred at, or shortly after the updating of the projection slope $\beta$ (in the figure $\beta$ was updated at iterations 225, 450, 675, 900, 1225, 1350, and 1575). We observed that asymmetry of the mesh with respect to the $\SI{45}{\degree}$ line from the origin to the center of the domain (see inset in the figure), as well as large updates to the projection slope $\beta$ can cause this issue. Although the optimizer still converged, these instabilities hindered its progress, as in the robust method we use the sensitivity of the design with the lowest objective function value to continue the optimization---and in the present case the (negative) objective based on spurious modes. This can steer the optimizer into local minima, and therefore hinders the overall optimization.

\section{Supporting images related to the hexagonal structure}\label{ap:supporting_hex}
\begin{figure}[!ht]
	\centering
	\begin{minipage}{0.45\textwidth}
		\centering
		\includegraphics[width=\textwidth]{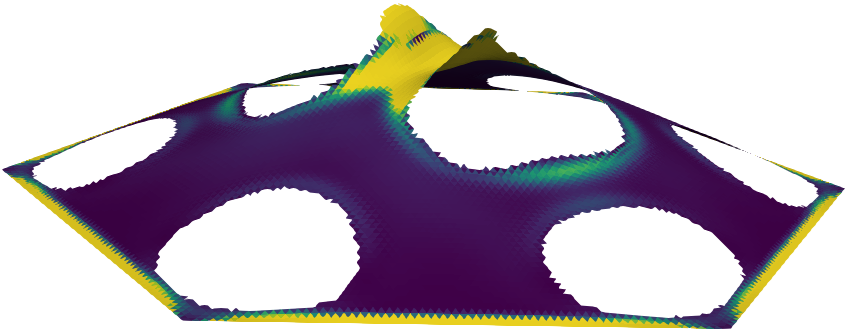}
		\caption*{Iter. 650}
	\end{minipage}
	\hfill
	\begin{minipage}{0.45\textwidth}
		\centering
		\includegraphics[width=\textwidth]{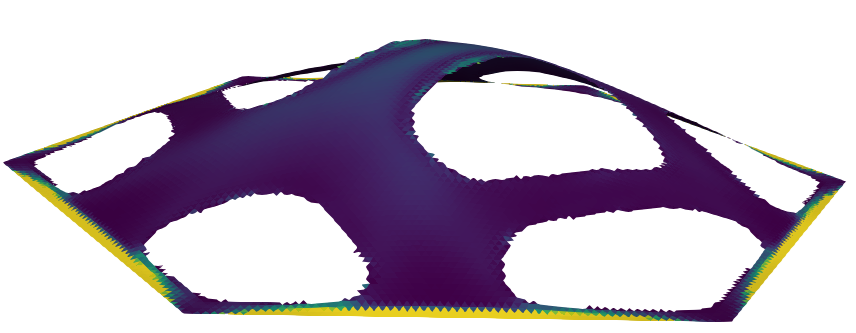}
		\caption*{Final Design}
	\end{minipage}
	\caption{3D rendering of the linear modal stiffnesses depicted in Figure \ref{tbl:hex}, for iteration 650 and the final design. A threshold of $\bar{\rho}_I^e\geq0.5$ and a domain deformation corresponding to the fundamental mode shape was applied for clarity. }
	\label{fig:hexagonK}
\end{figure}

We include two images showcasing the linear modal stiffness of the hexagonal structure at iterations 650 and 1519 (i.e., the final design). These were added to clarify the large decrease in linear modal stiffness at the center of the domain, which can be seen in the bottom row of images in Figure \ref{tbl:hex}. For clarity, prescribe a threshold only render elements with a projected density of $\bar{\rho}_I^e\geq0.5$, and warp the rendering by the fundamental vibration mode, scaled with the same parameter for both images. The images can be found in Figure~\ref{fig:hexagonK}.

\bibliographystyle{unsrt}
\bibliography{cas-refs}

\begin{thebibliography}{10}

\bibitem{nguyuen07}
Clark T.-c. Nguyen.
\newblock Mems technology for timing and frequency control.
\newblock {\em IEEE Transactions on Ultrasonics, Ferroelectrics, and Frequency Control}, 54(2):251--270, 2007.

\bibitem{mamin01}
H.~J. Mamin and D.~Rugar.
\newblock Sub-attonewton force detection at millikelvin temperatures.
\newblock {\em Applied Physics Letters}, 79(20):3358--3360, 11 2001.

\bibitem{chaste12}
J.~Chaste, A.~Eichler, J.~Moser, G.~Ceballos, R.~Rurali, and A.~Bachtold.
\newblock A nanomechanical mass sensor with yoctogram resolution.
\newblock 7:301--304, 2012.

\bibitem{burg07}
Thomas Burg, Michel Godin, Scott Knudsen, Wenjiang Shen, Greg Carlson, John Foster, Ken Babock, and Scott Manalis.
\newblock Weighing of biomolecules, single cells and single nanoparticles in fluid.
\newblock 446:1066--1069, 2007.

\bibitem{norte16}
R.~Norte, J.~Moura, and S.~Gr{\"o}blacher.
\newblock Mechanical resonators for quantum optomechanics experiments at room temperature.
\newblock 116(147202), 2016.

\bibitem{oconnell10}
A.~O'Connell, M.~Hofheinz, M.~Ansmann, R.~Bialczak, M.~Leander, E.~Lucero, M.~Neeley, D.~Sank, H.~Wang, M.~Weides, J.~Wenner, J.~Martinis, and A.~Cleland.
\newblock Quantum ground state and single-phonon control of a mechanical resonator.
\newblock 464:697--703, 2010.

\bibitem{zichao24}
Z.~Li, M.~Xu, R.A. Norte, A.M. Arag{\'o}n, P.G. Steeneken, and F.~Alijani.
\newblock Strain engineering of nonlinear nanoresonators from hardening to softening.
\newblock 7(53), 2024.

\bibitem{steeneken20}
M.~{\v S}i{\v s}kins, M.~Lee, S.~Ma{\~n}as-Valero, E.~Coronado, Y.M. Blanter, H.S.J.~Van der Zant, and P.G. Steeneken.
\newblock Magnetic and electronic phase transitions probed by nanomechanical resonators.
\newblock 11(2698), 2020.

\bibitem{ata23}
Makars {\v S}i{\v s}kins, Ata Ke{\c s}kekler, Maurits J.~A. Houmes, Samuel Ma{\~n}as-Valero, Eugenio Coronado, Yaroslav~M. Blanter, Herre S.~J. van~der Zant, Peter~G. Steeneken, and Farbod Alijani.
\newblock Nonlinear dynamics and magneto-elasticity of nanodrums near the phase transition, 2023.

\bibitem{fedorov21}
S.~Fedorov.
\newblock {\em Mechanical resonators with high dissipation dilution in precision and quantum measurements}.
\newblock PhD thesis, {\'E}cole Polytechnique F{\'e}d{\'e}rale de Lausanne, 1 2021.

\bibitem{shin21}
D.~Shin, A.~Cupertino, M.~De Jong, P.~Steeneken, and M.~Bessa.
\newblock Spiderweb nanomechanical resonators via bayesian optimization: Inspired by nature and guided by machine learning.
\newblock 34(3), 2021.

\bibitem{zichao23}
Z.~Li, M.~Xu, R.~Norte, F.~Alijani, A.~Arag{\'o}n, F.~van Keulen, and P.~Steeneken.
\newblock Tuning the q-factor of nanomechanical string resonators by torsion support design.
\newblock 122(1), 2023.

\bibitem{farbod23}
Tom\'as Manzaneque, Murali~K. Ghatkesar, Farbod Alijani, Minxing Xu, Richard~A. Norte, and Peter~G. Steeneken.
\newblock Resolution limits of resonant sensors.
\newblock {\em Phys. Rev. Appl.}, 19:054074, May 2023.

\bibitem{schmid11}
S.~Schmid, K.~D. Jensen, K.~H. Nielsen, and A.~Boisen.
\newblock Damping mechanisms in high-$q$ micro and nanomechanical string resonators.
\newblock {\em Phys. Rev. B}, 84:165307, 2011.

\bibitem{pratt22}
J.R. Pratt, A.~Agrawal, C.~Condos, C.~Pluchar, S.~Sclamminger, and D.~Wilson.
\newblock Nanoscale torsional dissipation dilution for quantum experiments and precision measurement.
\newblock {\em American Physical Society}, 13(011018), 2023.

\bibitem{fedorov20}
S.~Fedorov, A.~Beccari, N.~Engelsen, and T.~Kippenberg.
\newblock Fractal-like mechanical resonators with a soft-clamped fundamental mode.
\newblock 124(025502), 2020.

\bibitem{beccari21}
A.~Beccari, M.~Beryehi, R.~Groth, S.~Fedorov, A.~Arabmohegi, T.~Kippenberg, and N.~Engelsen.
\newblock Hierarchical tensile structures with ultralow mechanical dissipation.
\newblock 13(3097), 2021.

\bibitem{fedorov19}
S.~Fedorov, N.~Engelsen, A.~Ghadimi, M.~Bereyhi, R.~Schilling, D.~Wilson, and T.~Kippenberg.
\newblock Generalized dissipation dilution in strained mechanical resonators.
\newblock 99(054107), 2019.

\bibitem{beccari22}
A.~Beccari, D.~Visani, S.~Fedorov, M.~Beryehi, V.~Boureau, N.~Engelsen, and T.~Kippenberg.
\newblock Strained crystalline nanomechanical resonators with quality factors above 10 billion.
\newblock 18:436--441, 2022.

\bibitem{ghadimi17}
A.~H. Ghadimi, D.~Wilson, and T.~Kippenberg.
\newblock Radiation and internal loss engineering of high-stress silicon nitride nanobeams.
\newblock 17:3501--3505, 2017.

\bibitem{gisler22}
T.~Gisler, M.~Helal, D.~Sabonis, U.~Grob, M.~H{\'e}nritier, C.~Degen, A.~Ghadimi, and A.~Eichler.
\newblock Soft-clamped silicon nitride string resonators at millikelvin temperatures.
\newblock 129(104301), 2022.

\bibitem{tsaturyan17}
Y.~Tsaturyan, A.~Barg, E.~Polzik, and A.~Schliesser.
\newblock Ultracoherent nanomechanical resonators via soft clamping and dissipation dilution.
\newblock 12:776--784, 2017.

\bibitem{sigmund2004}
M.P. Bends{\o}e and O.~Sigmund.
\newblock {\em Topology Optimization Theory, Methods and Applications}.
\newblock Springer, 1 edition, 2004.

\bibitem{aage17}
Niels Aage, Erik Andreassen, Boyan~S. Lazarov, and Ole Sigmund.
\newblock Giga-voxel computational morphogenesis for structural design.
\newblock {\em Nature}, 550(7674):84--86, October 2017.

\bibitem{gao20}
Wenjun Gao, Fengwen Wang, and Ole Sigmund.
\newblock Systematic design of high-q prestressed micro membrane resonators.
\newblock {\em Computer Methods in Applied Mechanics and Engineering}, 361:112692, 2020.

\bibitem{hoj21article}
D.~H{\o}j, F.~Wang, W.~Gao, U.~Hoff, O.~Sigmund, and U.~Andersen.
\newblock Ultra-coherent nanomechanical resonators based on inverse design.
\newblock 12(5766), 2021.

\bibitem{thesishendrik}
H.J. Algra.
\newblock Optimizing nanomechanical resonators.
\newblock Master's thesis, Delft University of Technology, 2024.

\bibitem{keulen93}
F.~Van Keulen.
\newblock A geometrically nonlinear curved shell element with constant stress resultants.
\newblock 106(3):315--352, 1993.

\bibitem{pedersen01}
N.L. Pedersen.
\newblock On topology optimization of plates with prestress.
\newblock (51):225--239, 2001.

\bibitem{BRUNS20013443}
Tyler~E. Bruns and Daniel~A. Tortorelli.
\newblock Topology optimization of non-linear elastic structures and compliant mechanisms.
\newblock {\em Computer Methods in Applied Mechanics and Engineering}, 190(26):3443--3459, 2001.

\bibitem{guest04}
J.K.~Guest et~al.
\newblock Achieving minimum length scale in topology optimization using nodal design variables and projection functions.
\newblock 61:238--254, 2004.

\bibitem{stolpe01}
K.~Svanberg M.~Stolpe.
\newblock An alternative interpolation scheme for minimum compliance topology optimization.
\newblock 22:116--124, 2001.

\bibitem{pedersen00}
N.L. Pedersen.
\newblock Maximization of eigenvalues using topology optimization.
\newblock 20:2--11, 2000.

\bibitem{wang14}
F.~Wang, B.~Lazarov, O.~Sigmund, and J.~Jensen.
\newblock Interpolation scheme for fictitious domain techniques and topology optimization of finite strain elastic problems.
\newblock 276:453--472, 2014.

\bibitem{tsai13}
T.~Tsai and C.~Cheng.
\newblock Structural design for desired eigenfrequencies and mode shapes using topology optimization.
\newblock (27):673--686, 2013.

\bibitem{wang11}
O.~Sigmund F.~Wang, B. Stefanov~Lazarov.
\newblock On projection methods, convergence and robust formulations in topology optimization.
\newblock 43:767--784, 2011.

\bibitem{svanberg87}
K.~Svanberg.
\newblock The method of moving asymptotes - a new method for structural optimization.
\newblock 24:359--373, 1987.

\bibitem{arXivSigmund24}
Yincheng Shi, Fengwen Wang, Dennis H{\o}j, Ole Sigmund, and Ulrik~Lund Andersen.
\newblock Topology optimization of high-performance optomechanical resonator, 2024.

\bibitem{Engelsen2021proceeding}
Nils~J. Engelsen, Aman~R. Agrawal, and Dalziel~J. Wilson.
\newblock Ultra-high-q nanomechanics through dissipation dilution: Trends and perspectives.
\newblock In {\em 2021 21st International Conference on Solid-State Sensors, Actuators and Microsystems (Transducers)}, pages 201--205, 2021.

\bibitem{bereyhi19}
M.~J. Bereyhi, A.~Beccari, S.~Fedorov, A.~Ghadimi, R.~Schilling, D.~Wilson, N.~Engelsen, and T.~Kippenberg.
\newblock Clamp-tapering increases the quality factor of stressed nanobeams.
\newblock 19:2329--2333, 2019.

\bibitem{carstensen24}
D.Q. Ha and J.V. Carstensen.
\newblock Automatic hyperparameter tuning of topology optimization algorithms using surrogate optimization.
\newblock {\em Struc Multidisc Optim}, 67(157), 2024.

\end{thebibliography}

\end{document}